\documentclass[12pt, oneside]{article}   	
\usepackage[
  margin=3cm,
  includefoot,
  footskip=30pt,
]{geometry}     

\usepackage{graphicx}
\usepackage{newtxtext}
\usepackage{newtxmath}
\usepackage{float}
\usepackage{natbib}
\usepackage{hyperref}
\usepackage{enumitem}
\hypersetup{
    colorlinks = true,
    urlcolor   = blue,
    citecolor  = black,
}
\usepackage{subfigure}
\usepackage{csquotes}

\usepackage{mathtools,tikz,caption}
\DeclareRobustCommand\sampleline[1]{%
  \tikz\draw[#1] (0,0) (0,\the\dimexpr\fontdimen22\textfont2\relax)
  -- (2em,\the\dimexpr\fontdimen22\textfont2\relax);%
}

\newcommand{\de}[2]{\frac{\partial #1}{\partial #2}}

\newcommand{\RomanNumeralCaps}[1]
\linenumbers

\graphicspath{{IMAGES/}}

\title{\textbf{Assessment of heat transfer and Mach number effects on high-speed turbulent boundary layers}}

\begin{document}
\newgeometry{margin=1in}
\maketitle
\date{\vspace{-3ex}}

{Michele Cogo$^{1}$, Umberto Baù$^{2}$, Mauro Chinappi$^{3}$, Matteo Bernardini$^{4}$, and Francesco Picano$^{1,5}$}\\

\vspace{0.3 cm}

{\small$^{1}$Centro di Ateneo di Studi e Attività Spaziali “Giuseppe Colombo”, via Venezia 15, 35131 Padova, Italy, michele.cogo.1@phd.unipd.it \\$^{2}$Institute of Fluid Mechanics and Heat Transfer, TU Wien, 1060 Vienna, Austria\\
$^{3}$Department of Industrial Engineering, University of Rome Tor Vergata, via del Politecnico 1, 00133 Rome, Italy\\
$^{4}$Department of Mechanical and Aerospace Engineering, Sapienza University of Rome, via Eudossiana 18, 00184 Rome, Italy\\
$^{5}$Department of Industrial Engineering, Università degli Studi di Padova, via Venezia 1, 35131 Padova, Italy}
\vspace{1.5cc}

\begin{abstract}
High-speed vehicles experience a highly challenging environment in which the free-stream Mach number and surface temperature greatly influence aerodynamic drag and heat transfer.
The interplay of these two parameters strongly affects the near-wall dynamics of high-speed turbulent boundary layers in a non-trivial way, breaking similarity arguments on velocity and temperature fields, typically derived for adiabatic cases.
In this work, we present direct numerical simulations of flat-plate zero-pressure-gradient turbulent boundary layers spanning three free-stream Mach numbers [2,4,6] and four wall temperature conditions (from adiabatic to very cold walls), emphasising the choice of the diabatic parameter $\mathit{\Theta}$ \citep{zhang2014generalized} to recover a similar flow organisation at different Mach numbers.
We link qualitative observations on flow patterns to first- and second-order statistics to explain the strong decoupling of temperature-velocity fluctuations that occurs at reduced wall temperatures and high Mach numbers. For these cases, we find that the mean temperature gradient in the near-wall region can reach such a strong intensity that it promotes the formation of a secondary peak of thermal production in the viscous sublayer, which is in direct contrast with the monotonic behaviour of adiabatic profiles.
We propose different physical mechanisms induced by wall-cooling and compressibility that result in apparently similar flow features, such as a higher peak in the streamwise velocity turbulence intensity,  and distinct ones, such as the separation of turbulent scales. 
\end{abstract}
\thispagestyle{empty}
\clearpage
\restoregeometry

\section{Introduction}
The study of highly compressible turbulent boundary layers is of major importance for high-speed turbulence research.
Compressibility acts upon the flow by influencing the mean and fluctuating fields of thermodynamic quantities, which are in turn coupled to the momentum, promoting the energy exchange between kinetic and thermal fields.
This poses several difficulties in the prediction of drag and wall heat transfer, which makes engineering design more and more difficult as higher speeds are attained.

In recent decades, supersonic turbulent boundary layers have been extensively studied and compared to their compressible counterparts at the same Reynolds number, mainly focusing on the prediction of drag, e.g. \citet{bernardini2011inner,wenzel2018dns,Duan2011}. In fact, at supersonic speeds the wall temperature can be considered for practical purposes very close to the recovery temperature of the flow, implying a very low heat exchange at the wall.
The recovery temperature indicates the temperature that is attained by the flow when it is brought to rest in a non-isentropic manner, defined as
\begin{equation}
    T_r=T_{\infty}\left(1+r \frac{\gamma-1}{2}M_{\infty}^2\right),
\end{equation}
being $r = Pr^{1/3}$ the recovery factor \citep{Zhang2018}, where $Pr$ is the Prandtl number. 
However, in hypersonic boundary layers, the recovery temperatures are so high that the wall temperature is usually lower \citep{urzay2021engineering}, generating large heat fluxes to the wall.
This affects the flow dynamics in concurrency with the Mach number, enriching the physical effects that have to be accounted for when developing theoretical relations and reduced order models.

A renewed interest in hypersonic flight, along with the computational advancements that render Direct Numerical Simulations more feasible, sparked the attention on these problems, e.g. \citet{Zhang2018,wenzel2022influences}, but there is still a lack of understanding of the individual effect of different flow parameters.

The framework of theoretical relations applied to compressible flow for mean velocity and fluctuating fields aims at mapping compressible profiles onto incompressible reference by taking into account variations of mean properties such as density and viscosity.
When applied to the mean velocity field, these relations are called compressibility transformations, first introduced by \citet{van1956problem} by accounting for mean density variations in the wall-normal velocity profile.
Among the plethora of relations proposed in recent years, \citet{volpiani2020data} and \citet{griffin2021velocity} stand out as capable of efficiently collapsing velocity profiles even at high Mach numbers.
While \citet{volpiani2020data} uses a mixed physical and data-driven approach to determine the optimal parameters that define the weight of density and viscosity, \citet{griffin2021velocity} base their arguments on the total-stress equation, allowing for separate assumptions for the viscous sublayer and the log layer. 

Theoretical relations have also been derived to describe the interaction between kinetic and thermal fields, classically referred to as Strong Reynolds Analogy (SRA).
First proposed by \citet{morkovin1962effects}, SRA establishes a framework based on the similarity between the momentum and total enthalpy equations, from which a direct proportionality between velocity and total enthalpy can be inferred. 
Under the more restrictive condition of wall adiabaticity, a set of relations coupling velocity and temperature can be derived for both mean and fluctuating fields, in which the temperature resembles a passive scalar field. 

These relations have been extensively validated for adiabatic TBLs at different Mach numbers, e.g. \citet{bernardini2011inner,wenzel2018dns}, although at hypersonic speeds ($M_{\infty}>5$) discrepancies start to arise \citep{Zhang2018}.
Subsequent extensions of the SRA accounting for diabatic walls have been recently proposed (e.g. \citet{zhang2014generalized}), which obtained promising results for different flow conditions, even when thermochemical effects are present \citep{passiatore2021,passiatore2022thermochemical,DiRenzo2021}.

A cold wall imposes a change in the sign of the mean temperature gradient near the wall, affecting the production of temperature fluctuations, which may result in a severe loss of similarity between velocity and temperature fields, a building block of SRA, clearly visible in instantaneous snapshots of turbulent structures \citep{cogo2022direct,wan2022wall}.
However, these studies also noted that comparing cases with different Mach numbers at a fixed wall-to-recovery temperature ratio $T_w/T_r$ ($<1$) resulted in vastly different near-wall dynamics for temperature fluctuations, in a way that cold cases at high $M_{\infty}$ seemed \enquote{more adiabatic} than their low $M_{\infty}$ counterparts.
Recently, other definitions of the wall temperature condition have been proposed, such as the Eckert number $Ec=(\gamma-1)M_{\infty}^2 T_{\infty}/(T_r-T_w)$ \citep{wenzel2022influences} or the diabatic parameter $\mathit{\Theta}=(T_w-T_{\infty})/(T_r-T_{\infty})$ \citep{zhang2014generalized}, which are capable to account for the indirect effect of Mach number on the wall temperature condition. 
Although progress has been made to incorporate the effects of compressibility and heat transfer on these relations, their individual influence is still not well understood.
While compressibility effects induced by the increase in Mach number can be similar to a change in wall temperature condition (and vice versa) for certain mechanisms, such as redistribution of turbulent kinetic energy \citep{duan2011direct}, their relative role is still unclear in other aspects, such as separations of turbulent scales \citep{huang2022direct}. 
In this regard, wall-cooling has been shown to reduce the separation between the large and small turbulence scales in hypersonic flows \citep{huang2022direct,fan2022energy}, but the specific role of the Mach number is still debated.
Furthermore, while wall-cooling has been shown to be dominant in regulating energy exchanges in the near-wall region \citep{fan2022energy}, the effect of the Mach number is still not clear.
These and other authors called upon the need for additional computations to determine their individual effects.

The aim of this study is to unveil the physical mechanisms that yield similarities and differences between the effect of compressibility and wall-cooling.
To pursue this objective, an extensive DNS database consisting of 12 simulations of zero-pressure-gradient TBLs has been computed fixing the friction Reynolds number ($Re_\tau=443$), while spanning three Mach numbers $M_{\infty} = [2,4,6]$ and four diabatic parameters $\mathit{\Theta}=[0.25,0.5,0.75,1.0]$, going from extremely cold walls, $\mathit{\Theta}=0.25$, to adiabatic case, $\mathit{\Theta}=1$.
The computed database is discussed in the present paper with the aim to be used for the development of simplified models for high-speed wall-bounded flows with strong heat flux.

The remainder of the paper is organised as follows.
The numerical method and details on the simulation setup are outlined in section \S \ref{sec:setup}.
In section \S \ref{sec:slice}, a general visualisation of instantaneous velocity and temperature fields is given, describing the individual effect of Mach and wall temperature conditions on the flow dynamics and turbulent structures.
Then, first-order statistics for mean velocity and temperature are presented in section \S \ref{sec:mean}, which also compares different SRA formulations.
Finally, second-order statistics are presented in section \S \ref{sec:fluc}, focusing on the effect of wall-cooling on thermal production, and its implications on velocity-temperature correlations.

\section{Simulation parameters and computational setup}\label{sec:setup}
The three-dimensional compressible Navier-Stokes equations are numerically solved for a viscous, heat-conducting gas
\begin{equation}\label{eq:n-s}
    \begin{split}
        \de{\rho}{t}+\de{(\rho u_j)}{x_j}=0,\\
        \de{(\rho u_i)}{t}+\de{(\rho u_i u_j)}{x_j}+\de{p}{x_i}-\de{\sigma_{ij}}{x_j}=0,\\
        \de{(\rho E)}{t}+\de{(\rho E u_j+pu_j)}{x_j}-\de{(\sigma_{ij}u_i-q_j)}{x_j}=0,
    \end{split}
\end{equation}
where $\rho$ is the density, $u_i$ denotes the velocity component in the \textit{i}th Cartesian direction ($i=1,2,3$), $p$ is the thermodynamic pressure, $E=c_v T+u_i u_i/2$ the total energy per unit mass and
\begin{equation}
 \sigma_{ij}=\mu \left( \de{u_i}{x_j}+\de{u_j}{x_i}-\frac{2}{3}\de{u_k}{x_k}\delta_{ij}\right), \qquad
 q_j=-k\de{T}{x_j}
\end{equation}
represents the viscous stress tensor and the heat flux vector, respectively. The molecular viscosity $\mu$ is assumed to follow Sutherland's law 
\begin{equation}
    \frac{\mu}{\mu_{\infty}}=\left(\frac{T}{T_{\infty}}\right)^{1/2} \frac{1+C/T_{\infty}}{1+C/T},
\end{equation}
where $C=110.4$ K and $T_{\infty}=220.0$ K, representing the typical conditions that are met in the stratosphere. 
The thermal conductivity $k$ is related to the viscosity by the expression $k=c_p \mu/Pr$, where $c_p$ is the specific heat at constant pressure and the Prandtl number is $Pr=0.72$.
The thermodynamic variables are correlated to each other by means of the equation of state for a calorically perfect gas. 
This choice was also assumed for cases at $M_{\infty}=6$, after having verified that by introducing a dependence of specific heat with temperature $c_p=f(T)$ differences in all statistics were negligible.
Moreover, gas dissociation effects are also not expected in the present database, according to the observations of \citet{passiatore2022thermochemical} which observed negligible effects with $T_w=1800K$ (our highest imposed value is $T_w=1640 K$ for $M_{\infty}=6$).
The system of equations is solved on a Cartesian grid using the in-house code STREAmS \citep{bernardini2021streams,bernardini2023streams}, which has been extensively validated in numerous canonical configurations \citep{bernardini2011wall,bernardini2011wallp,cogo2022direct}.
Convective terms are discretised using high-order, energy-preserving schemes applied in shock-free regions, while a high-order shock capturing scheme (WENO) is applied when shock waves are identified by the Ducros sensor \citep{ducros1999large}. 
Diffusive terms are discretised using a locally conservative formulation \citep{de2021high}, expanded to Laplacian to ensure finite molecular dissipation at all resolved wavelengths.
The solver takes advantage of a multi-GPU architecture by means of the CUDA Fortran paradigm.
The domain is a rectangular box of length $L_x=100 \, \delta_{in}$, $L_y=15 \, \delta_{in}$, $L_z=9 \, \delta_{in}$, where $\delta_{in}$ is the boundary layer thickness at the inflow station, based on the $99\%$ of the freestream velocity $u_{\infty}$ (which is referred for other stations as $\delta_{99}$).
For each spatial direction, the number of computational points employed for all cases is $N_x=5120$,  $N_y=320$, and $N_z=512$. 
Periodic boundary conditions are enforced in the spanwise direction, purely non-reflecting boundary conditions are employed for the outflow and the top boundary, and unsteady characteristic boundary conditions are used at the bottom wall \citep{poinsot1992boundary},
where the isothermal wall temperature condition is enforced.
The recycling-rescaling procedure~\citep{pirozzoli2010direct} is applied at the inflow to reach a fully developed state, the recycling length being placed at a distance of $80 \, \delta_{in}$ from the inlet, ensuring a complete decorrelation of the fluctuations between the recycling station and the inflow plane~\citep{morgan2011improving}.

Table~\ref{Tab:table1} summarises the flow conditions and grid resolutions for each run, where $M_{\infty}$
is the free-stream Mach number and $Re_{\tau}$ is the friction Reynolds number, defined as the
ratio between the boundary layer thickness $\delta_{99}$ and the viscous length scale $\delta_{\nu}=\nu_w/u_{\tau}$,
where $u_{\tau}=\sqrt{\tau_w/\rho_w}$ is the friction velocity, $\tau_w$ is the mean wall shear stress, and $\nu_w$ is the kinematic viscosity at the wall.
$\Delta x^+=\Delta x / \delta_{\nu}$ and $\Delta z^+=\Delta z / \delta_{\nu}$  are the uniform grid spacings in the streamwise and spanwise directions
and $\Delta y^+=\Delta y / \delta_{\nu}$ represents the non-uniform wall-normal grid spacing (the wall and edge values are reported). 
In the wall-normal direction, the stretching function of \citet{pirozzoli2010direct} is employed, which provides a more favourable scaling of the number of grid points with the Reynolds number. Furthermore, this function has the natural property of yielding precisely constant resolution in terms of the local Kolmogorov length scale $\eta$ in the outer part of the wall layer while maintaining a uniform near-wall spacing.\\
%
The present database is composed of a total of 12 simulations, spanning three Mach numbers $M_{\infty} = [2,4,6]$ and four diabatic parameters $\mathit{\Theta}=[0.25,0.5,0.75,1.0]$ (see Table \ref{Tab:table1}).
As discussed previously, the way in which the wall temperature condition is defined is of great importance to discern it from the Mach number effect.
However, a formulation in which $T_w$ is not a function of $T_r$ (which is, in turn, a quadratic function of $M_\infty$) is not possible, because $T_r$ is the value reached at the wall under adiabatic conditions ($T_w=T_r$).
Therefore, the goal of a suitable parameter is not to be independent of $M_\infty$, but to incorporate it in order to have \enquote{the same integral behaviour between different cases, regardless of whether its variation is caused by the change of the Mach number or of the wall temperature} \citep{wenzel2022influences}.
This is the rationale with which \citet{wenzel2022influences} argued that the Eckert number $Ec=(\gamma-1)M_{\infty}^2 T_{\infty}/(T_r-T_w)$ represents a more suitable option than the conventional $T_w/T_r$ ratio.
The Eckert number happens to be directly related to the diabatic parameter $\mathit{\Theta}=(T_w-T_{\infty})/(T_r-T_{\infty})$ proposed by \citet{zhang2014generalized}, since it can be shown that $Ec = 2/\left[r(1-\mathit{\Theta})\right]$.
This parameter shows more clearly the improvement over the conventional ratio $T_w/T_r$, showing that $T_{\infty}$ needs to be subtracted from both $T_w$ and $T_r$ to compare only the $\Delta T$ that is recovered when the flow is brought at rest, being the only one responsible for kinetic-internal energy exchanges.
In this study, we choose to use the diabatic parameter $\mathit{\Theta}$ over $Ec$ given its simplicity, but we also report the latter in Table \ref{Tab:table1}.
Table~\ref{tab:stat} summarises the boundary layer parameters at selected locations where turbulence statistics are gathered.

Throughout this study, we use the symbols $u$, $v$, and $w$ to denote the streamwise, wall-normal and spanwise velocity components and the decomposition of any variable is conducted
using either the standard Reynolds decomposition ($f=\bar{f}+f'$) or the density-weighted (Favre) representation ($f=\tilde{f}+f^{''}$), being $\tilde{f}=\overline{\rho f}/\bar{\rho}$. Here, the averaging is conducted using multiple samples and along the periodic direction $z$.

\begin{table}  
\centering
\begin{tabular}{ccccccccc}
\hline 
\hline
Run & $M_{\infty}$ & $Re_{\tau}$& $\mathit{\Theta}$ & $T_w /T_r$ & Ec  & $\Delta x^+$ &$\Delta y^+_{w,edge}$ & $\Delta z^+$ \rule{0pt}{2.6ex} \rule[-1.2ex]{0pt}{0pt}\\
\hline
M2T025 & 2.00 & 436 -- 579 & 0.25  & 0.69  &2.975    & 4.51& 0.71-4.64 & 4.07         \rule{0pt}{2.6ex}\\
M2T050 & 2.00 & 427 -- 564 & 0.5   & 0.79  &4.463    & 4.52 & 0.71-4.64 & 4.07         \rule{0pt}{2.6ex}\\
M2T075 & 2.00 & 424 -- 561 & 0.75  & 0.9   &8.926    & 4.53 & 0.71-4.65 & 4.08        \rule{0pt}{2.6ex}\\
M2T100 & 2.00 & 415 -- 584 & 1.0   & 1.0   &$\infty$ & 4.52&0.71-4.64 & 4.07    \rule{0pt}{2.6ex}\\
\hline
M4T025 & 4.00 & 404 -- 535 & 0.25  & 0.44  &2.975    & 4.36&0.68-4.52 & 3.93       \rule{0pt}{2.6ex}\\
M4T050 & 4.00 & 391 -- 521 & 0.5   & 0.63  &4.463    & 4.38& 0.68-4.53 & 3.94       \rule{0pt}{2.6ex}\\
M4T075 & 4.00 & 379 -- 507 & 0.75  & 0.81  &8.926    & 4.37&0.68-4.53 & 3.94       \rule{0pt}{2.6ex}\\
M4T100 & 4.00 & 371 -- 494 & 1.0   & 1.0.  &$\infty$ & 4.38 &0.68-4.53 &3.95     \rule{0pt}{2.6ex}\\
\hline
M6T025 & 6.00 & 376 -- 500 & 0.25  & 0.35  &2.975    & 4.24 & 0.66-4.42 & 3.82       \rule{0pt}{2.6ex}\\
M6T050 & 6.00 & 351 -- 470 & 0.5   & 0.57  &4.463    & 4.21&0.66-4.40 &3.80       \rule{0pt}{2.6ex}\\
M6T075 & 6.00 & 343 -- 462 & 0.75  & 0.78  &8.926    & 4.24&0.66-4.43 &3.83        \rule{0pt}{2.6ex}\\
M6T100 & 6.00 & 337 -- 451 & 1.0   & 1.0   &$\infty$ & 4.26&0.67-4.44 &3.84       \rule{0pt}{2.6ex}\\
\hline
\hline
\end{tabular}
\caption{Summary of parameters for DNS study.
	Grid spacings are given in wall-units according to the stations selected in table \ref{tab:stat}. The values of $\Delta y^+_{w}$ and $\Delta y^+_{edge}$ refer to the wall-normal spacing at the wall and at the boundary layer edge, respectively. The range of $Re_{\tau}$ is representative of the statistical growth of the boundary layer's thickness along $x$. \label{Tab:table1}}
\end{table}

\begin{table}  
\centering
\begin{tabular}{cccccccccc}
\hline 
\hline
Station &  $Re_{\tau}$& $Re_{\theta}$ & $Re_{\delta2}$ & $Re_{\tau}^*$  &$\delta^*/\delta$ & $\theta/\delta$&$H$ &$-B_q (\cdot 10^{-2})$ & $C_f(\cdot 10^{-3})$ \rule{0pt}{2.6ex} \rule[-1.2ex]{0pt}{0pt}\\
\hline
M2T025 &  443 & 1179  & 1043  &541    &  0.210 &  0.092& 2.291  &2.29 &  3.40      \rule{0pt}{2.6ex}\\
M2T050 &  443 & 1404  & 1155  &645    &  0.224 &  0.088& 2.528  &1.34 &  3.16     \rule{0pt}{2.6ex}\\
M2T075 &  443 & 1613  & 1169  &750    &  0.234 &  0.085& 2.756  &0.57 &  2.99   \rule{0pt}{2.6ex}\\
M2T100 &  443 & 1848  & 1239  &856    &  0.246 &  0.083& 2.979  &0.08   &  2.79  \rule{0pt}{2.6ex}\\
\hline                                                           
M4T025 &  443 & 1607  & 1101  &825    &  0.314 &  0.067& 4.669  &6.10 &  2.19   \rule{0pt}{2.6ex}\\
M4T050 &  443 & 2331  & 1267  &1228   &  0.331 &  0.061& 5.460  &3.07 &  1.85   \rule{0pt}{2.6ex}\\
M4T075 &  443 & 3090  & 1429  &1638   &  0.355 &  0.056& 6.297  &1.17 &  1.61   \rule{0pt}{2.6ex}\\
M4T100 &  443 & 3718  & 1530  &2024   &  0.367 &  0.052& 7.030  &0.17  &  1.43  \rule{0pt}{2.6ex}\\
\hline                                                           
M6T025 &  443 & 2169  & 1156  &1273   &  0.399 &  0.048& 8.279  &8.89 &  1.39  \rule{0pt}{2.6ex}\\
M6T050 &  443 & 3586  & 1449  &2103   &  0.426 &  0.043& 9.813  &4.05 &  1.09   \rule{0pt}{2.6ex}\\
M6T075 &  443 & 4835  & 1629  &2951   &  0.443 &  0.039& 11.485 &1.49 &  0.93     \rule{0pt}{2.6ex}\\
M6T100 &  443 & 5893  & 1764  &3701   &  0.454 &  0.036& 12.800 &0.16  &  0.81    \rule{0pt}{2.6ex}\\
\hline
\hline
\end{tabular}
\caption{Boundary layer properties averaged at the selected station. $Re_{\tau}=\rho_w u_{\tau} \delta/\mu_w$; $Re_{\theta}=\rho_{\infty}u_{\infty}\theta/\mu_{\infty}$; $Re_{\delta_2}=\rho_{\infty}u_{\infty}\theta/\mu_{w}$; $Re_{\tau}^*=\sqrt{\rho_{\infty} \tau_w}\delta/\mu_{\infty}$; $H=\delta^*/\theta$ ($\delta^*$ and $\theta$ are the boundary layer displacement and momentum thickness, respectively). $B_q=q_w/(\rho_w C_p u_{\tau}T_w)$ and $C_f=\tau_w/(1/2 \rho_{\infty} u_{\infty}^2)$ are the nondimensional wall heat transfer $q_w=-\bar{k} \ \partial \tilde{T}/\partial y$ and skin friction coefficient $\tau_w = \bar{\mu} \ \partial \tilde{u} / \partial y$, respectively.\label{tab:stat}}
\end{table}


\section{Instantaneous visualisation} \label{sec:slice}
\begin{figure}  
	\centering
	\subfigure[PARAMETRI-1][$M_{\infty}=2, \mathit{\Theta}=0.25$]{\includegraphics[width=0.48\textwidth]{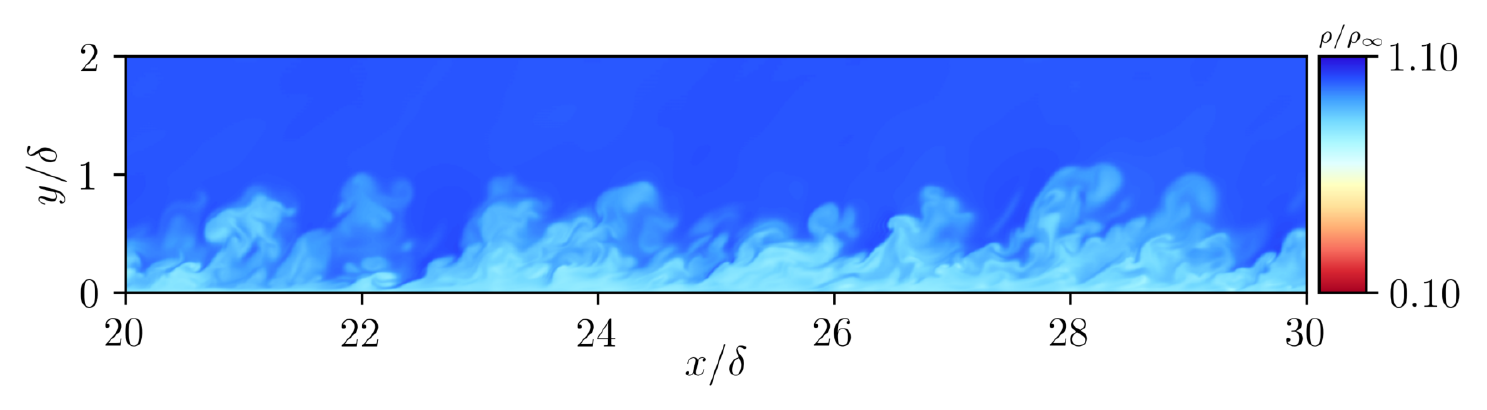}}
	\subfigure[PARAMETRI-1][$M_{\infty}=2, \mathit{\Theta}=1.0 $]{\includegraphics[width=0.48\textwidth]{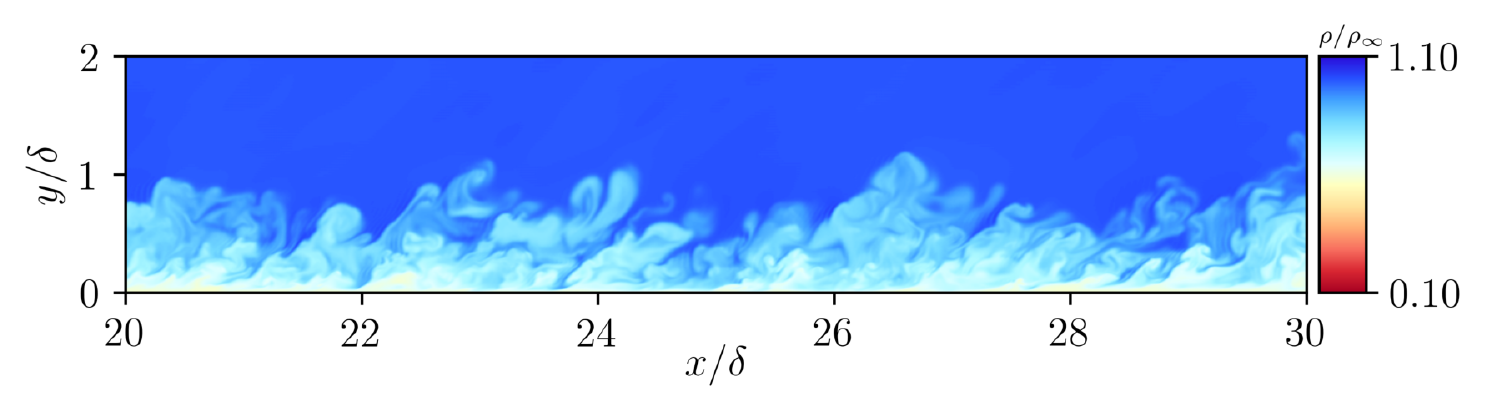}}
	\subfigure[PARAMETRI-1][$M_{\infty}=4, \mathit{\Theta}=0.25$]{\includegraphics[width=0.48\textwidth]{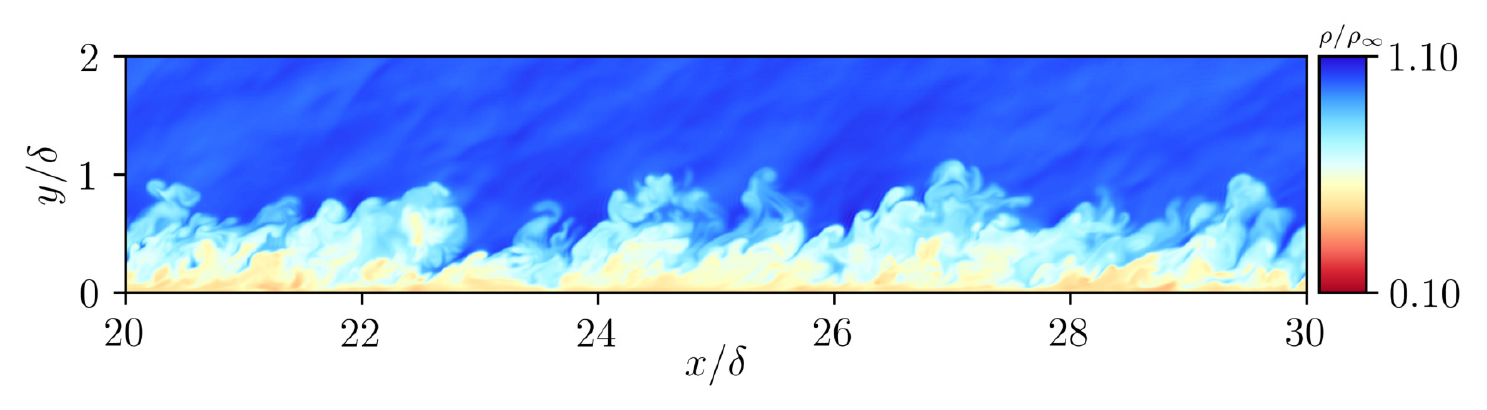}}
	\subfigure[PARAMETRI-1][$M_{\infty}=4, \mathit{\Theta}=1.0 $]{\includegraphics[width=0.48\textwidth]{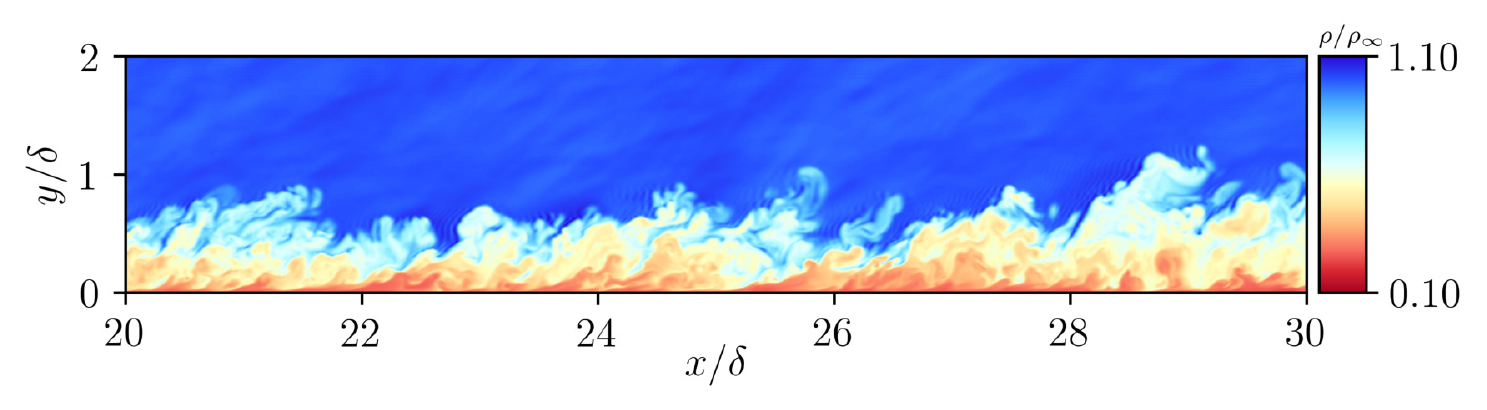}}
	\subfigure[PARAMETRI-1][$M_{\infty}=6, \mathit{\Theta}=0.25$]{\includegraphics[width=0.48\textwidth]{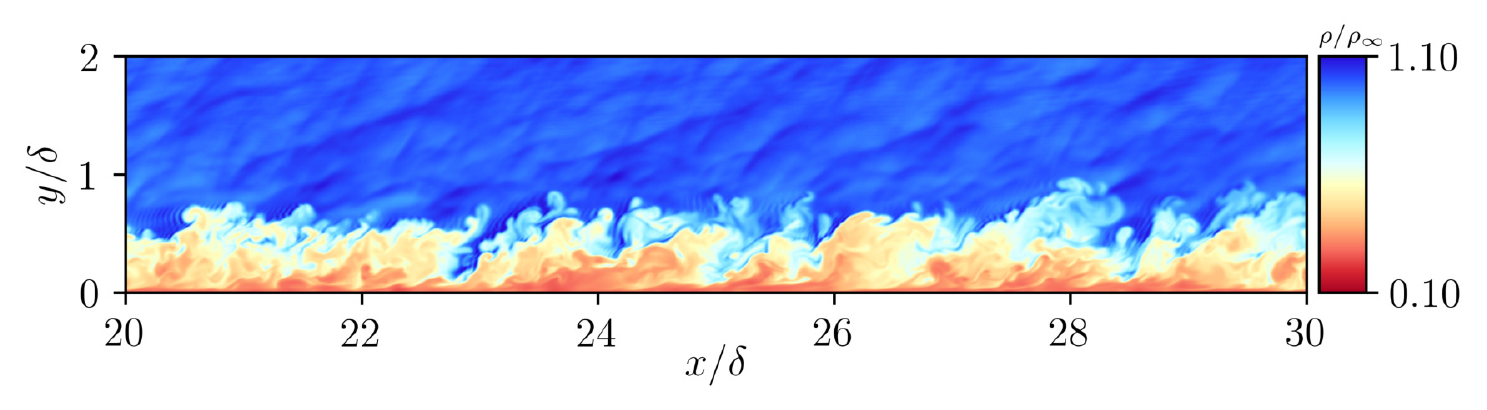}}
	\subfigure[PARAMETRI-1][$M_{\infty}=6, \mathit{\Theta}=1.0 $]{\includegraphics[width=0.48\textwidth]{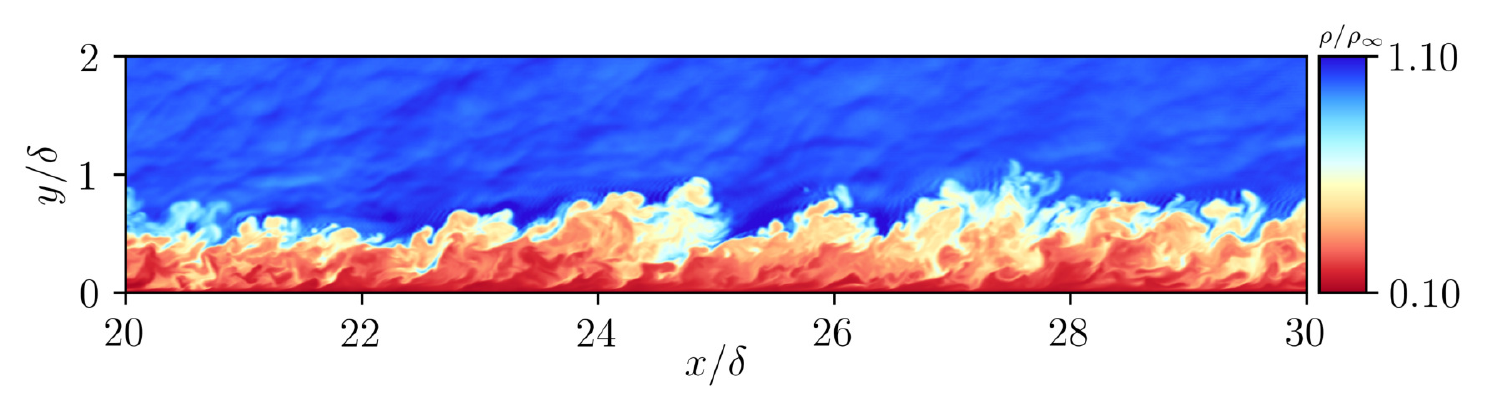}}
	\vspace{-0.2cm}
  \caption{Instantaneous density in wall-normal slices ($x$-$y$ plane), with a window size of $\Delta x =20\delta-30 \delta$ and $\Delta y =0\delta-2 \delta$. Here, all Mach numbers are shown while the two extremes are chosen with regard to wall-cooling ($\mathit{\Theta}=0.25$ and $\mathit{\Theta}=1.0$).\label{fig:slicexy}}
\end{figure}

To highlight the emerging features of the flow in a qualitative way, we selected the two extreme cases with regard to the wall-cooling condition, $\mathit{\Theta}=0.25$ and $\mathit{\Theta}=1.0$, for each Mach number in our database.  Figure \ref{fig:slicexy} shows a portion of wall-normal $x$-$y$ planes coloured with the instantaneous density, whose variability is a clear sign of the degree of compressibility. The effect of Mach number is clearly apparent for all cases moving from top to bottom with a decrease of the minimum value of density and an increase of the general level of acoustic disturbances, generated in the boundary layer and emanated towards the far field.
However, a stronger wall-cooling (lower $\mathit{\Theta}$, left column of figure \ref{fig:slicexy}) attenuates this effect, since lower wall temperatures generate higher density fields in the near wall region.

\begin{figure}  
	\centering
  \subfigure{\includegraphics[width=0.48\textwidth]{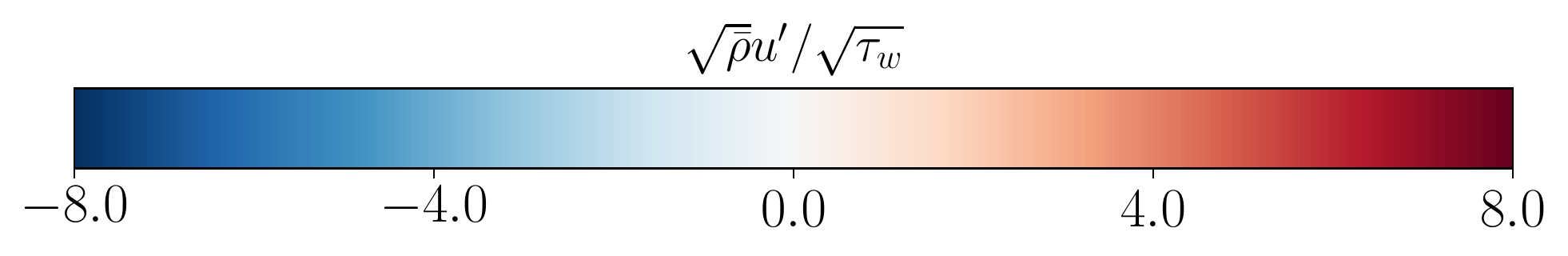}}
  \subfigure{\includegraphics[width=0.48\textwidth]{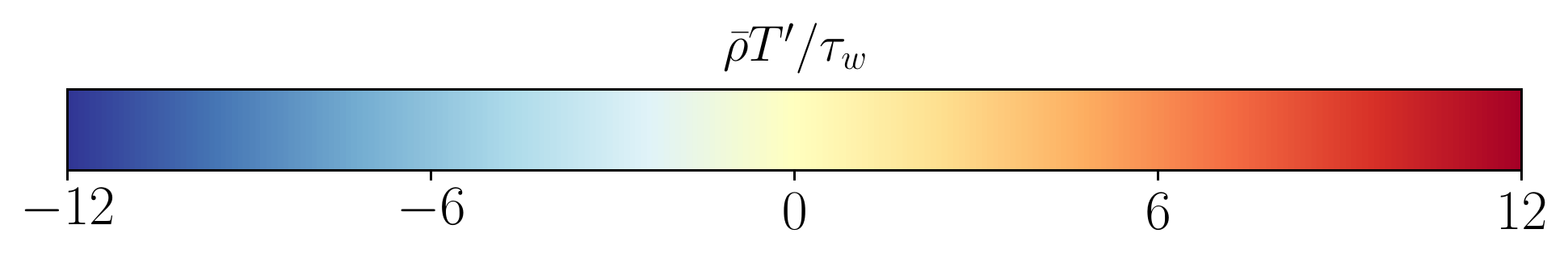}}
	\subfigure[PARAMETRI-1][$M_{\infty}=2, \mathit{\Theta}=0.25$]{\includegraphics[width=0.49\textwidth]{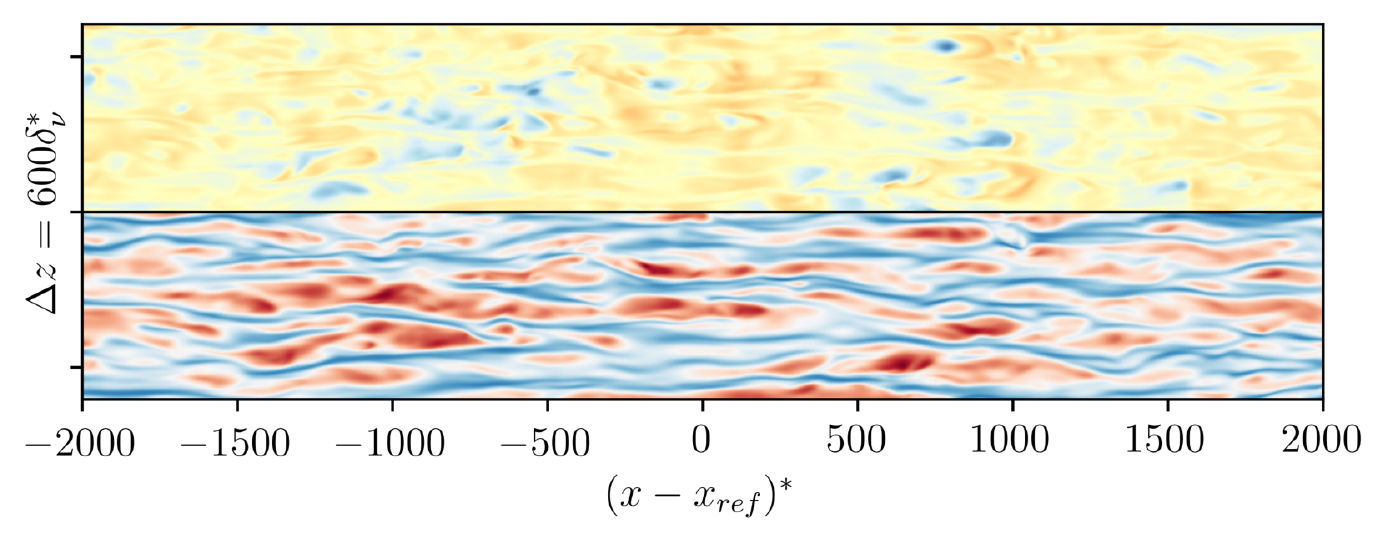}}
	\subfigure[PARAMETRI-1][$M_{\infty}=2, \mathit{\Theta}=1.0 $]{\includegraphics[width=0.49\textwidth]{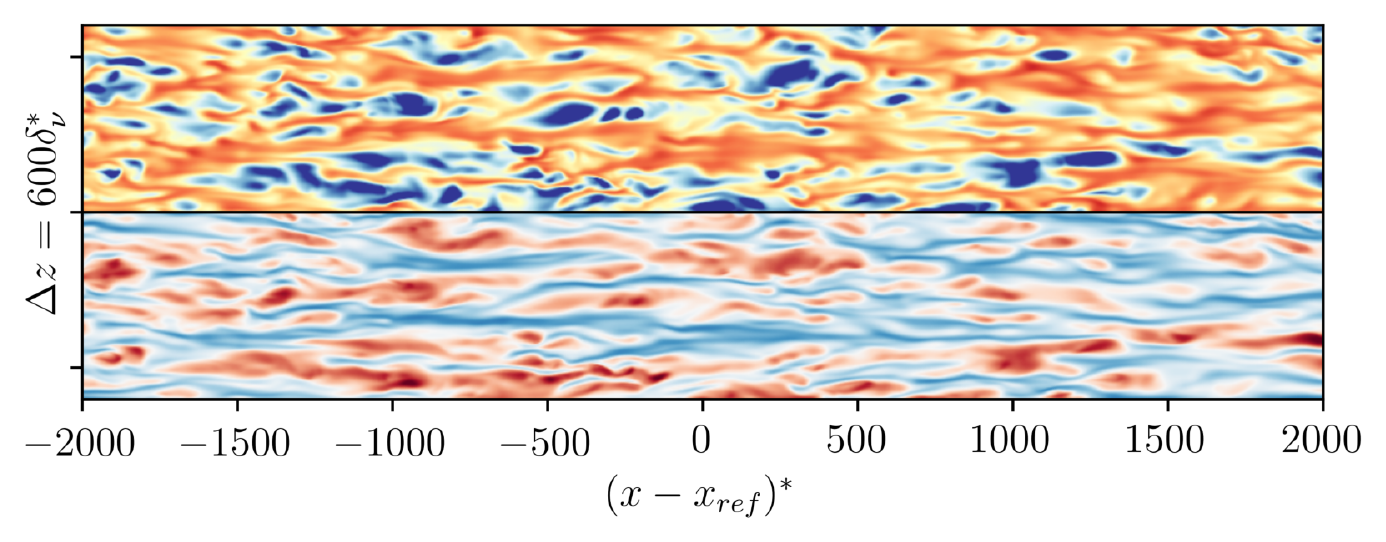}}
	\subfigure[PARAMETRI-1][$M_{\infty}=4, \mathit{\Theta}=0.25$]{\includegraphics[width=0.49\textwidth]{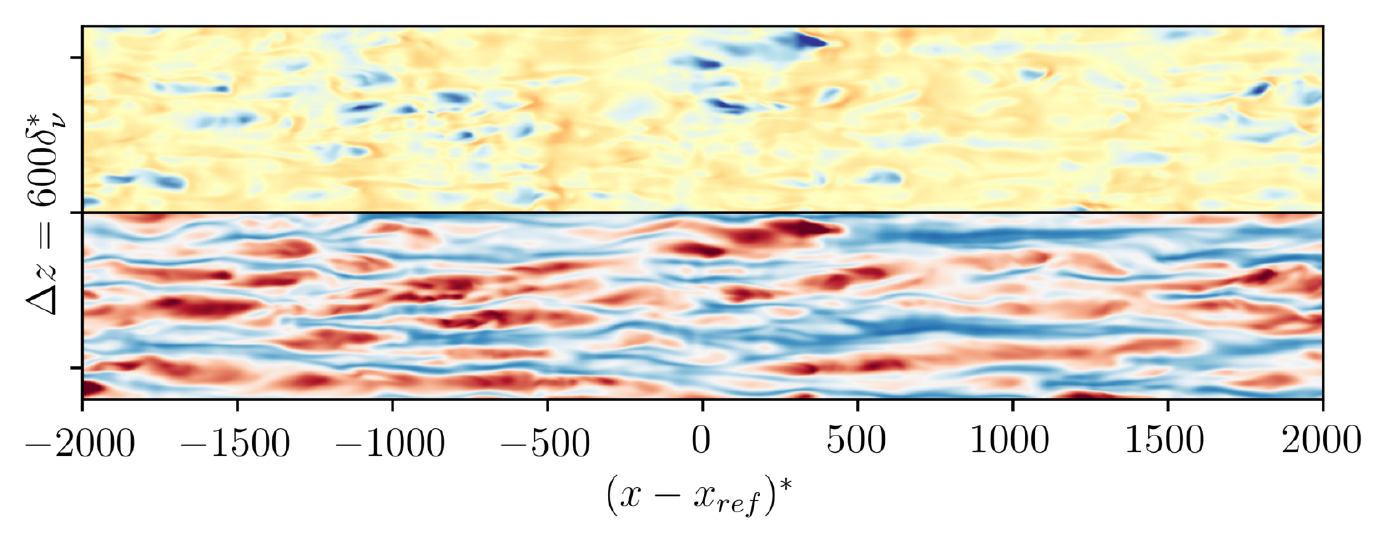}}
	\subfigure[PARAMETRI-1][$M_{\infty}=4, \mathit{\Theta}=1.0 $]{\includegraphics[width=0.49\textwidth]{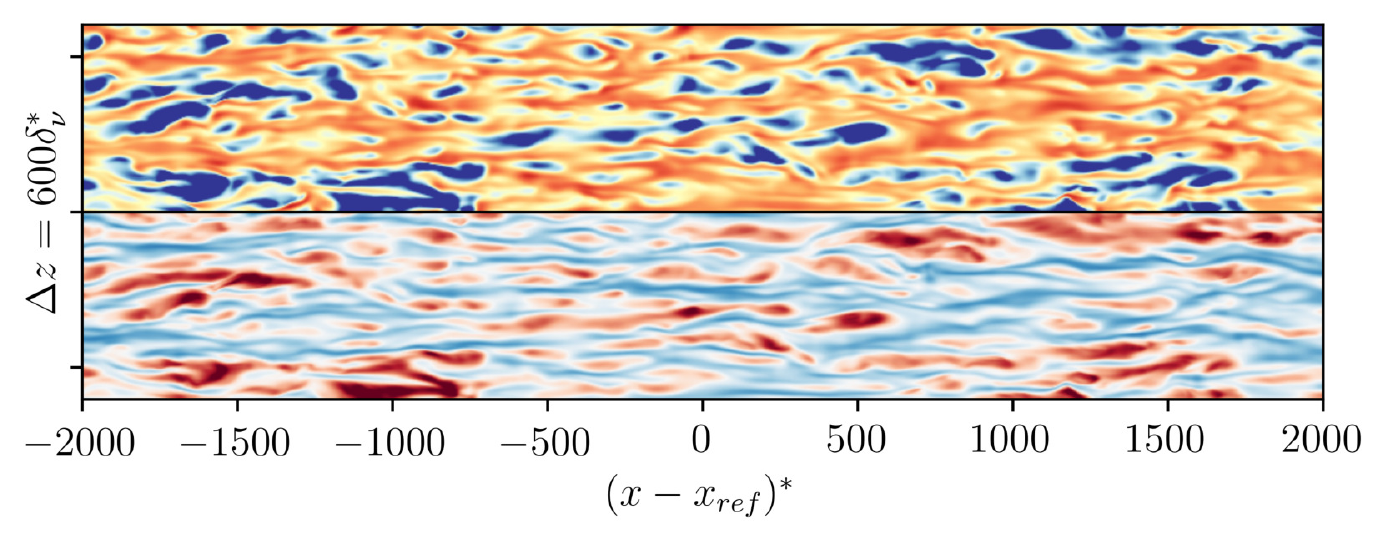}}
	\subfigure[PARAMETRI-1][$M_{\infty}=6, \mathit{\Theta}=0.25$\label{fig:xzM6025}]{\includegraphics[width=0.49\textwidth]{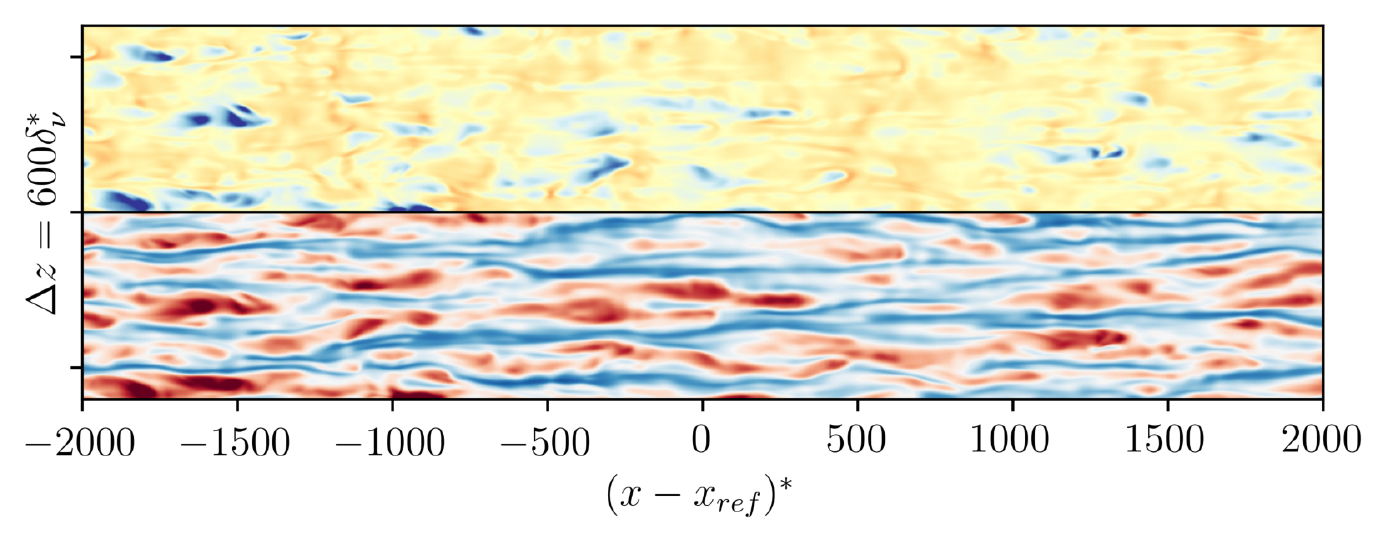}}
	\subfigure[PARAMETRI-1][$M_{\infty}=6, \mathit{\Theta}=1.0 $\label{fig:xzM6100}]{\includegraphics[width=0.49\textwidth]{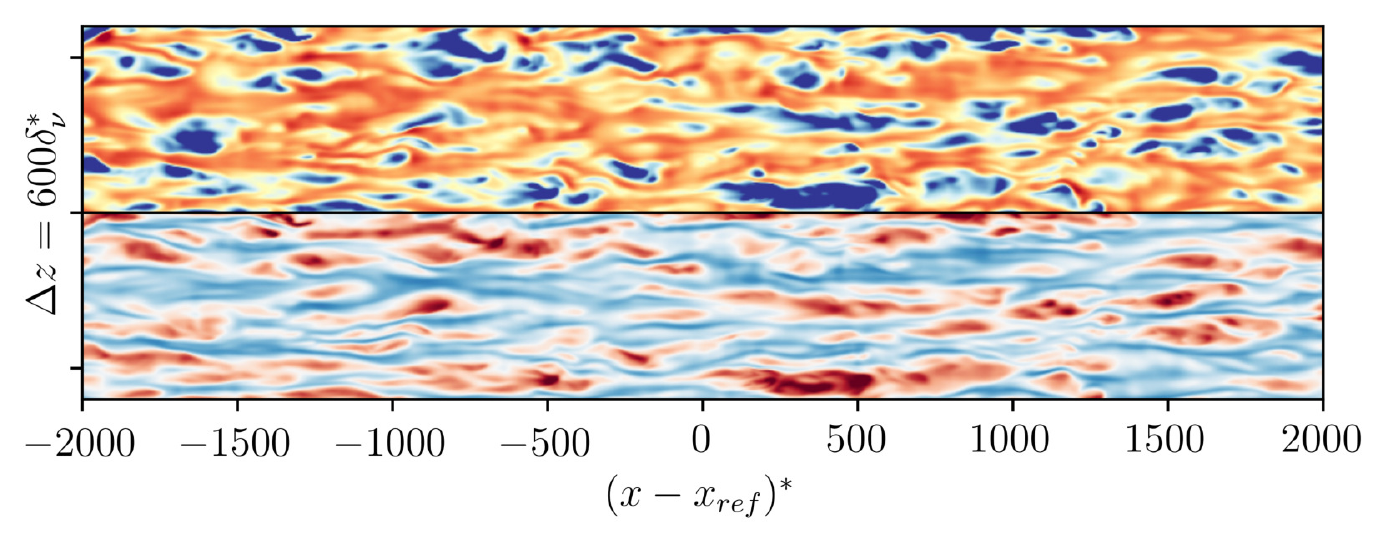}}
	\vspace{-0.2cm}
  \caption{Temperature fluctuations $\bar{\rho} T'/ \tau_w$ (top) and streamwise velocity fluctuations $\sqrt{\bar{\rho} } u'/ \sqrt{\tau_w}$ (bottom) in wall-parallel slices ($x$-$z$ plane) selected at $y^*\approx10$. Here, all Mach numbers are shown while the two extremes are chosen with regard to wall-cooling ($\mathit{\Theta}=0.25$ and $\mathit{\Theta}=1.0$). Here, $x_{ref}$ is the streamwise location of the selected station. \label{fig:slice_xz}}
\end{figure}

The intensity of wall-cooling strongly affects the coupling between velocity and temperature fluctuations, especially in the near-wall region. This is apparent in figure \ref{fig:slice_xz}, which compares these quantities in wall-parallel slices located at approximately $y^*\approx 10$, representing the onset turbulence activity after the viscous sub-layer.
Here, $y^*=y/ \delta_{\nu,SL}$ is the semilocal scaled wall-normal coordinate, with $\delta_{\nu,SL}= \bar{\nu}/\sqrt{\tau_w/\bar{\rho}}$.
The chosen $x$-$z$ planes are centred at the selected stations of table \ref{Tab:table1} spanning a window of $\Delta x^*=4000$ and $\Delta z^*=600$.
Velocity fluctuations $\sqrt{\bar{\rho}} u^{'}/\sqrt{\tau_w}$ are scaled according to the Morkovin's transformation \citep{morkovin1962effects} (also used in section \ref{sec:velfluc}), which enables comparison across different Mach numbers and wall temperature values by accounting for the variation of the mean properties of the flow. In other words, velocity fluctuations are scaled by the semilocal friction velocity $u_{\tau,SL}=\sqrt{\tau_w/\bar{\rho}}$, which differs from the conventional one by employing the mean density $\bar{\rho}$ instead of the wall density $\rho_w$.
Temperature fluctuations $\bar{\rho} T^{'}/ (R \tau_w)$ are scaled in a similar fashion, assuming $\tau_w$ a proper parameter to scale pressure fluctuations, then $\tau_w/(R \bar{\rho})$ can be used to scale temperature (for further details refer to section \ref{sec:thermo_fluc}).
A general look at the velocity fluctuations shows the presence of near-wall streaks for all cases, representative of the near-wall self-sustaining cycle of turbulence.
Similar values of intensities appear across all cases. This result is not shared with temperature fluctuations, where cold cases ($\mathit{\Theta}<1$) show reduced intensity and a clear breakdown of elongated streaks, appearing more isotropic when compared to their adiabatic counterpart. Adiabatic cases maintain a streaky pattern, which shows a clear coupling with the velocity field. 
Although this behaviour will be further discussed in the following sections by analysing temperature fluctuations and thermal production profiles, these qualitative results are consistent with the discussion of \citet{wenzel2022influences}, which states that the same general behaviour due to the effect of wall-cooling is expected when comparing flows with the same Eckert number (or diabatic parameter $\mathit{\Theta}$).

\section{Mean flow statistics}\label{sec:mean}
In this section, we present the wall-normal profiles of averaged quantities such as velocity and temperature, selected at stations listed in table \ref{tab:stat}. We consider mean velocity profiles in the framework of compressibility transformations, which aim at incorporating compressibility effects in wall-bounded flow statistics in order to recover the incompressible behaviour. 
Since the pioneering work of \citet{van1951turbulent}, several relations have been proposed to account for the variations of mean fluid properties, such as density and viscosity. These relations can be cast in terms of mapping functions $f_I$ and $g_I$ for wall distance $y_I$ and mean velocity $u_I$, which denote the equivalent incompressible distributions obtained from the transformation $I$:
\begin{equation}\label{eq:trasf}
    y_{I}=\int_0^yf_I dy ,\quad     \quad u_I=\int_0^{\Tilde{u}} g_I d\Tilde{u}.
\end{equation}
Table \ref{Tab:table_trasf} shows the relative values of $f_I$ and $g_I$ for \citet{van1951turbulent} and the recent transformation of \citet{volpiani2020data}, which employs a partially data-driven approach to derive the mapping exponents.
\begin{table}  
\centering
\begin{tabular}{ccc}
\hline 
\hline
Transformation & Wall distance ($f_I$) & Mean velocity ($g_I$) \\
\hline
\citet{van1951turbulent} & $f_{VD}=1$ & $g_{VD}=R^{1/2}$   \rule{0pt}{2.6ex}  \rule[-1.2ex]{0pt}{0pt}\\
\citet{volpiani2020data} & $f_{VI}=\frac{R^{1/2}}{M^{3/2}}$ & $g_{VI}=\frac{R^{1/2}}{M^{1/2}}$ \rule{0pt}{3.6ex} \rule[-1.2ex]{0pt}{0pt}  \\
\hline
\hline
\end{tabular}
\caption{Compressibility transformations for the wall distance and the mean velocity according to Eq. \eqref{eq:trasf}, where $R=\bar{\rho}/\bar{\rho}_w$ and $M=\bar{\mu}/\bar{\mu}_w$.}\label{Tab:table_trasf}
\end{table}

\citet{griffin2021velocity} transformation, instead, is based on the total stress equation, which reads:
\begin{equation}
	\tau^+=S_t^+ \left( \frac{\tau_v^+}{S_{TL}^+}+ \frac{\tau_R^+}{S_{eq}^+}\right)
\end{equation}
where $\tau_v^+$ and $\tau_R^+$ are the scaled viscous and Reynolds shear stresses (whose sum is equal to $\tau^+$), while $S_{TL}^+=\partial U_{TL}^+ / \partial y^*$ and $S_{eq}^+=\partial U_{eq}^+ / \partial y^*$ are the generalised nondimensional mean shear stresses derived for the viscous region (the subscript $TL$ indicated the accordance with the \citet{trettel2016mean} velocity transformation) and for the log layer (the subscript  $eq$ indicates the assumption of turbulence quasi-equilibrium). 
The generalised nondimensional mean shear $S_t^{+}=\partial U_t^+ / \partial y^*$ is the unknown and once computed it can be integrated with respect to the semilocal wall-normal coordinate $y^*$, leading to the transformed velocity $u^+_{GR} = \int S_t^+ dy^*$.

\begin{figure}  
	\centering
	\subfigure[PARAMETRI-2][Van Driest et al. \label{fig:vandriest}]{\includegraphics[width=0.95\textwidth]{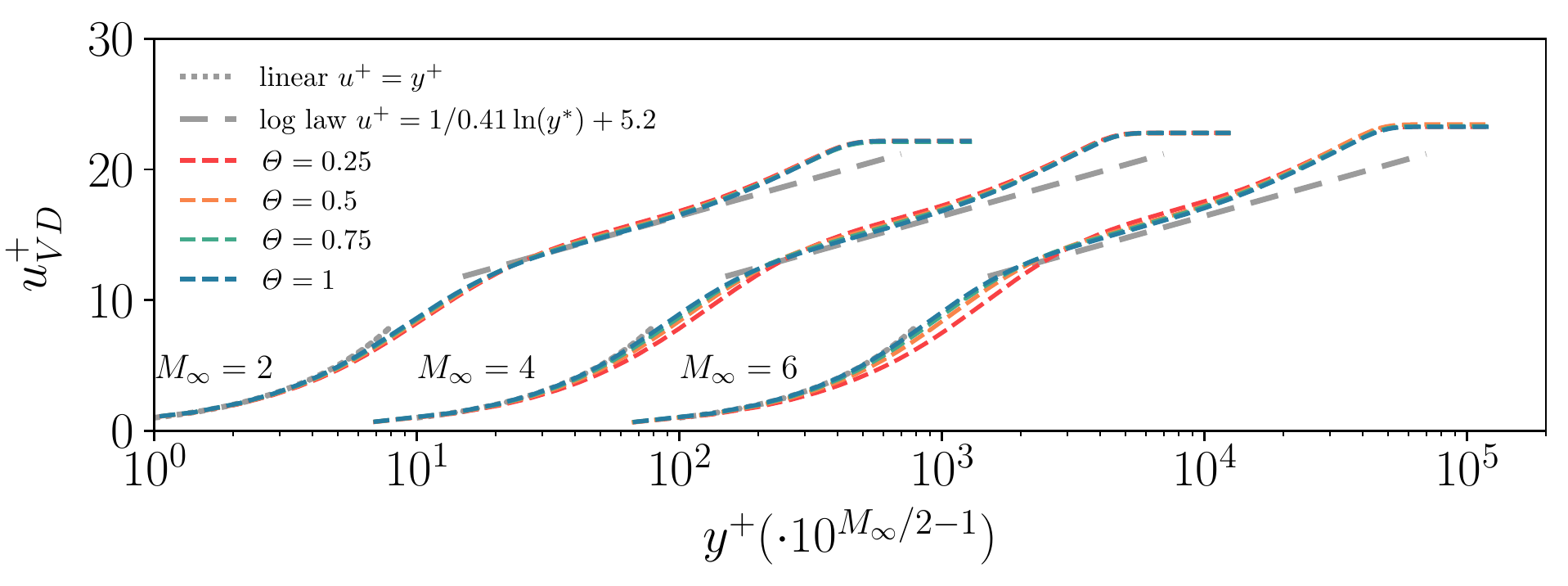}}
		\subfigure[PARAMETRI-2][Volpiani et al. \label{fig:volpiani}]{\includegraphics[width=0.95\textwidth]{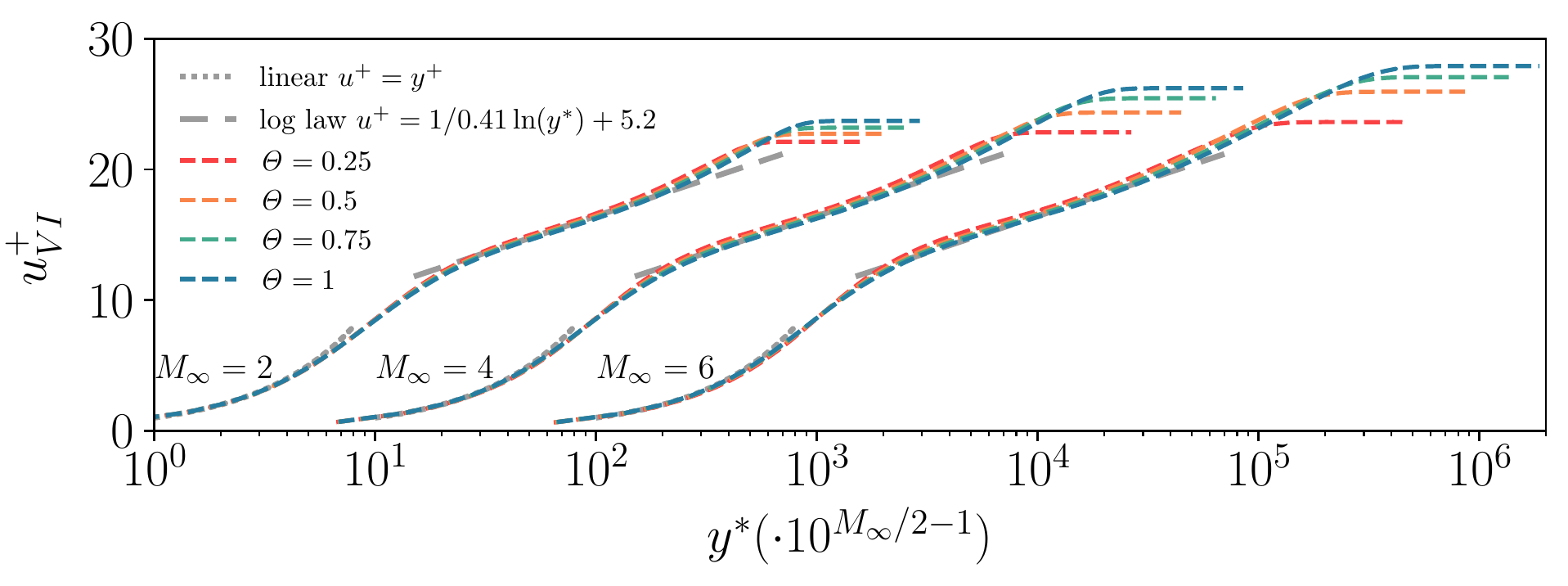}}
	\subfigure[PARAMETRI-2][Griffin et al. \label{fig:griffin}]{\includegraphics[width=0.95\textwidth]{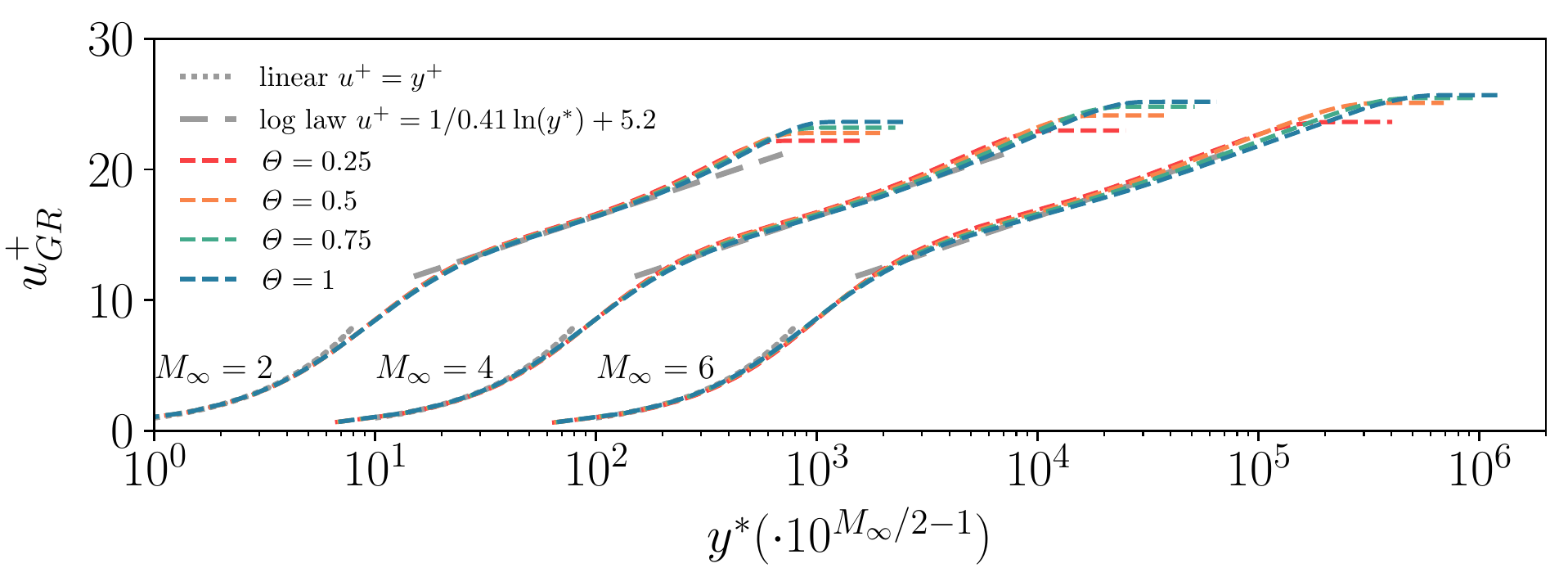}}
	\vspace{-0.2cm}
	
	\caption{Mean velocity profiles at stations listed in table \ref{tab:stat} scaled according to (a) \citet{van1951turbulent}, (b) \citet{volpiani2020data} and (c) \citet{griffin2021velocity} compressibility transformations. Profiles have been translated along the $x$ axis according to the law $10^{M_\infty/2-1}$ to enable better comparison. \label{fig:umean}}
\end{figure}

We report in figure \ref{fig:umean} the scaled profiles according to the classical law of \citet{van1951turbulent} (which has been the standard for several decades and widely employed in wall modelling) and the latest transformations of  \citet{volpiani2020data} and  \citet{griffin2021velocity}. 
Panel \ref{fig:vandriest} reveals the main weaknesses of the \citet{van1951turbulent} scaling, whose accuracy is affected both by the increase of the Mach number and wall-cooling. In particular, non-adiabatic cases at $M_{\infty}=4,6$ show a clear departure from the linear law of the wall, while even adiabatic cases show a positive shift from the log-law as compressibility increases.
Panels \ref{fig:volpiani} and \ref{fig:griffin} show a great improvement in collapsing all profiles to the laws of the wall, the only minor discrepancy being present in the log layer for extremely cold cases at high Mach numbers. Overall, our database supports \citet{volpiani2020data} and \citet{griffin2021velocity} transformations, proving their wide range of applicability.


\begin{figure}  
  \centering
  
  \subfigure[PARAMETRI-1][\label{fig:tmean_yd}]{\includegraphics[width=0.95\textwidth]{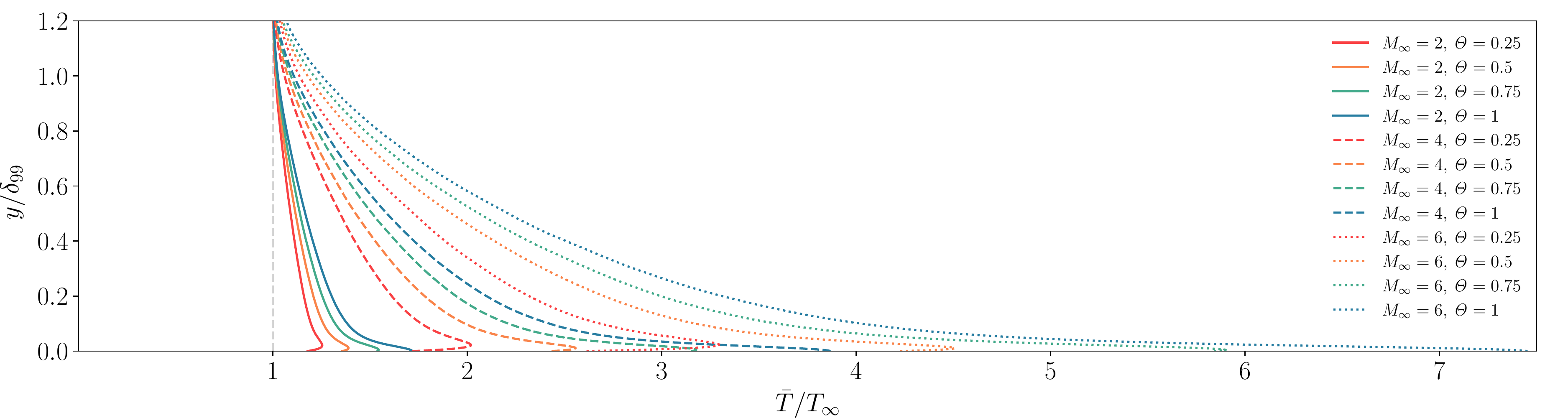}}\\
  \vspace{-0.4cm}
  \subfigure[PARAMETRI-1][$\mathit{\Theta}=0.25$\label{fig:tmean_yp_1}]{\includegraphics[height=0.36\textwidth]{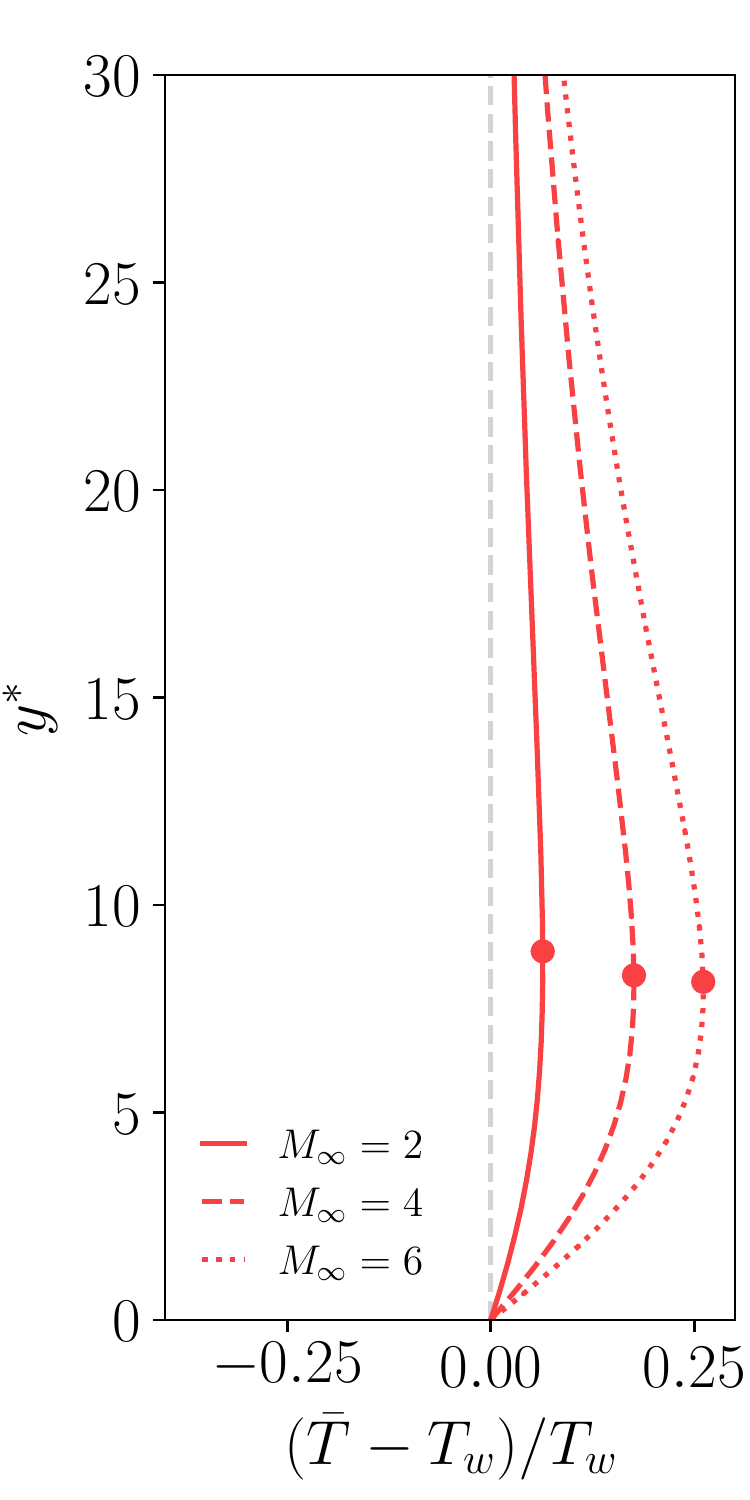}}
  \subfigure[PARAMETRI-1][$\mathit{\Theta}=0.5$\label{fig:tmean_yp_2}]{\includegraphics[height=0.36\textwidth]{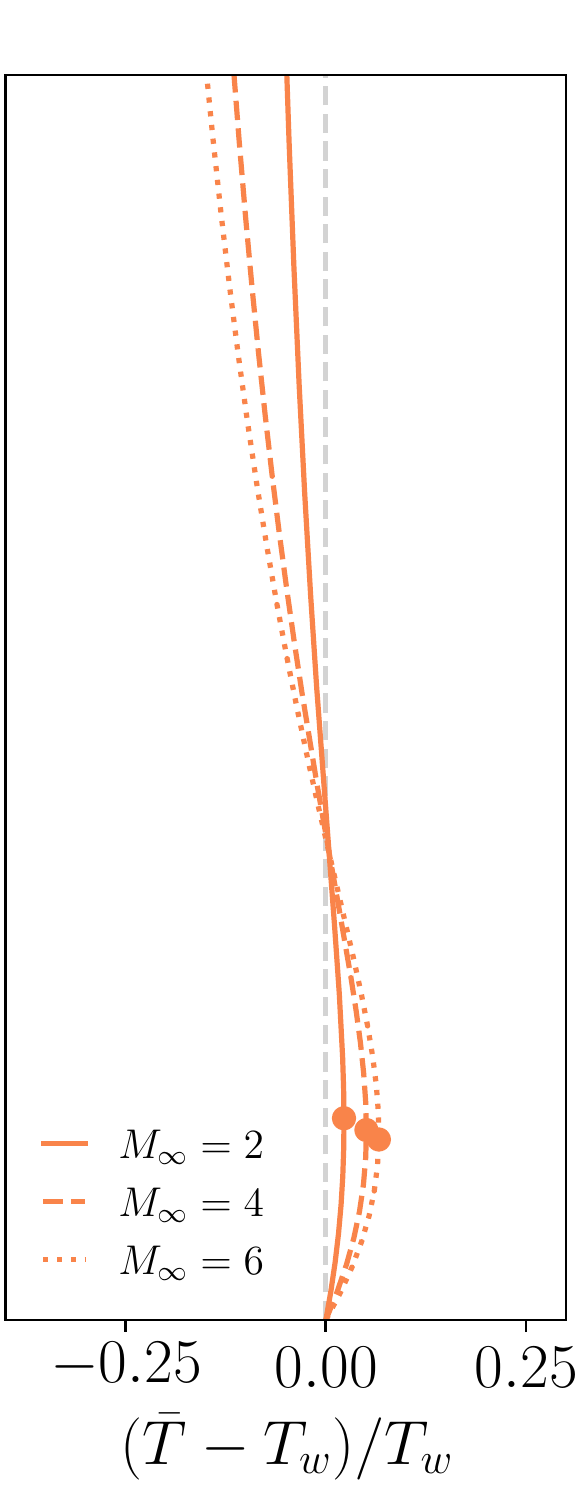}}
  \subfigure[PARAMETRI-1][$\mathit{\Theta}=0.75$\label{fig:tmean_yp_3}]{\includegraphics[height=0.36\textwidth]{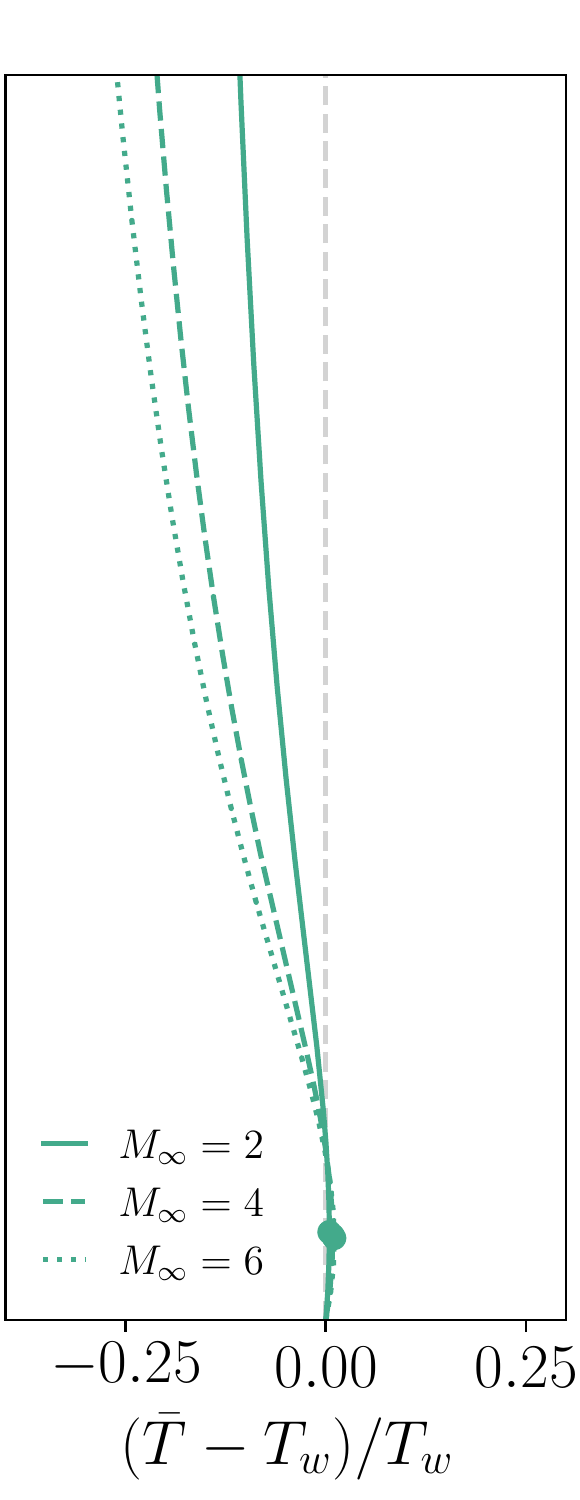}}
  \subfigure[PARAMETRI-1][$\mathit{\Theta}=1.$\label{fig:tmean_yp_4}]{\includegraphics[height=0.36\textwidth]{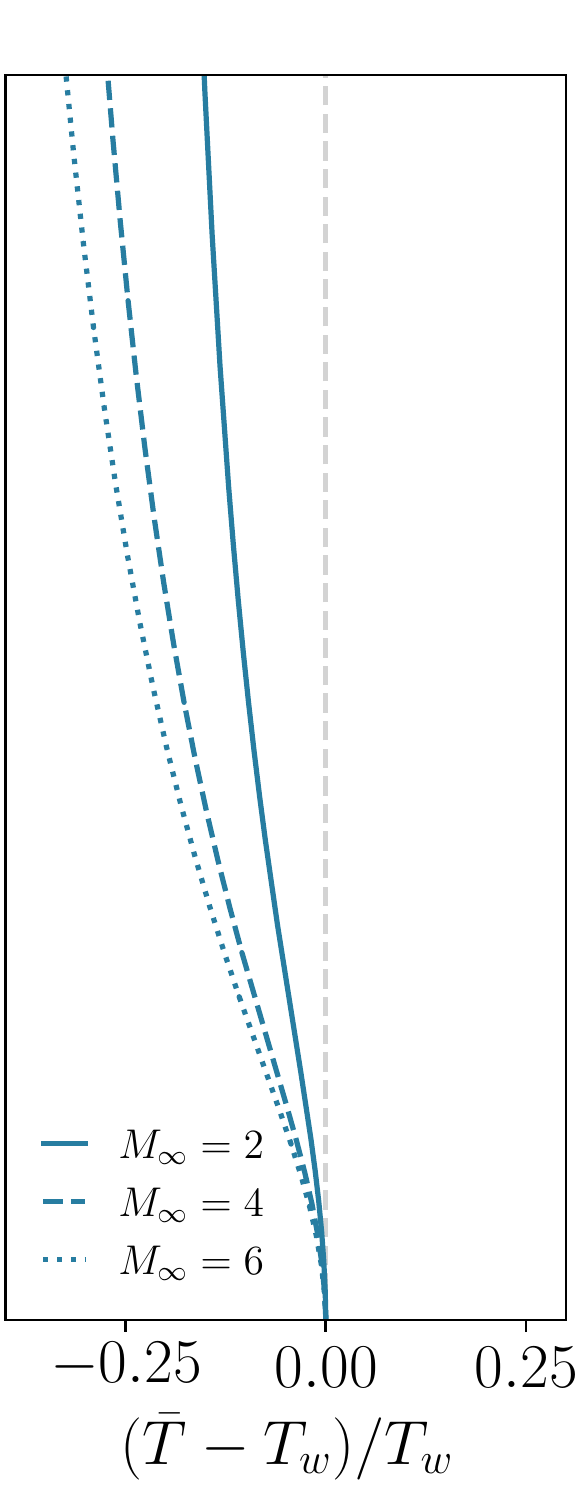}}
  \subfigure[PARAMETRI-1][\label{fig:t_peaks}]{\includegraphics[height=0.36\textwidth]{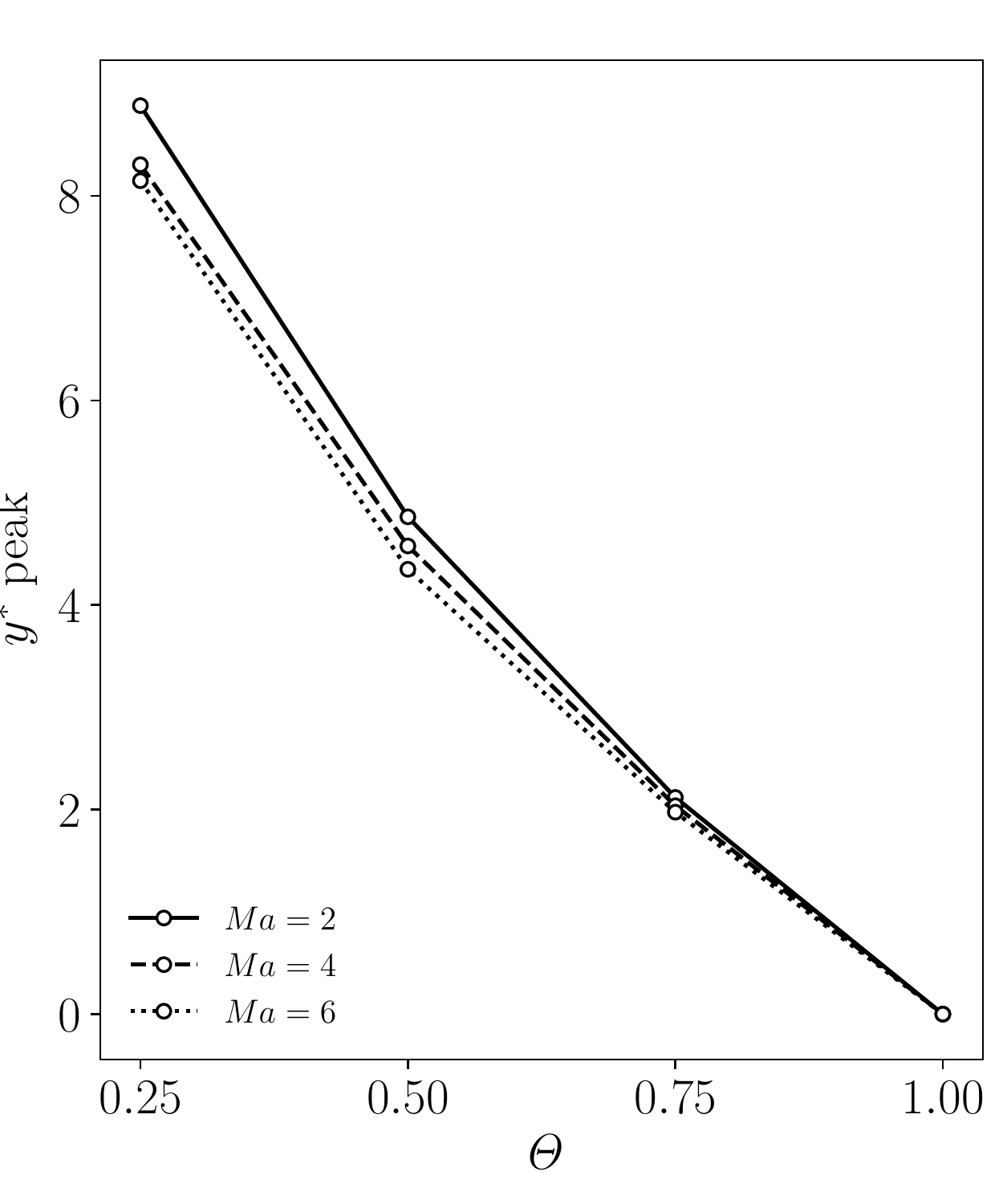}}
  \vspace{-0.2cm}
  \caption{Panel (a): Mean temperature profiles for all cases of table \ref{Tab:table1} as a function of the wall-normal coordinate $y/ \delta_{99}$. Panels (b-c-d-e): Mean temperature profiles and relative peaks as a function of the wall-normal coordinate $y^*$ scaled with $T_w$. Panel (f): Wall-normal position of mean temperature peaks as a function of the wall-cooling $\mathit{\Theta}$ parameter. 
  \label{fig:tmean}} 
\end{figure}

Figures \ref{fig:tmean_yd} to \ref{fig:tmean_yp_4} show the mean temperature profiles throughout the height of the boundary layer and in the near-wall region, respectively. In particular, the latter profiles are scaled using the wall temperature $T_w$. As expected, the adiabatic wall temperature greatly increases with the Mach number, while enhanced wall-cooling (lower $\mathit{\Theta}$) forces the mean temperature profiles to slant towards lower wall temperatures ($T_w<T_r$). 
The combination of these two conditions imposes a change in the sign of temperature gradient near the wall, which is necessary to adjust to a wall temperature lower than the recovery value. Thus, a local peak arises, whose prominence and location are directly connected to the phenomenon of aerodynamic heating, generating a net heat flux from the flow to the solid boundary.
Local temperature peaks are marked in figures \ref{fig:tmean_yp_1} to \ref{fig:tmean_yp_4} with dots. 
An increase in the Mach number generates more intense gradients and higher peak temperatures for non-adiabatic cases, enhancing aerodynamic heating. However, the wall-normal position of the peaks seems to be mainly affected by the change $\mathit{\Theta}$, and weakly dependent on the Mach number. This is apparent in Figure \ref{fig:t_peaks}, which shows a progressive departure from the wall of the peak location as the wall-cooling increases, with a mild downward shift at high Mach numbers.
As anticipated in section \S \ref{sec:slice}, the position of the local maximum of the temperature profile
has major implications in the generation of temperature fluctuations, which affect both their overall intensity and their spatial organisation (breakdown of near-wall streaks). Actually, the departure from a monotonic adiabatic profile to increasingly prominent local peaks of mean temperature profiles prevents the formation of organised temperature streaks that are generated by thermal production (see section \S \ref{sec:thermo_fluc}).
\subsection{Reynolds analogy}
In this section, the coupling between velocity and temperature is discussed for both mean and fluctuating fields.
The DNS results are compared with the classical relations of \citet{walz1969boundary} 
\begin{equation}\label{eq:walz}
    \frac{\bar{T}}{T_{\infty}}=\frac{T_w}{T_{\infty}}+\frac{T_r-T_w}{T_{\infty}}\frac{\bar{u}}{U_{\infty}}+ \frac{T_{\infty}-T_r}{T_{\infty}}\left(\frac{\bar{u}}{U_{\infty}}\right)^2
\end{equation}
and the modified relation of \citet{zhang2014generalized}
\begin{equation}\label{eq:zhang}
     \frac{\bar{T}}{T_{\infty}}=\frac{T_w}{T_{\infty}}+\frac{T_{rg}-T_w}{T_{\infty}}\frac{\bar{u}}{U_{\infty}}+ \frac{T_{\infty}-T_{rg}}{T_{\infty}}\left(\frac{\bar{u}}{U_{\infty}}\right)^2
\end{equation}
where $T_{rg}=T_{\infty}+r_g U_{\infty}^2/(2 c_p)$ and $r_g=2 c_p (T_w-T_{\infty})/U_{\infty}^2-2 \, Pr \, q_w/(U_{\infty} \tau_w)$.

\begin{figure}  
  \centering
  \subfigure[PARAMETRI-1][$M_{\infty}=2$]{\includegraphics[width=0.32\textwidth]{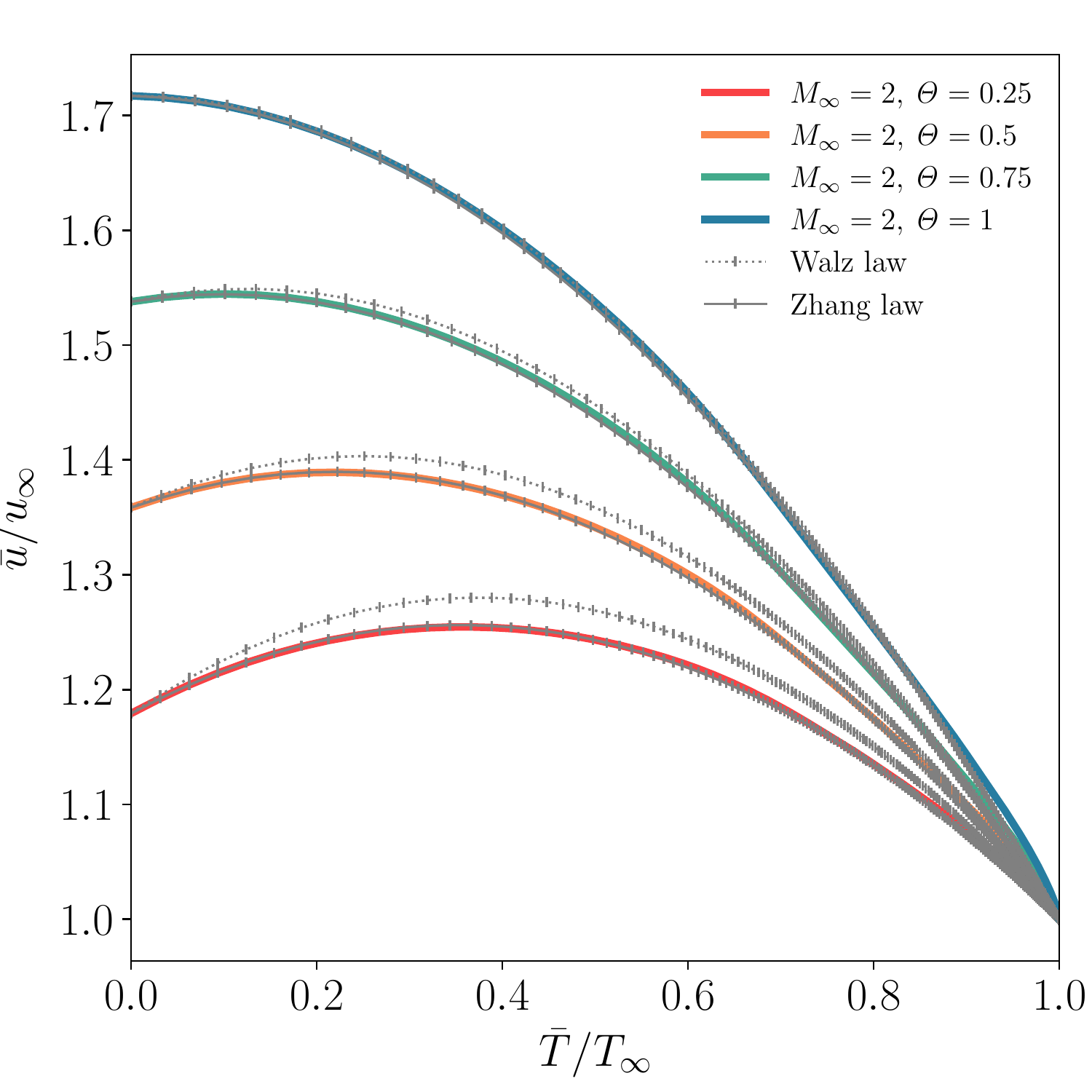}}
  \subfigure[PARAMETRI-1][$M_{\infty}=4$]{\includegraphics[width=0.32\textwidth]{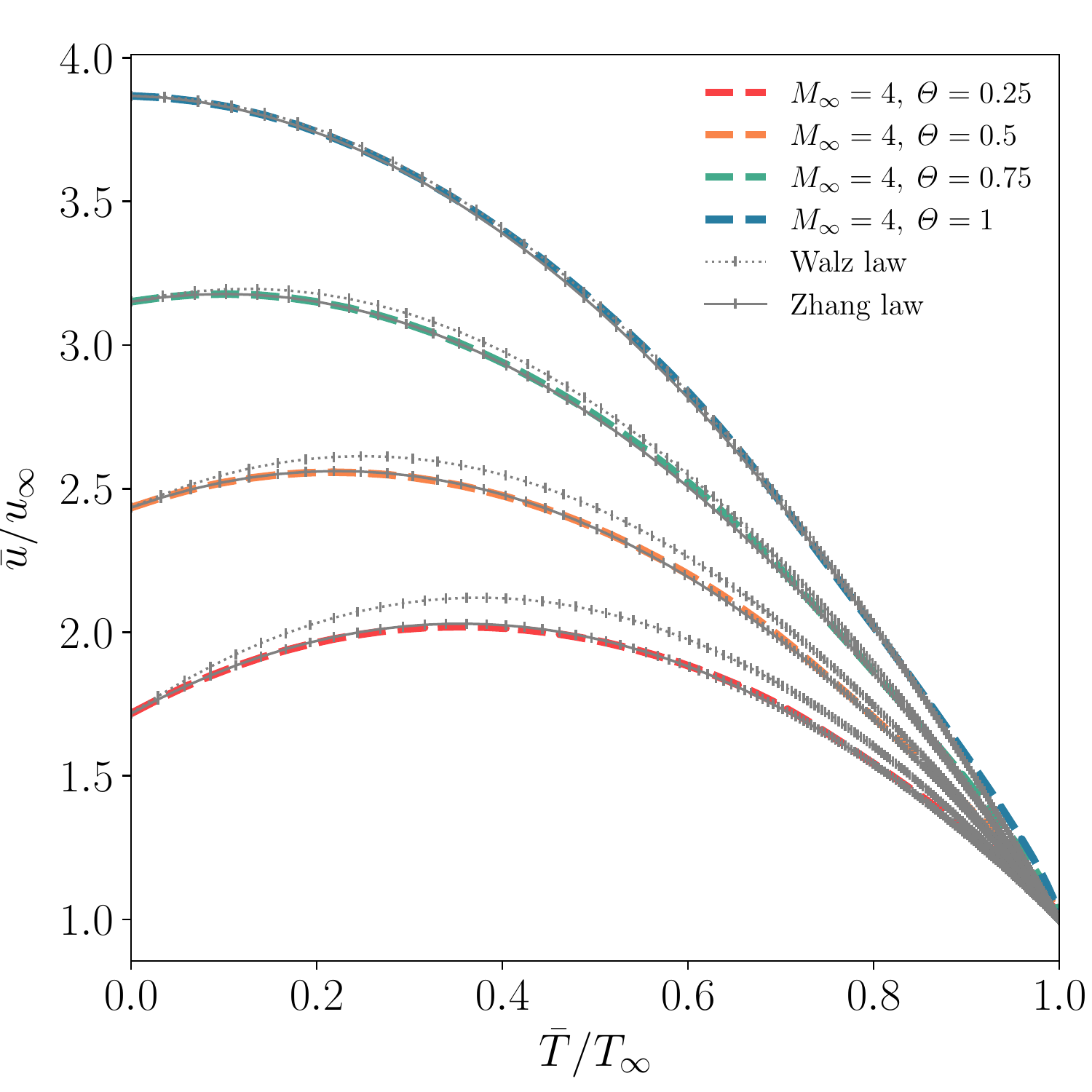}}
  \subfigure[PARAMETRI-1][$M_{\infty}=6$]{\includegraphics[width=0.32\textwidth]{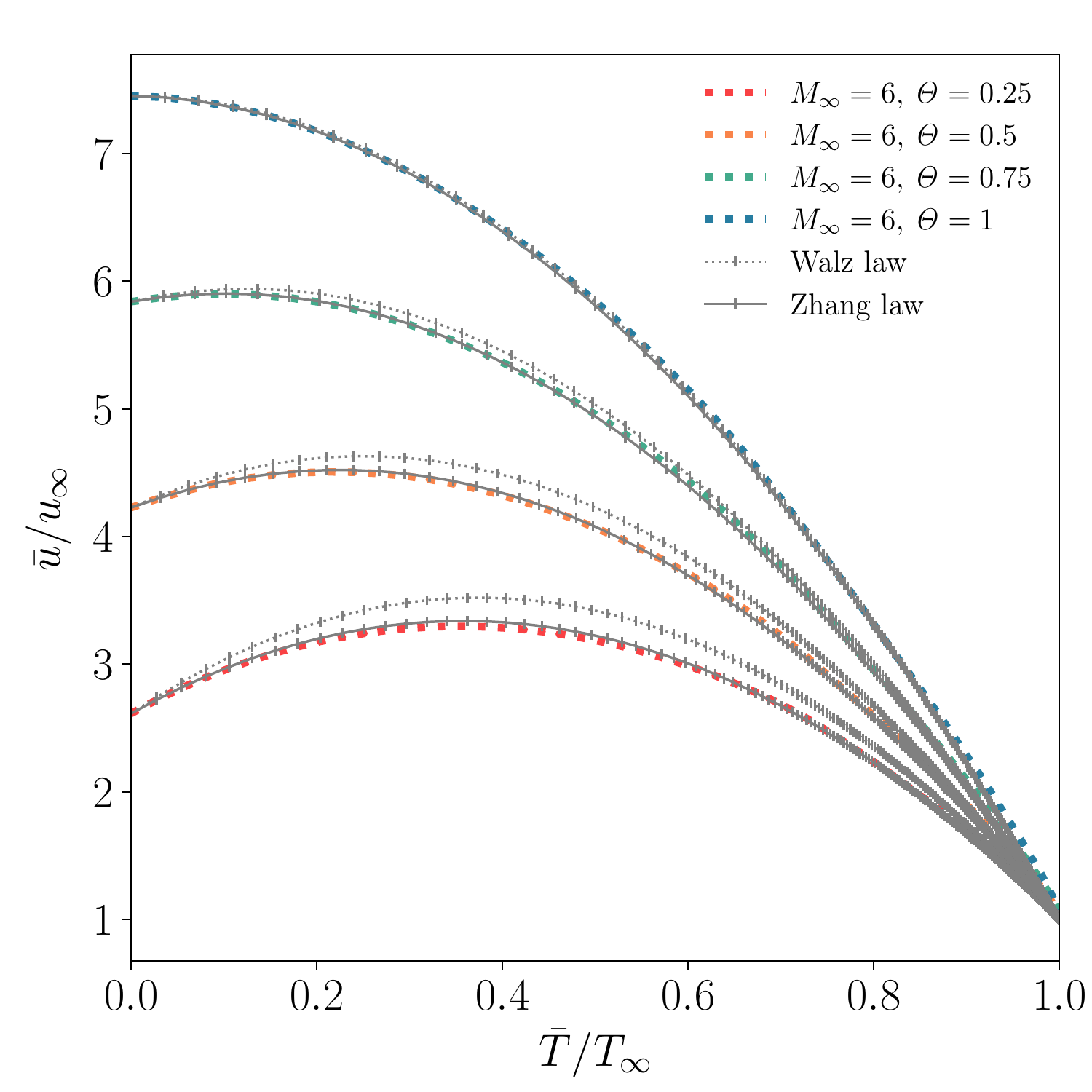}}
\caption{Mean temperature profiles against mean velocity compared with the classical law of \citet{walz1969boundary} (Eq. \eqref{eq:walz}) and the modified relation of \citet{zhang2014generalized} (Eq. \eqref{eq:zhang}). 
}\label{fig:temp_vel}
\end{figure}

Figure \ref{fig:temp_vel} compares the relations \eqref{eq:walz} and \eqref{eq:zhang} with the present database.
As expected, \citet{walz1969boundary} relation greatly degrades its accuracy when the wall-cooling is increased, while \citet{zhang2014generalized} is able to better perform under these conditions, the only minor deviations being present for the case M6T025. 
However, we note that \citet{walz1969boundary} law still excellently holds for adiabatic cases at high Mach numbers, while being incapable of correctly capturing wall-cooling effects.

For engineering design purposes, the value of $r_g$ can be difficult to evaluate given its dependence on the wall temperature $T_w$ and the ratio of the wall heat flux $q_w$ and the wall shear stress $\tau_w$. 
Following the discussion of \citet{zhang2014generalized}, the Reynolds analogy factor $s$ comes into play to greatly simplify the calculation, since $r_g$ can be rewritten in terms of $s$
\begin{equation}\label{Eq:rg_sPr}
r_g = r[s Pr + (1-s Pr) \mathit{\Theta}]
\end{equation}
being $s$ defined as
\begin{equation}
    s = \frac{2 C_h}{C_f} = \frac{q_w u_{\infty}}{\tau_w c_p (T_w-T_r)}
\end{equation}
where $C_f=\tau_w/(1/2 \rho_{\infty} u_{\infty}^2)$ is the skin friction coefficient and $C_h=q_w/(\rho_{\infty} u_{\infty} c_p (T_w-T_r))$ the Stanton number.
The simplification consists in the fact that several authors \citep{Duan2010,zhang2014generalized} identified the term $s Pr$ to be an empirical constant around the value of $0.8\pm0.03$ (data fitting of \citet{zhang2014generalized}) over several different flow cases, meaning that only $T_w$ would be needed to be evaluated to compute $r_g$.

\begin{figure}  
  \centering
\includegraphics[width=0.6\textwidth]{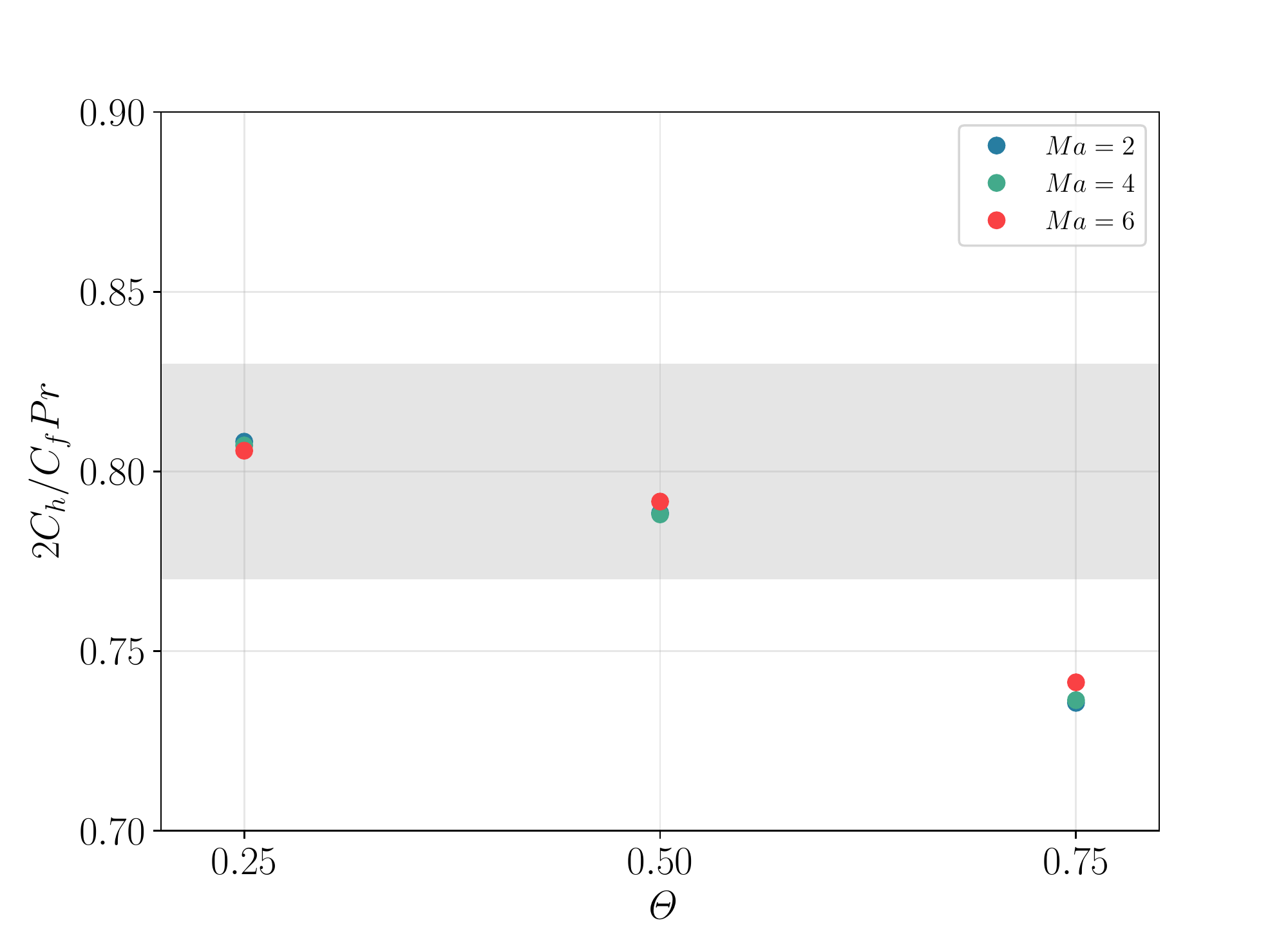}
\caption{Reynolds analogy factor $s=2 C_h/C_f Pr$ as a function of the diabatic parameter $\mathit{\Theta}$ for different Mach numbers. The grey band refers to the data fitting of $0.8\pm0.03$ of \citet{zhang2014generalized}.
}\label{fig:cf_ch}
\end{figure}

Figure \ref{fig:cf_ch} reports the computed values of $s Pr$ in our database showing a good agreement to \citet{zhang2014generalized} fit.
A slight decreasing trend with $\mathit{\Theta}$ can be observed, and it is interesting to note that at a given $\mathit{\Theta}$ the values appear to be independent of $M_{\infty}$.

The data reported in figure \ref{fig:cf_ch} have a mean value and standard deviation of $0.78\pm0.03$, which is close to the value reported by \citep{zhang2014generalized}. 
By approximating $r_g$ in Eq. \eqref{Eq:rg_sPr} with the mean value of $sPr$ and comparing it with DNS data, we obtain a maximum error of $5\%$ (for the case M6T025), which can be considered acceptable for engineering purposes.


Another important set of theoretical relations that couple the thermodynamic and kinetic fluctuating fields is given by the Strong Reynolds Analogy (SRA) \citep{morkovin1962effects}. Originally derived for an adiabatic case, the three main relations can be expressed as

\begin{equation}\label{eq:SRA}
\begin{gathered}
	\frac{\left(\widetilde{T^{''2}}\right)^{1/2}/\Tilde{T}}{(\gamma-1)\Tilde{M}^2\left(\widetilde{u^{''2}}\right)^{1/2}/\Tilde{u}} \approx 1, \\
   R_{u^{''} T^{''}} = \frac{\widetilde{u^{''} T^{''}}}{\sqrt{\widetilde{u^{'' 2}}} \sqrt{\widetilde{T^{'' 2}}}} \approx 1, \\
   Pr_t=\frac{\overline{\rho u^{\prime} v^{\prime}}(\partial \Tilde{T} / \partial y)}{\overline{\rho T^{\prime} v^{\prime}}(\partial \Tilde{u} / \partial y)} \approx 1 .
\end{gathered}
\end{equation}
where we remind the Favre average definition $\tilde{f}=\overline{\rho f}/ \bar{\rho}$ and that $f^{''}=f-\tilde{f}$.
\begin{figure}  
  \centering
\includegraphics[width=0.6\textwidth]{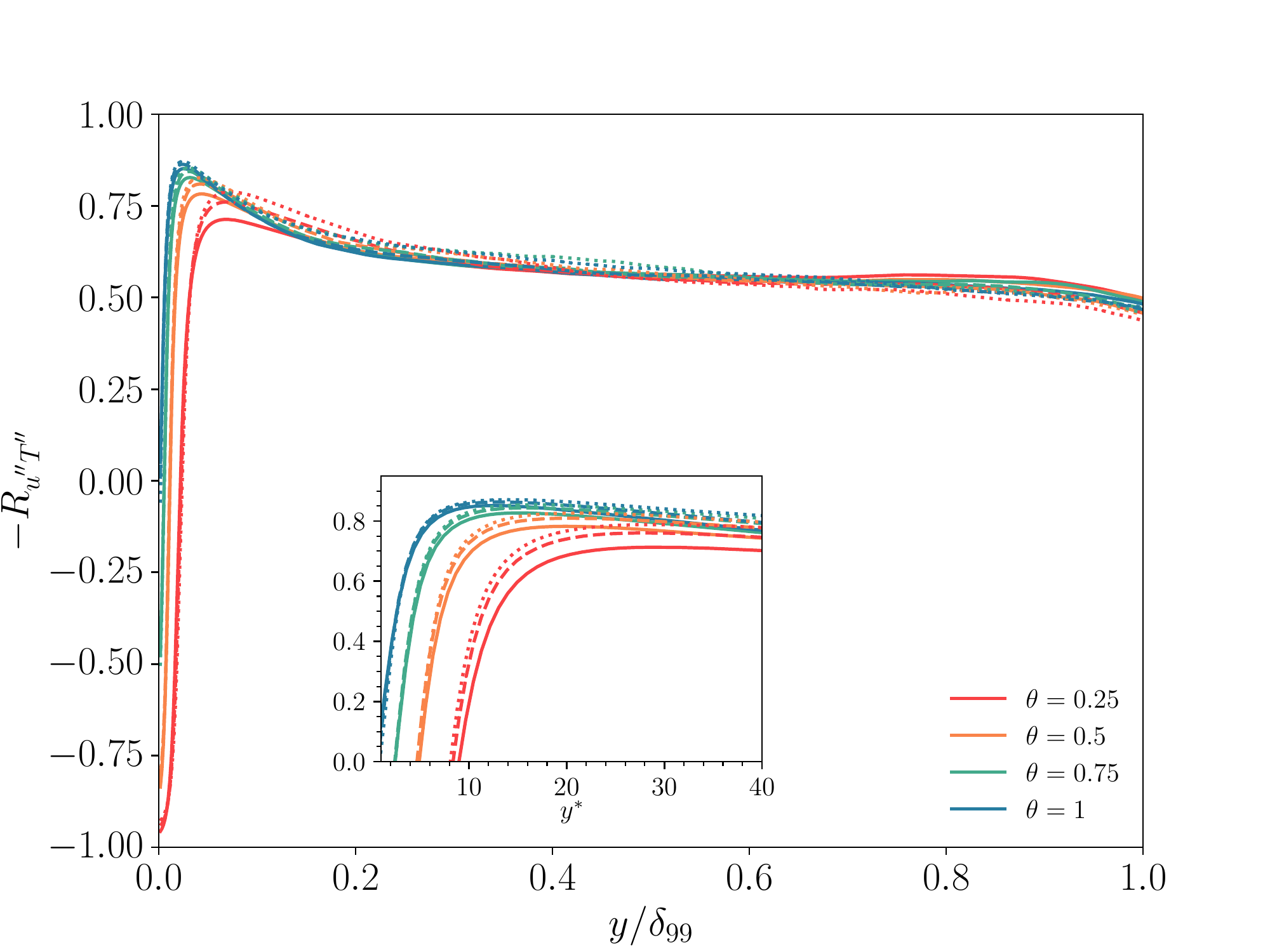}
\caption{Velocity and temperature correlation $R_{u^{''} T^{''}}$ as function of $y/\delta_{99}$. Full lines (\sampleline{}) indicate $M_{\infty}=2$, dashed lines (\sampleline{dashed}) $M_{\infty}=4$ and dotted lines (\sampleline{dotted}) $M_{\infty}=6$.}\label{fig:Rut}
\end{figure}
Figure \ref{fig:Rut} shows the profiles of $R_{u^{''} T^{''}}$ that clearly deviate from unity, which is expected since it was derived assuming zero total temperature fluctuation \citep{morkovin1962effects}.
All profiles collapse around the value $-R_{u^{''} T^{''}}=0.6$, except in the near-wall region, which is marked with an inset \citep{Duan2010}.
The inset of Figure \ref{fig:Rut} shows that the crossover location, where $R_{u^{''} T^{''}}=0$, corresponds approximately to the location of the maximum mean temperature.
Here, we observe that as the wall gets progressively cold, the crossover location moves at higher $y^*$ values, indicating a temperature-velocity decorrelation that is progressively moved farther from the wall.
Our database also shows that this location is almost independent of the Mach number when $\mathit{\Theta}$ is fixed, whereas distinct Mach and wall-cooling effects are visible on the near-wall peak intensity and position of $R_{u^{''} T^{''}}$.
\begin{figure}  
	\centering
	\subfigure[PARAMETRI-3][Original SRA\label{fig:oSRA}]{\includegraphics[width=0.48\textwidth]{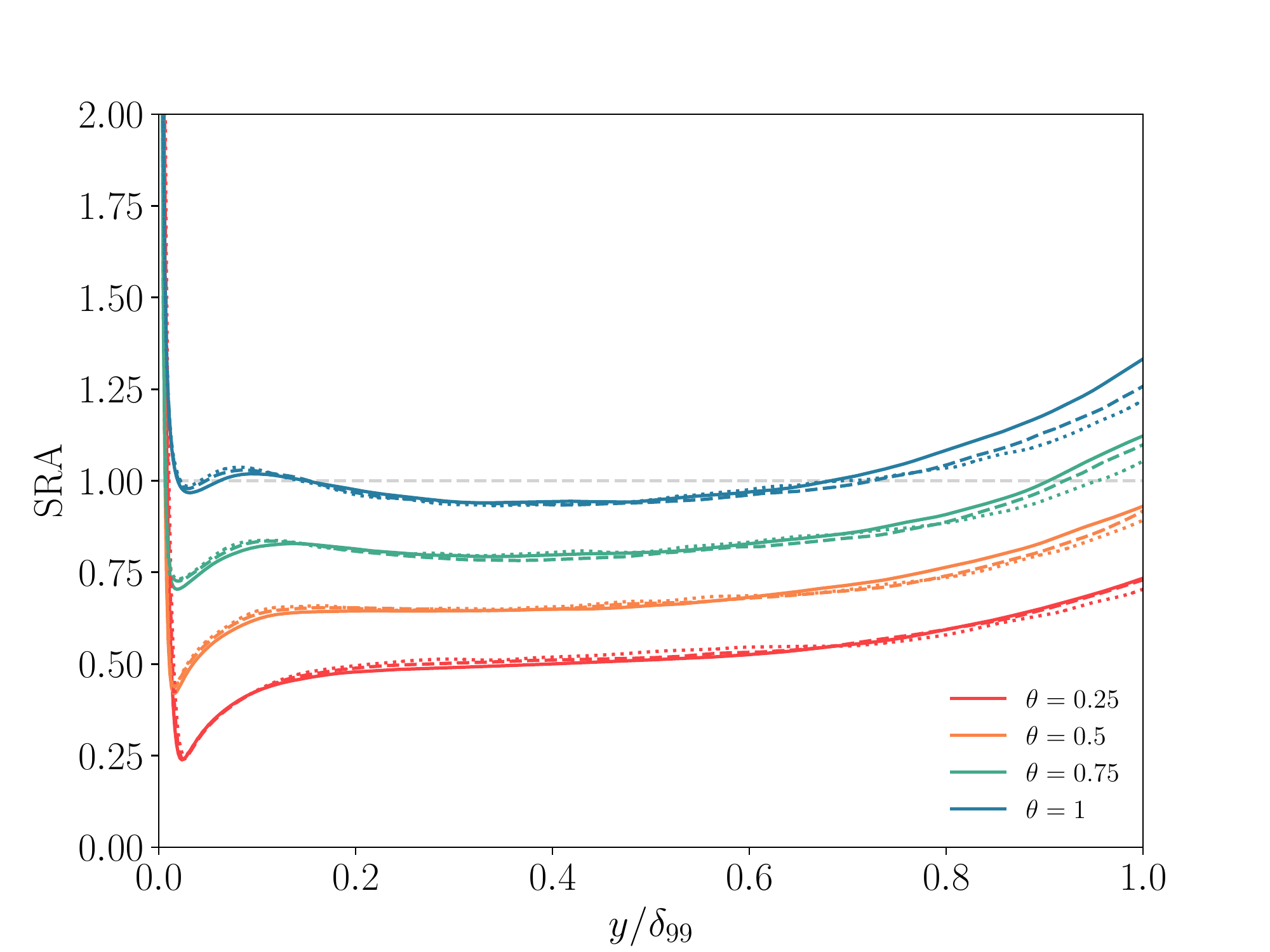}}
    \subfigure[PARAMETRI-3][]{\includegraphics[width=0.48\textwidth]{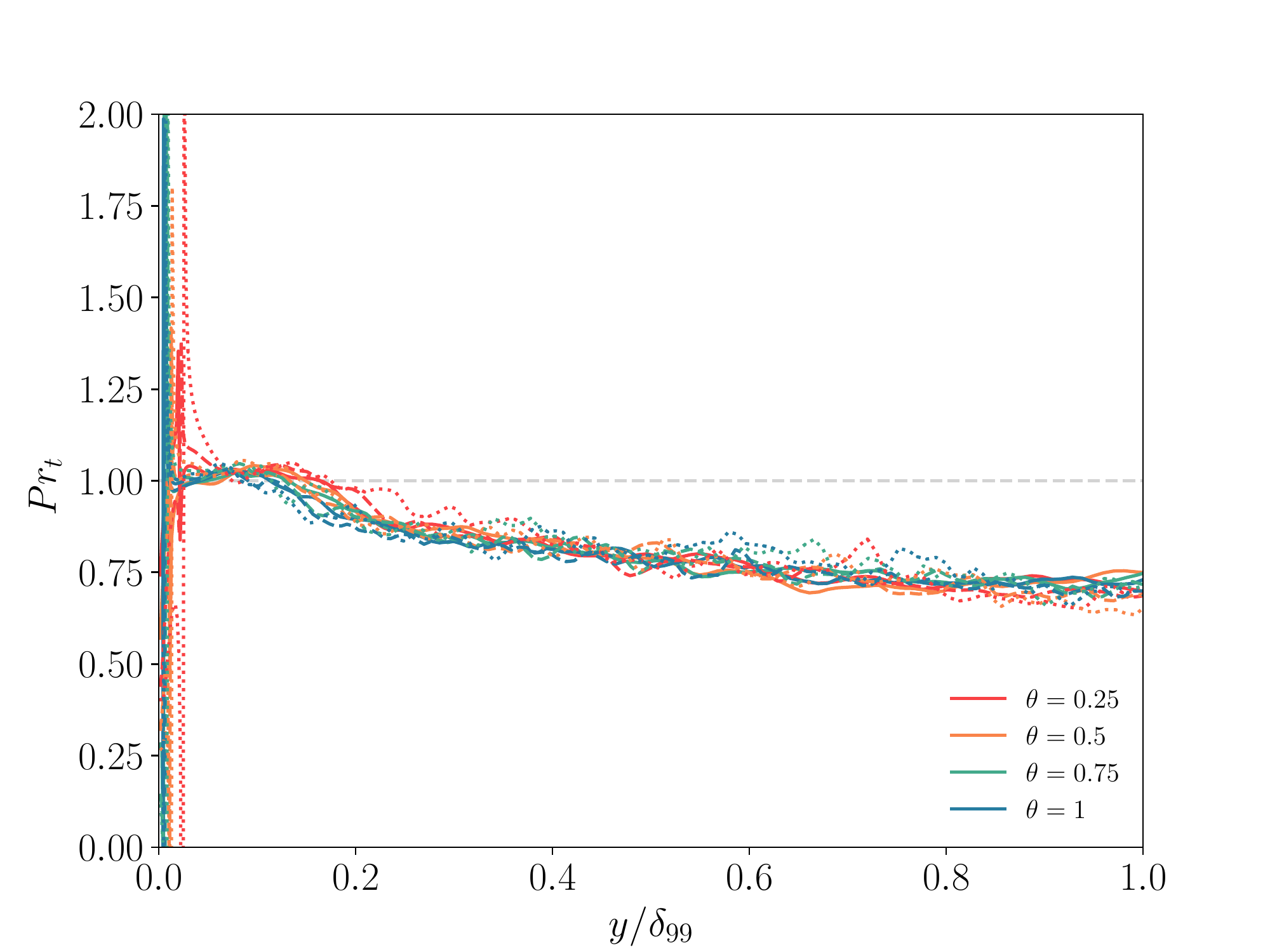}}
	\subfigure[PARAMETRI-3][Modified HSRA\label{fig:hsra}]{\includegraphics[width=0.48\textwidth]{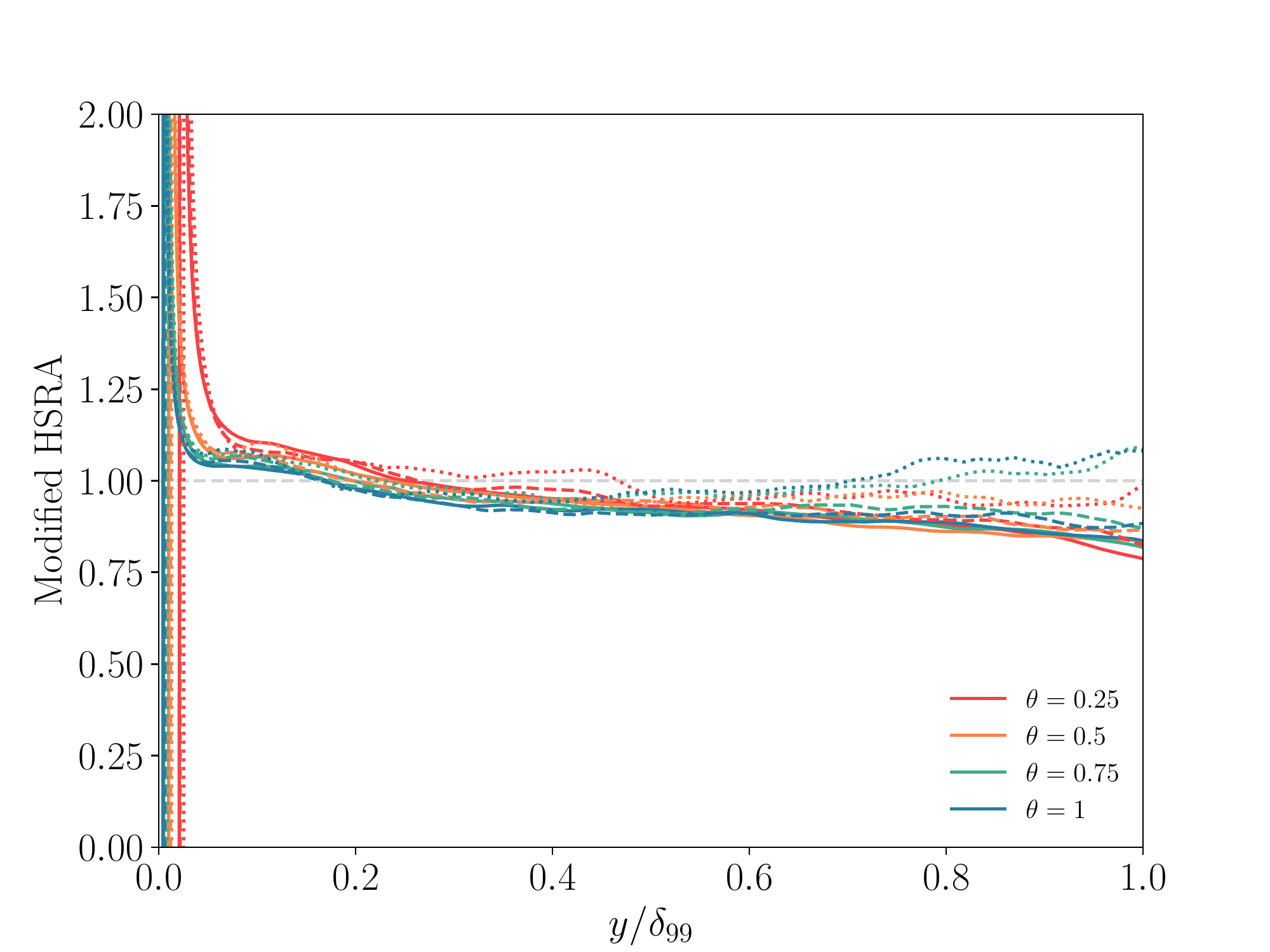}}
    \subfigure[PARAMETRI-3][]{\includegraphics[width=0.48\textwidth]{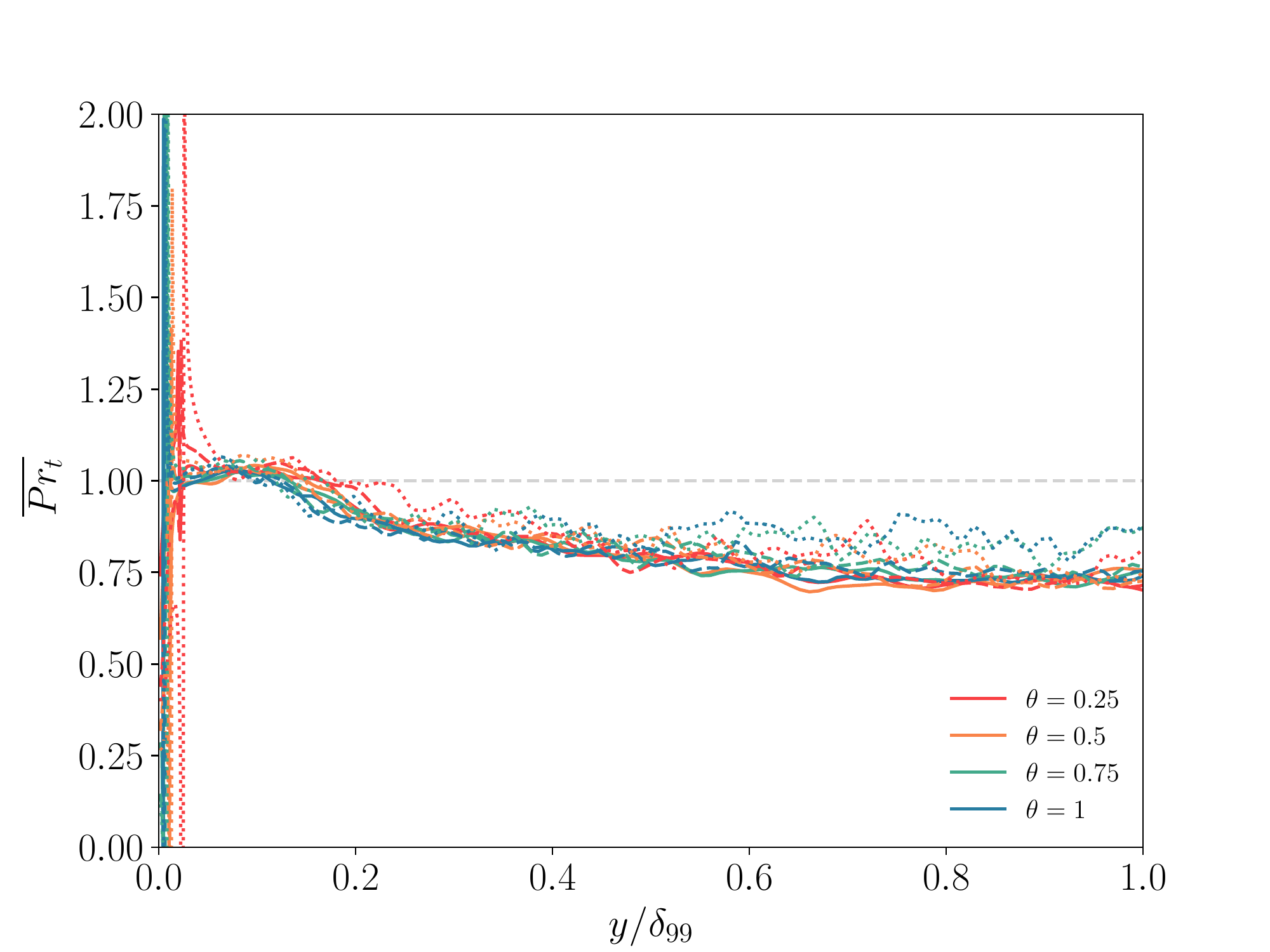}}
	\caption{Comparison of the Strong Reynolds Analogies in the (a,b) original form (Eq. \eqref{eq:SRA}), (c,d)  Modified HSRA \citep{zhang2014generalized}) (Eqs. \eqref{eq:hrsa}, \eqref{eq:Pr_av}). Full lines (\sampleline{}) indicates $M_{\infty}=2$, dashed lines (\sampleline{dashed}) $M_{\infty}=4$ and  dotted lines (\sampleline{dotted}) $M_{\infty}=6$.}\label{fig:SRA}
\end{figure}
The remaining two relations of equation \eqref{eq:SRA} have been modified over the years to account for finite heat flux at the wall and remove wall temperature dependence \citep{huang1995compressible} (HSRA). 
The most recent improvement has been made by \citet{zhang2014generalized}, who proposed another definition of the turbulent Prandtl number $\overline{Pr_t}$ which should perform better at high Mach numbers, yielding the following expression of the modified strong Reynolds analogy (modified HRSA):
\begin{equation}\label{eq:hrsa}
	\frac{\left(\widetilde{T^{''2}}\right)^{1/2}/\Tilde{T}}{(\gamma-1)\Tilde{M}^2\left(\widetilde{u^{''2}}\right)^{1/2}/\Tilde{u}}\overline{Pr_t} \left(1-(\partial\Tilde{T}_t / \partial\Tilde{T})\right)\approx 1
\end{equation}
where the proposed definition of $\overline{Pr_t}$ is
\begin{equation}\label{eq:Pr_av}
	\overline{Pr_t}=\frac{\overline{(\rho v)' u^{'}}\partial \Tilde{T}/\partial y}{\overline{ (\rho v^{'})T^{'}}\partial \Tilde{u}/\partial y}=Pr_t \frac{1+\bar{v}\overline{\rho' u'}/\overline{\rho v' u'}}{1+\bar{v}\overline{\rho' T'}/\overline{\rho v' T'}}
\end{equation}
in which the difference from the classical definition is notable when both $\bar{v}$ and $\rho'$ are nonzero. 

Figure \ref{fig:SRA} compares the wall-normal profiles obtained with the original SRA and the modified version of Eq. \eqref{eq:hrsa}, as well as the profiles of $Pr_t$ and the modified $\overline{Pr_t}$ of Eq. \eqref{eq:Pr_av}.
Figure \ref{fig:hsra} shows that the modified version of \citet{zhang2014generalized} clearly improves the insensitivity to the freestream Mach number and wall temperature condition, with only slight deviations at the edge of the boundary layer. 
It is also interesting to note the excellent collapse that the original SRA of panel \ref{fig:oSRA} exhibits for profiles at fixed $\mathit{\Theta}$, independently of the Mach number, highlighting the relevance of the diabatic parameter $\mathit{\Theta}$ in accounting for the effects of different wall temperatures independently of the Mach number.

\section{Fluctuation statistics}\label{sec:fluc}
\subsection{Velocity fluctuations and length scales}\label{sec:velfluc}
\begin{figure} 
	\centering
	\subfigure[PARAMETRI-1][$M_{\infty}=2$]{\includegraphics[width=0.48\textwidth]{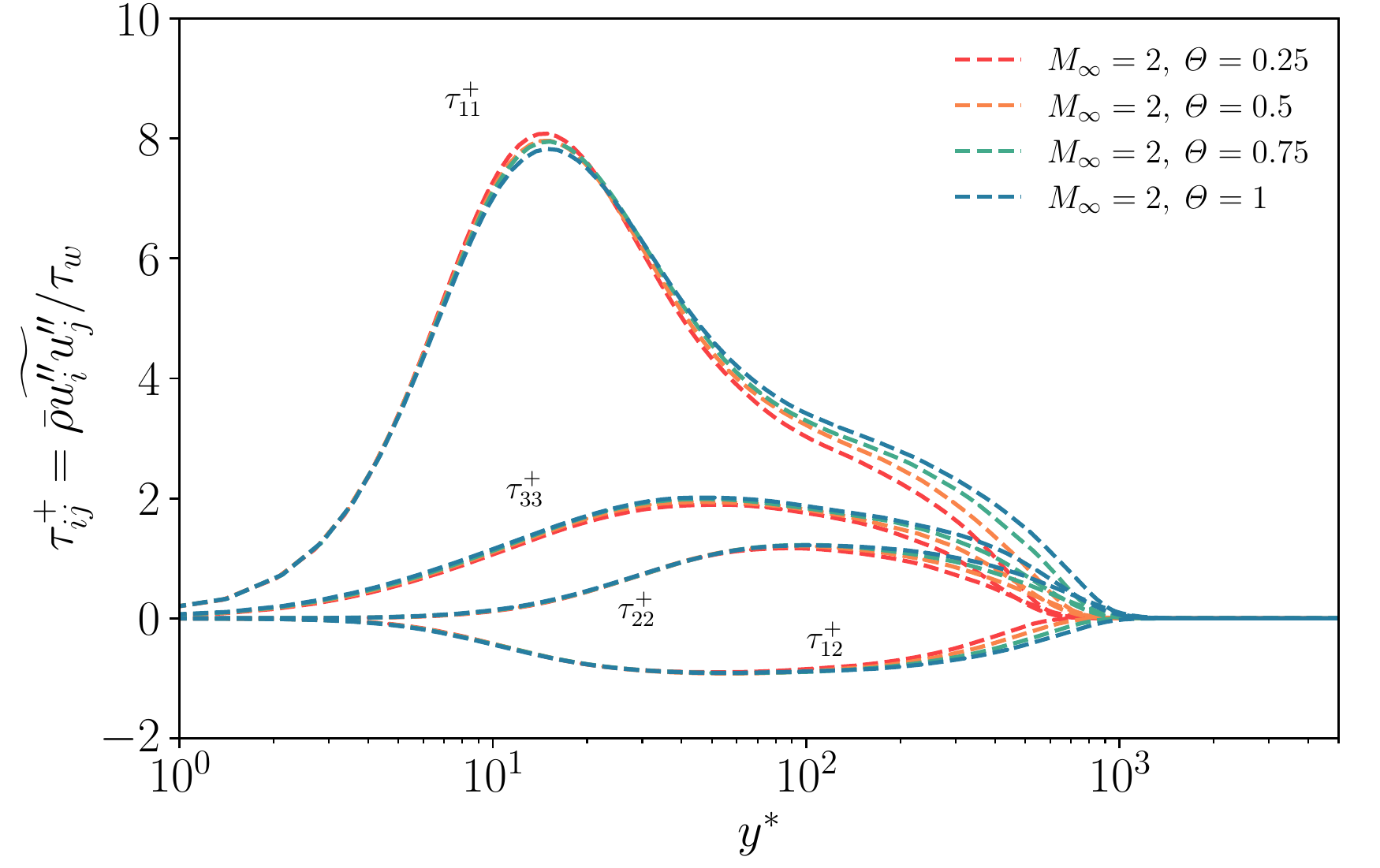}}
	\subfigure[PARAMETRI-1][$M_{\infty}=2$]{\includegraphics[width=0.48\textwidth]{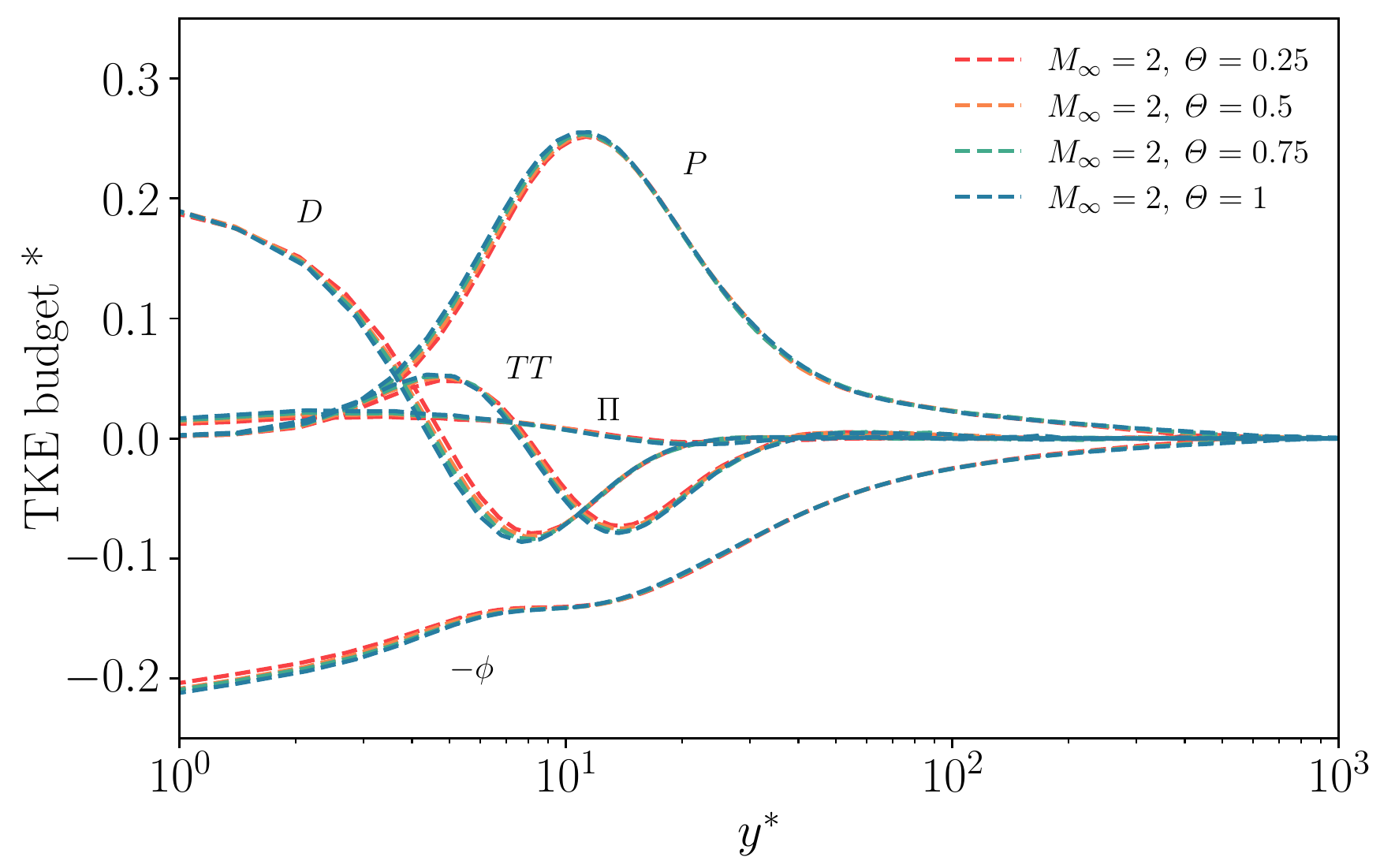}}\\
	\subfigure[PARAMETRI-2][$M_{\infty}=4$]{\includegraphics[width=0.48\textwidth]{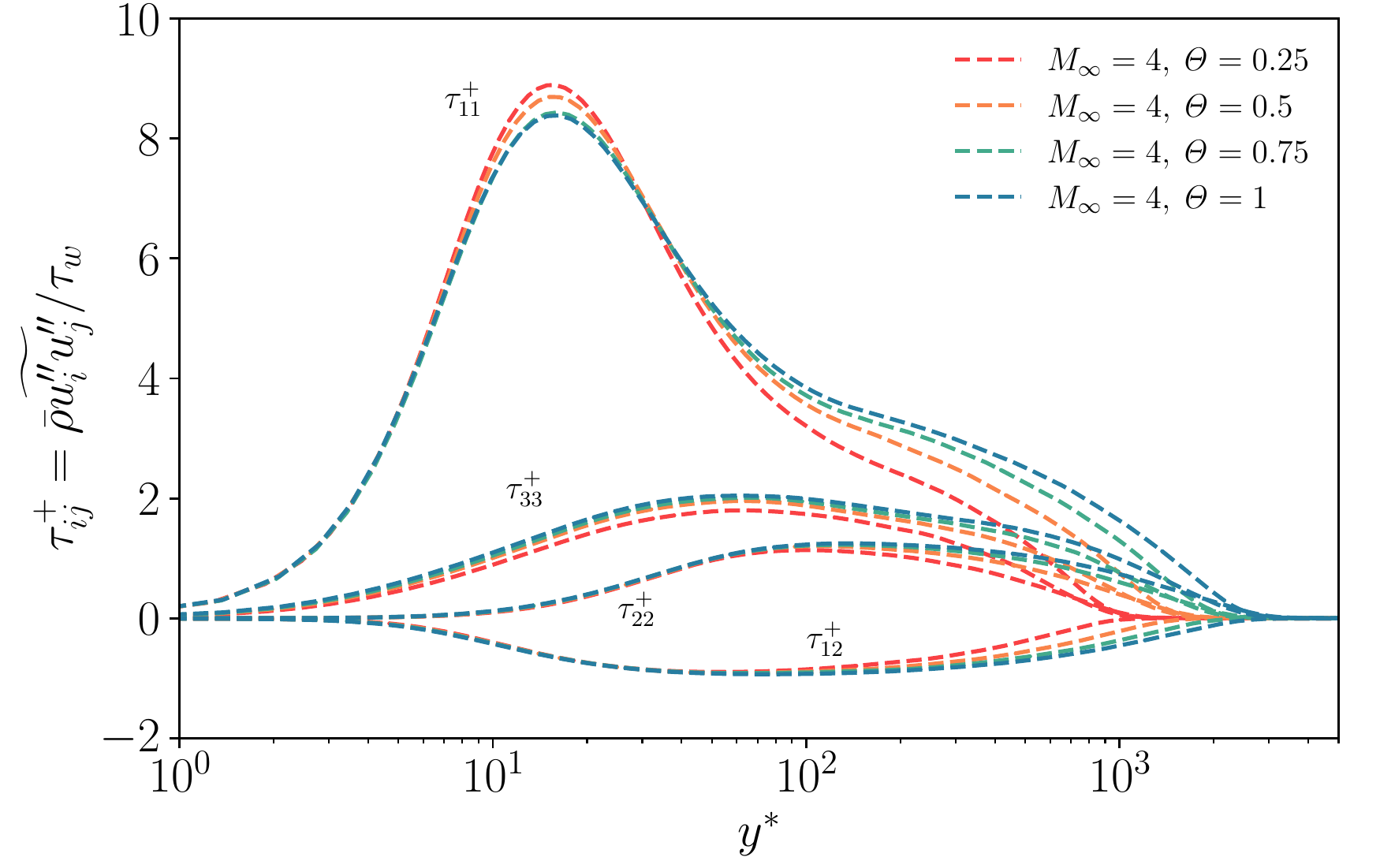}}
	\subfigure[PARAMETRI-2][$M_{\infty}=4$]{\includegraphics[width=0.48\textwidth]{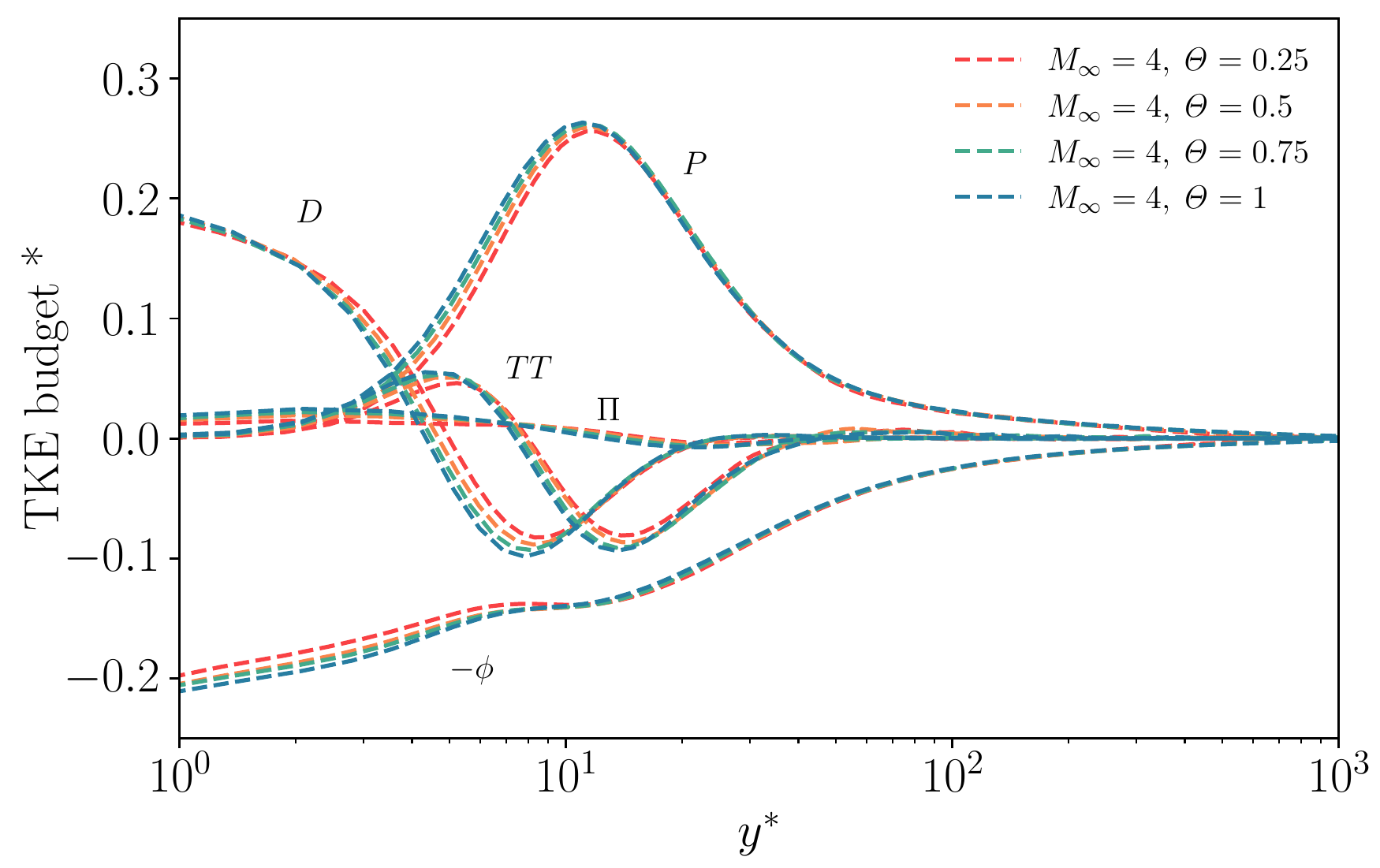}}\\
	\subfigure[PARAMETRI-2][$M_{\infty}=6$]{\includegraphics[width=0.48\textwidth]{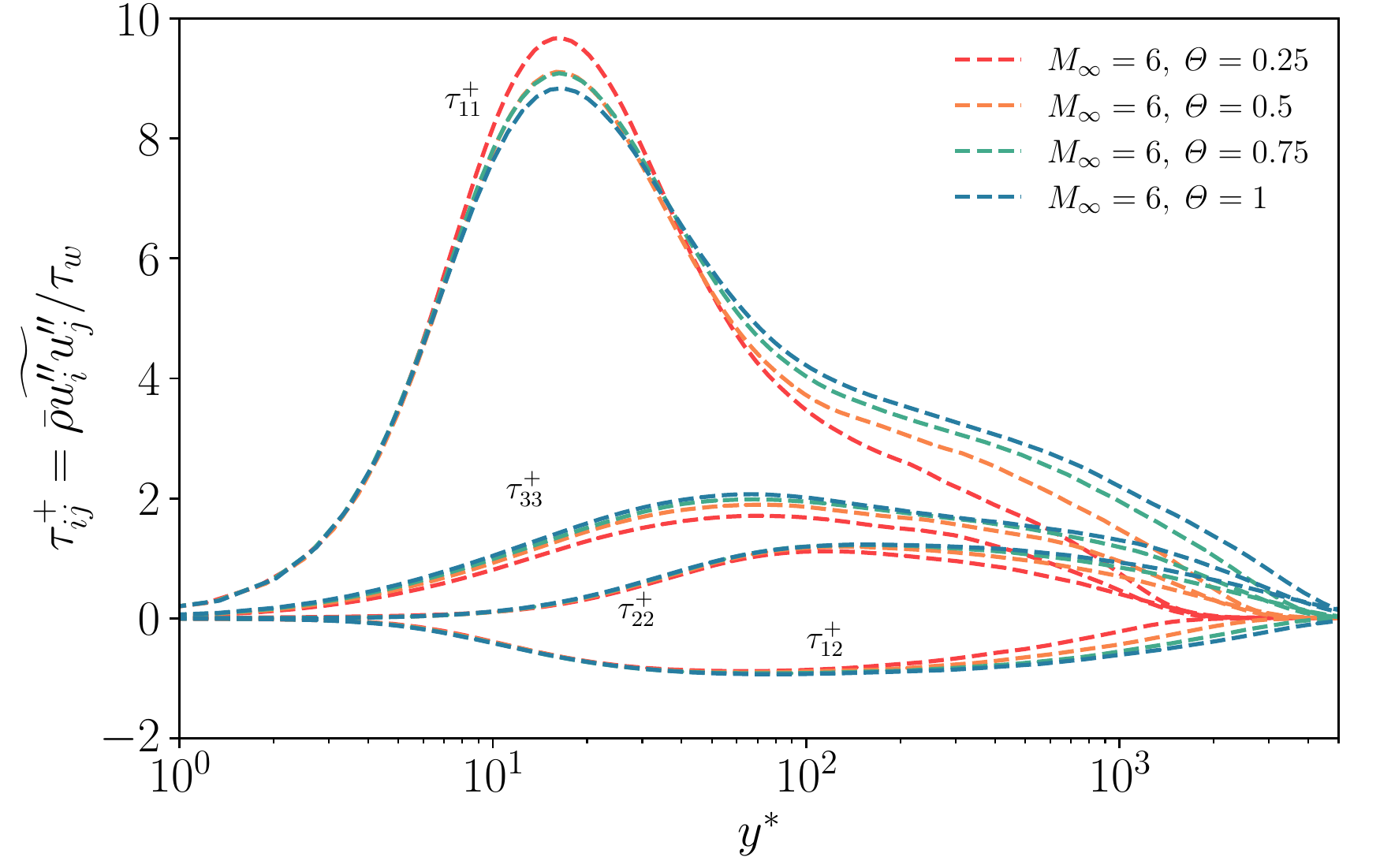}}
	\subfigure[PARAMETRI-2][$M_{\infty}=6$]{\includegraphics[width=0.48\textwidth]{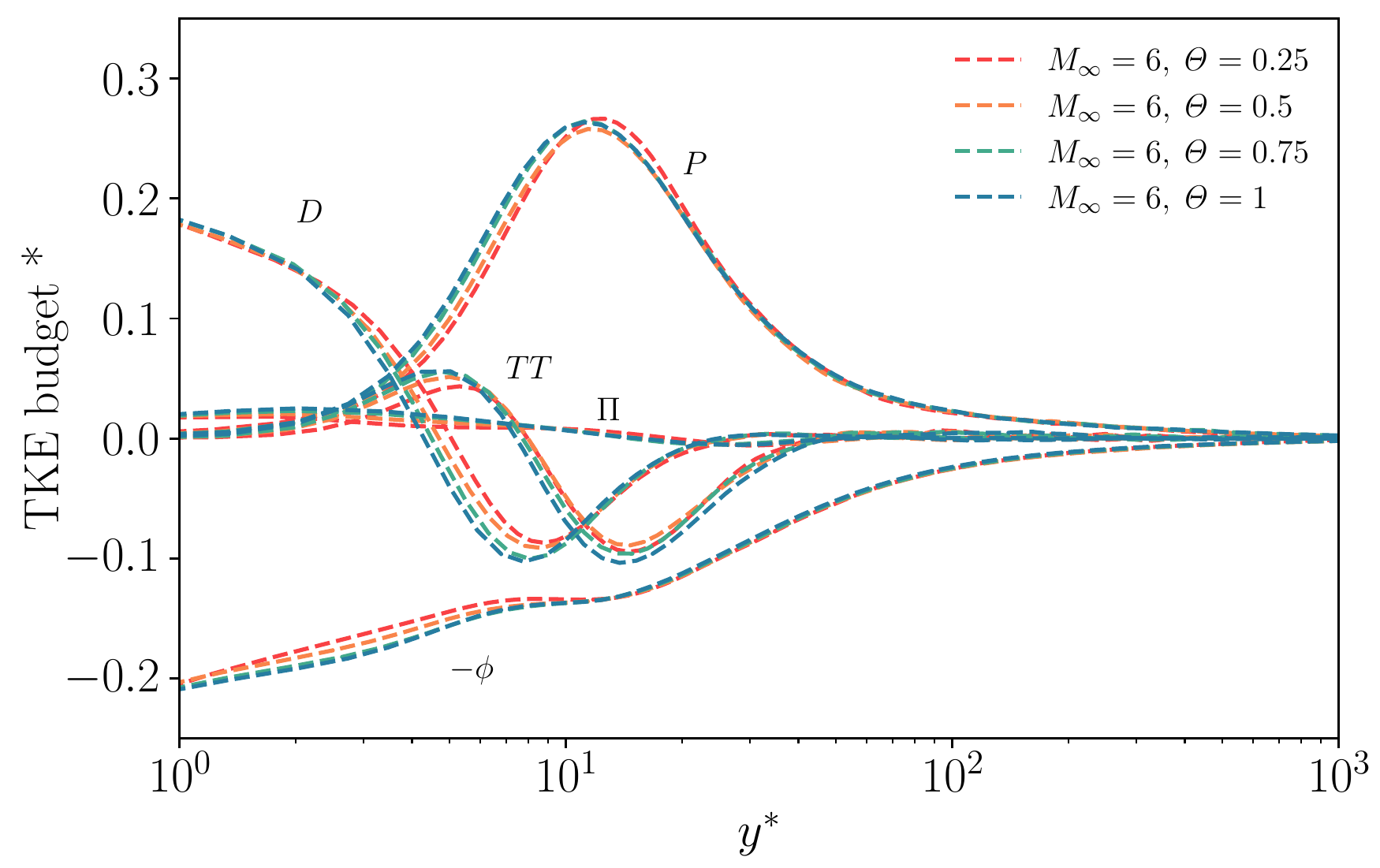}}
	\vspace{-0.2cm}
  \caption{Semilocal-scaled turbulent velocity fluctuations (a,c,e) and turbulent kinetic budget (b,d,f) as function of the wall-normal distance $y^*$. Here, different diabatic parameters $\mathit{\Theta}$ are compared at a given Mach number $M_{\infty}$. \label{fig:velfluc_t}}
\end{figure}

\begin{figure} 
	\centering
	\subfigure[PARAMETRI-1][$\mathit{\Theta}=0.25$]{\includegraphics[width=0.48\textwidth]{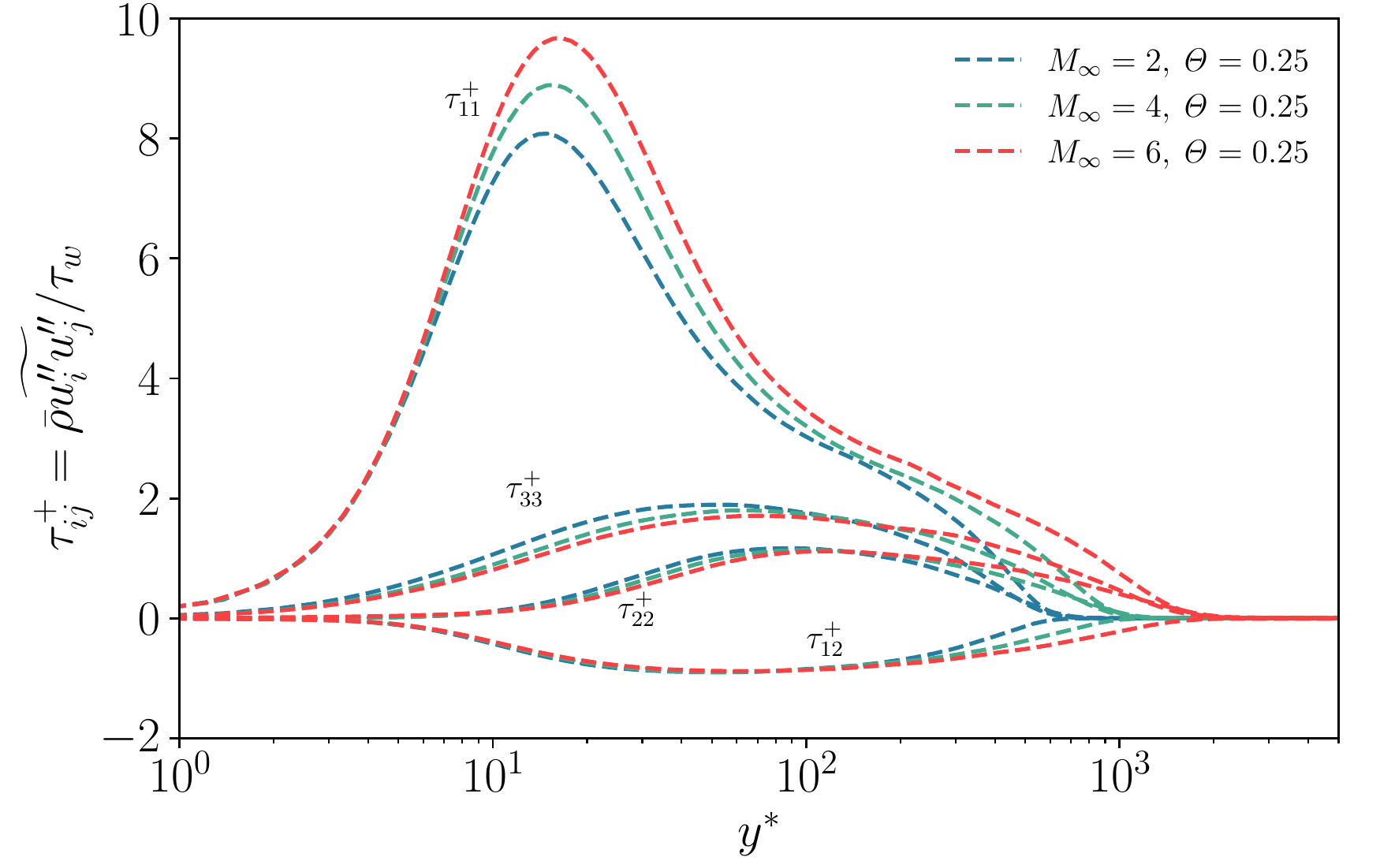}}
	\subfigure[PARAMETRI-1][$\mathit{\Theta}=0.25$]{\includegraphics[width=0.48\textwidth]{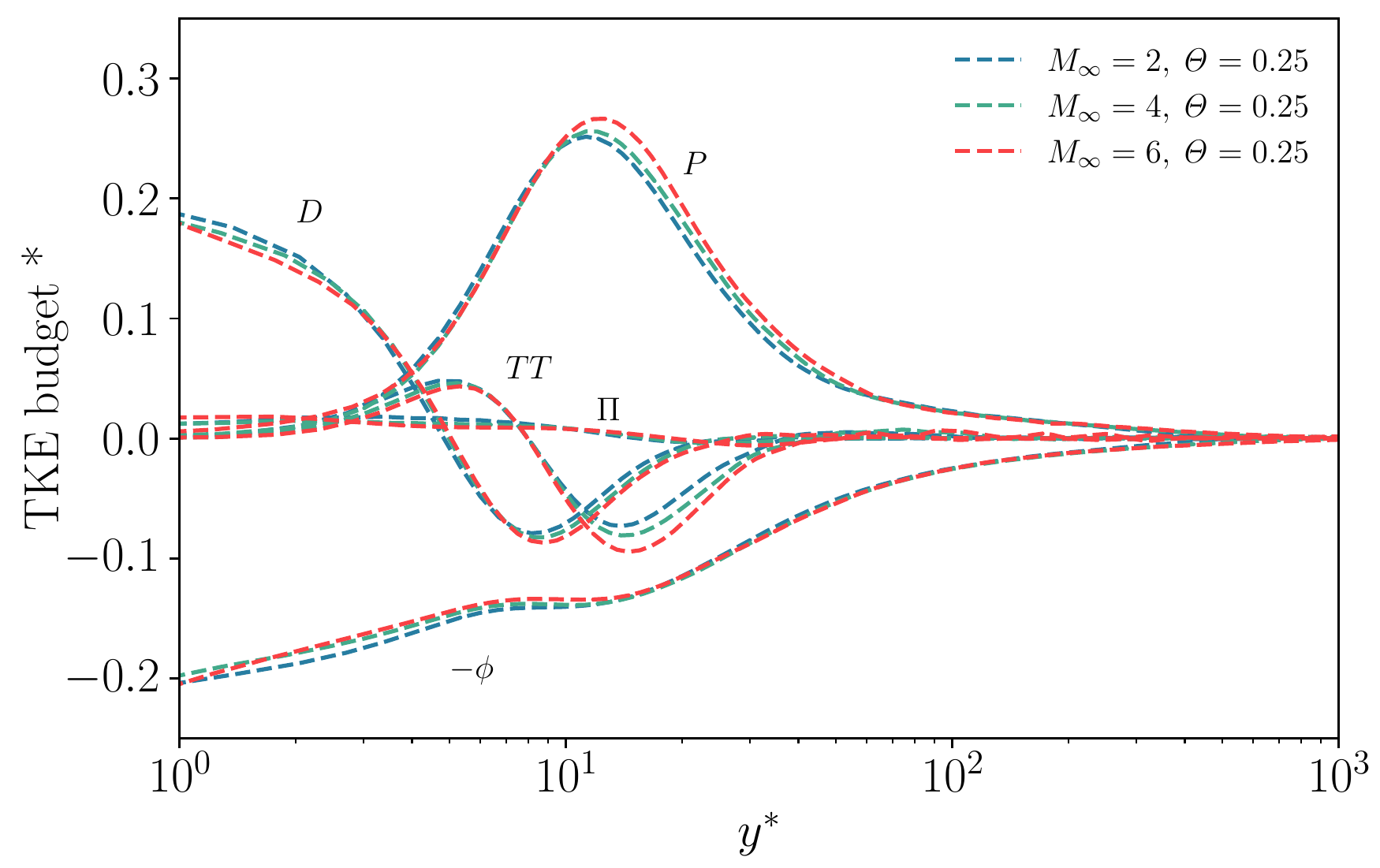}}\\
	\subfigure[PARAMETRI-2][$\mathit{\Theta}=0.5 $]{\includegraphics[width=0.48\textwidth]{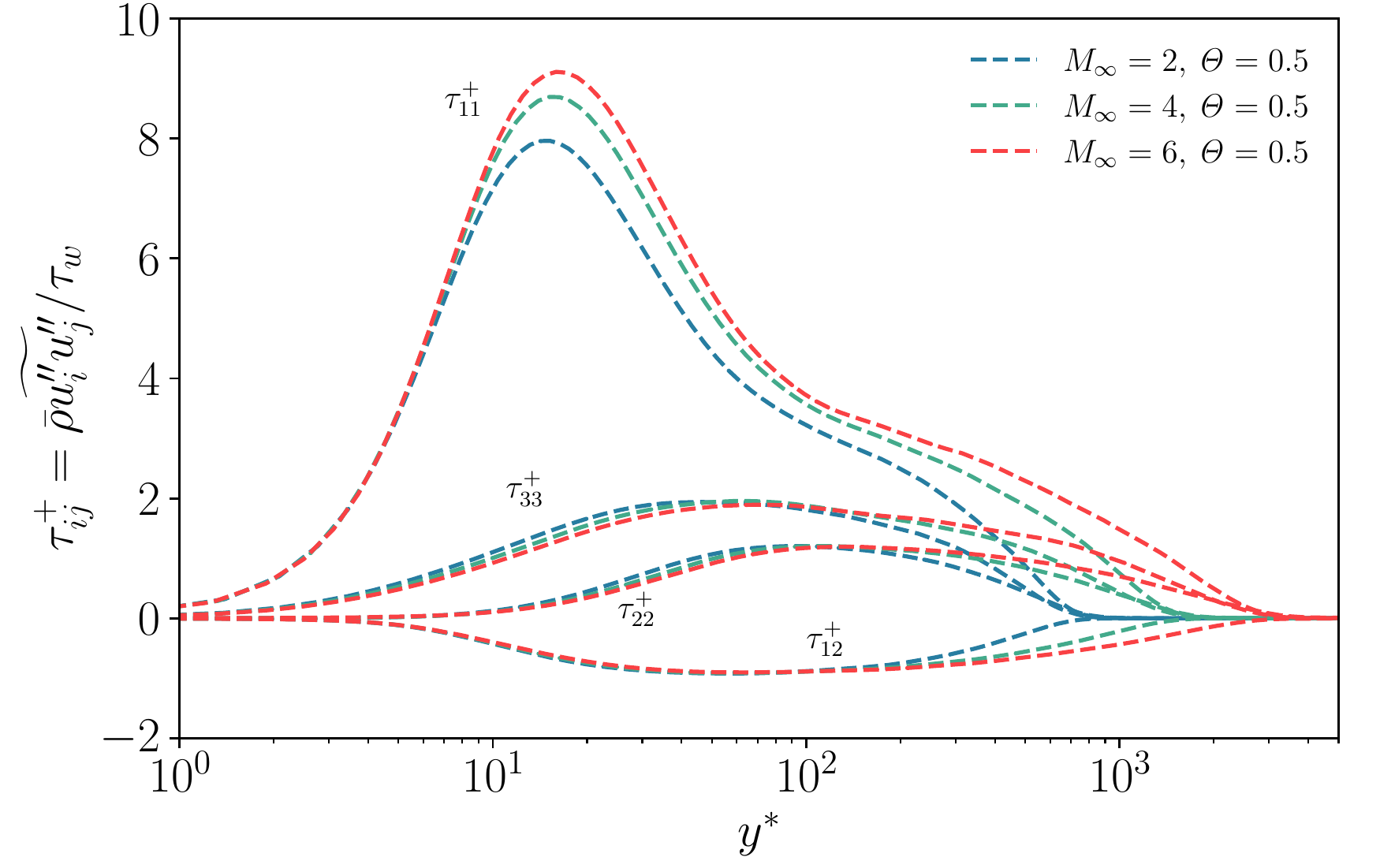}}
	\subfigure[PARAMETRI-2][$\mathit{\Theta}=0.5 $]{\includegraphics[width=0.48\textwidth]{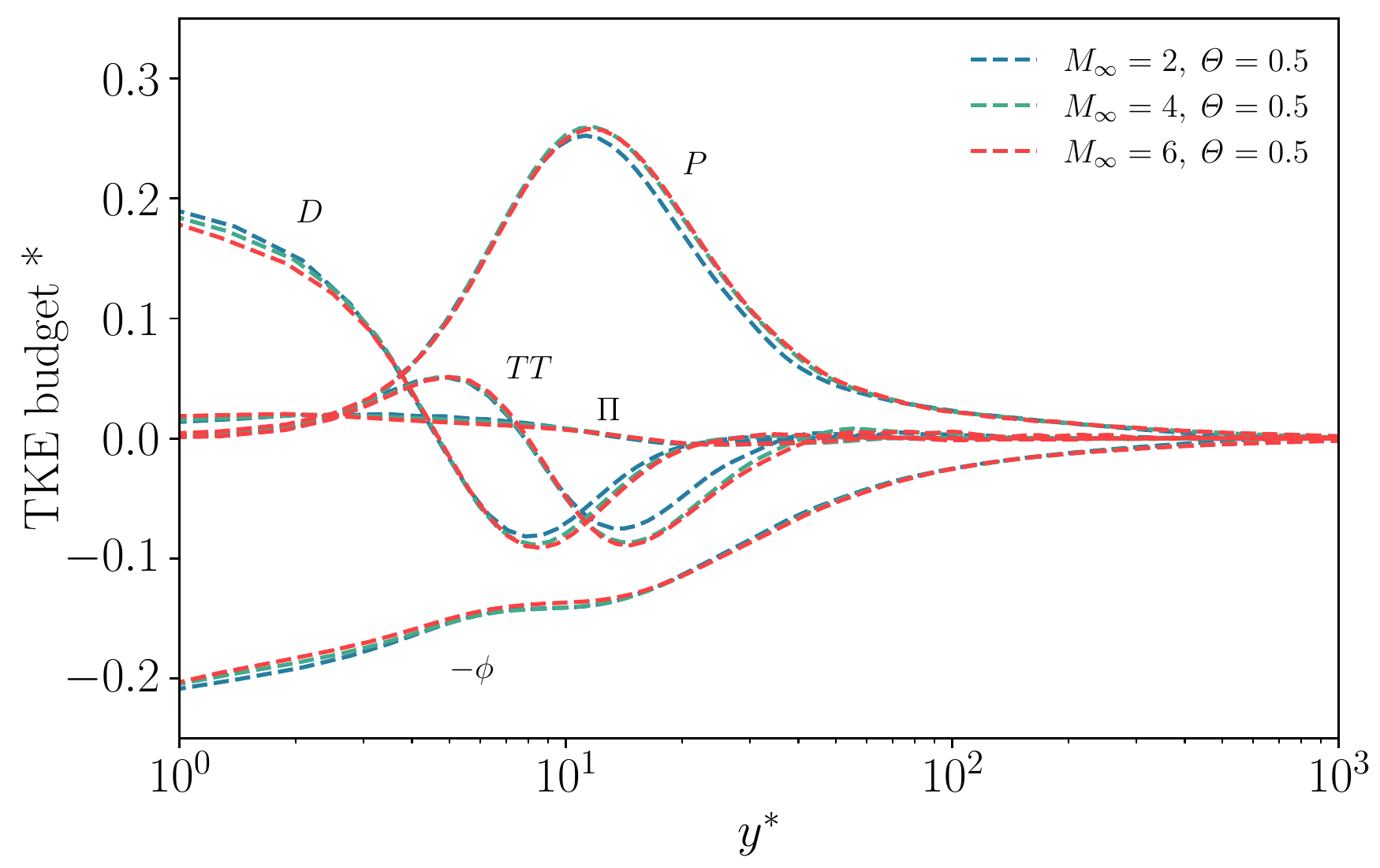}}\\
	\subfigure[PARAMETRI-2][$\mathit{\Theta}=0.75$]{\includegraphics[width=0.48\textwidth]{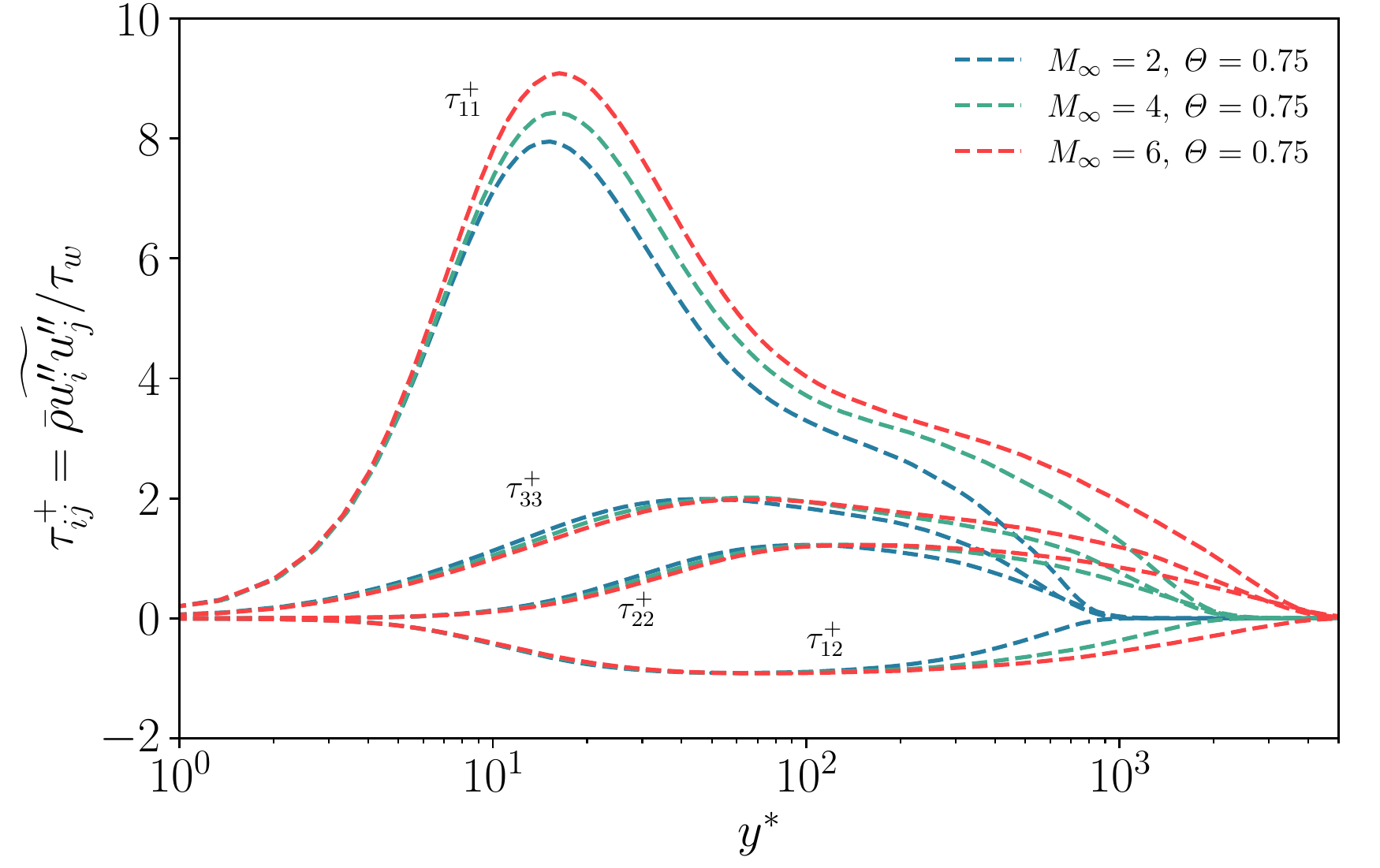}}
	\subfigure[PARAMETRI-2][$\mathit{\Theta}=0.75$]{\includegraphics[width=0.48\textwidth]{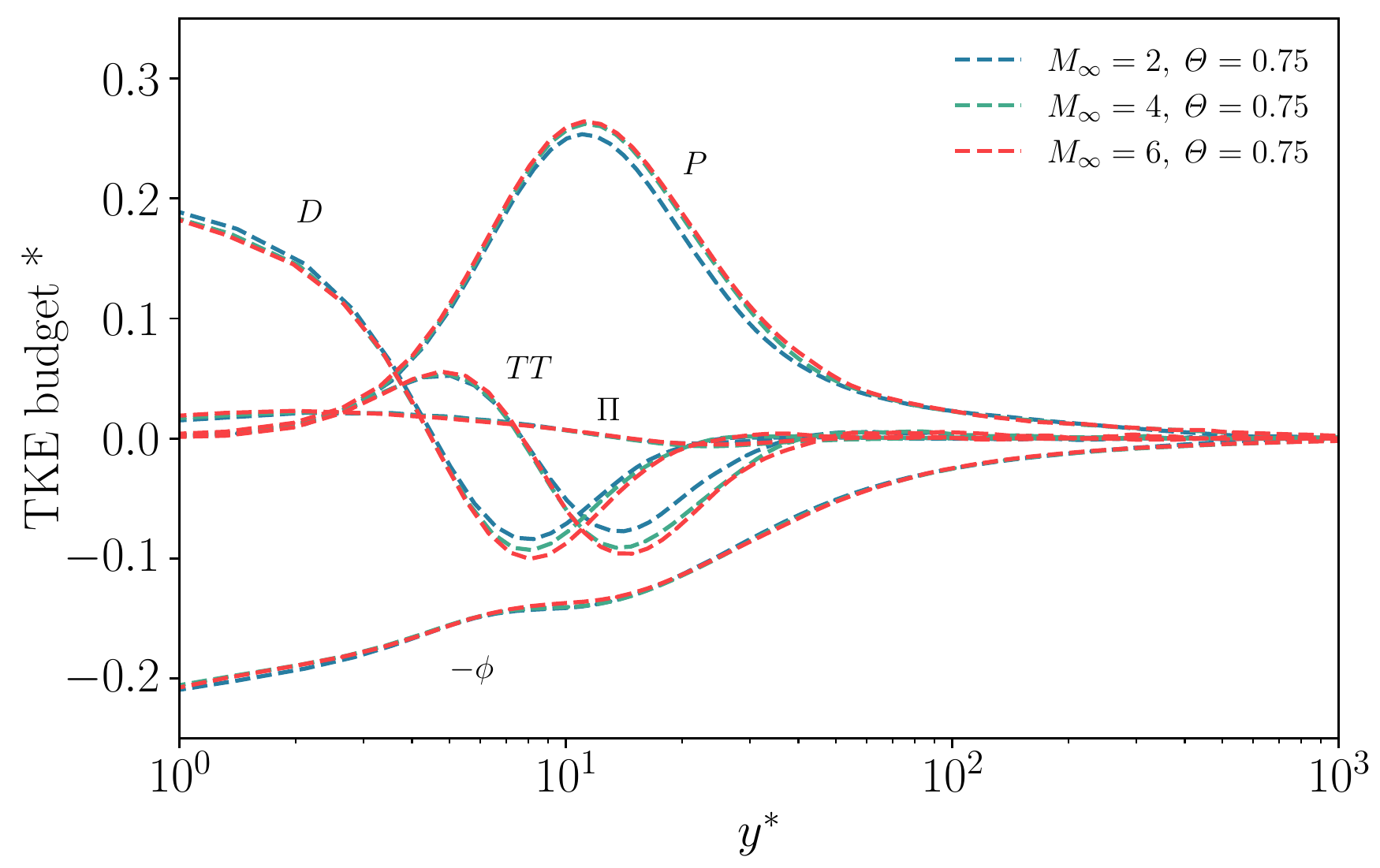}}\\
	\subfigure[PARAMETRI-2][$\mathit{\Theta}=1.0 $]{\includegraphics[width=0.48\textwidth]{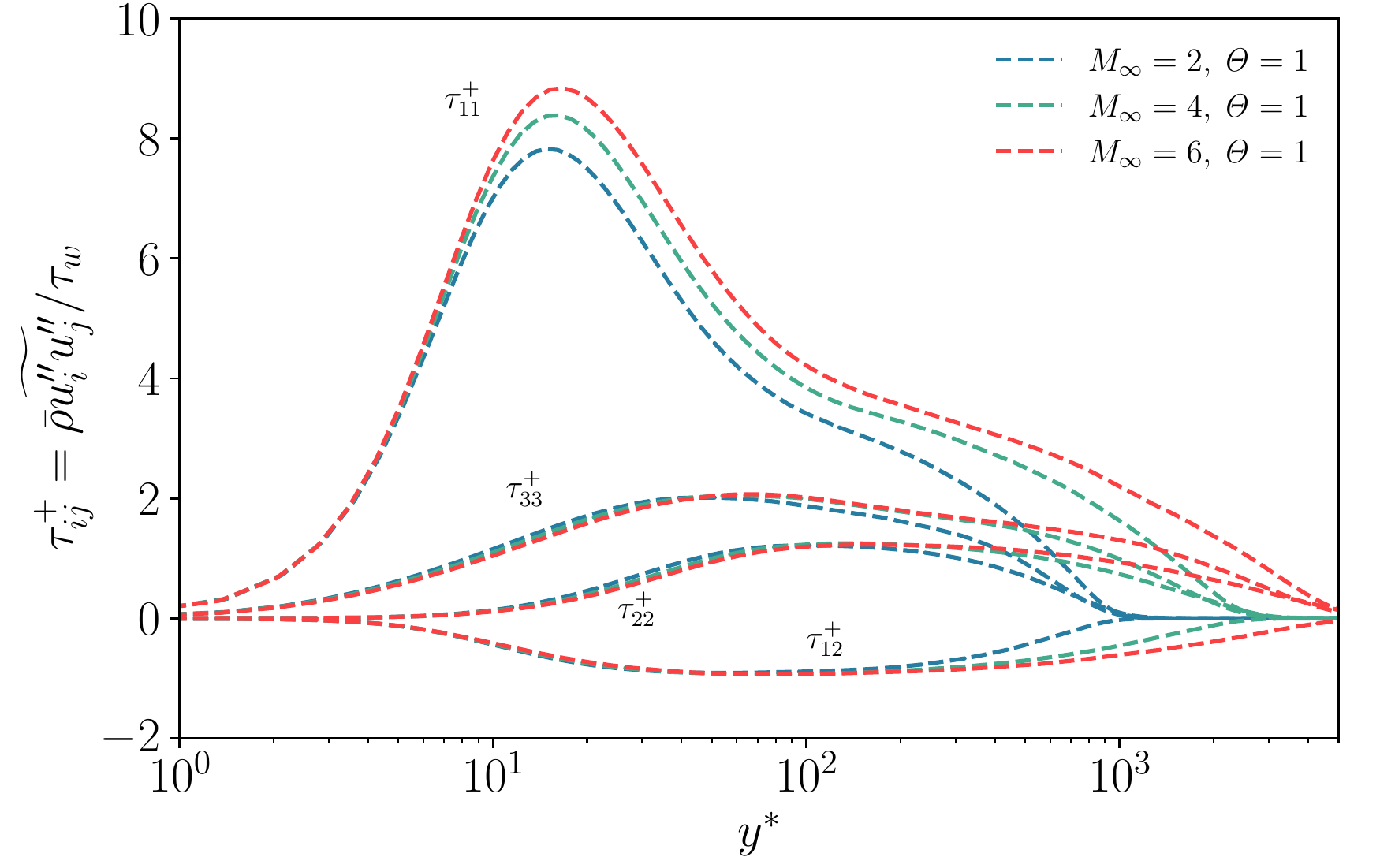}}
	\subfigure[PARAMETRI-2][$\mathit{\Theta}=1.0 $]{\includegraphics[width=0.48\textwidth]{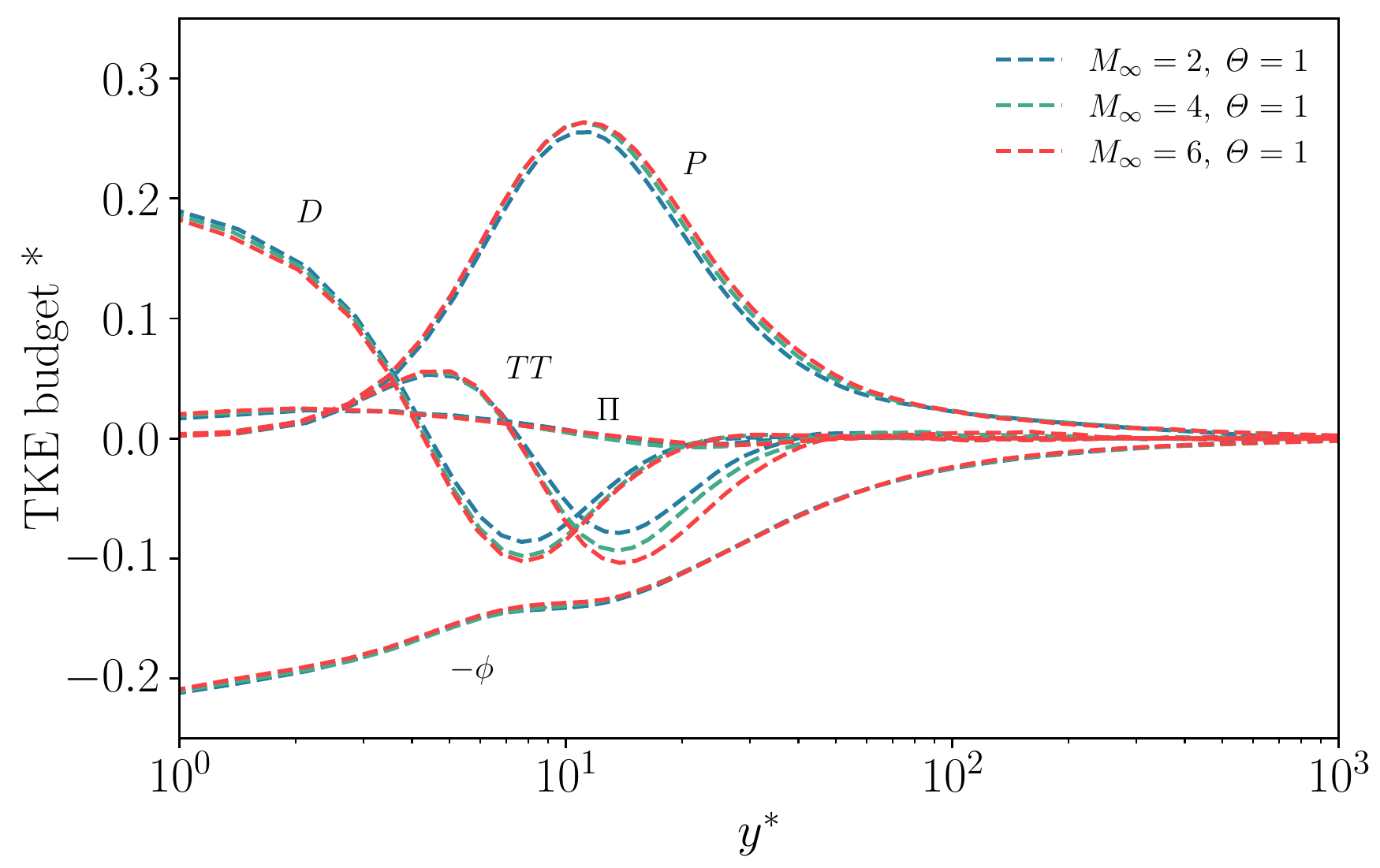}}
	\vspace{-0.2cm}
  \caption{Semilocal-scaled turbulent velocity fluctuations (a,c,e) and turbulent kinetic budget (b,d,f) as function of of the wall-normal distance $y^*$. 
  Here, different Mach number $M_{\infty}$ are compared at a given diabatic parameter $\mathit{\Theta}$. 
  \label{fig:velfluc_m}}
\end{figure}

The distribution of velocity fluctuation intensities and Reynolds shear stress is reported in the left panels of figures \ref{fig:velfluc_t} and \ref{fig:velfluc_m}, using the classical transformation of \citet{morkovin1962effects}:
\begin{equation}
\left(u_i^*\right)^2=\frac{\widetilde{u_i^{\prime \prime 2}}}{u_\tau^2} \frac{\bar{\rho}}{\bar{\rho}_w}, \quad(u v)^*=\frac{\widetilde{u^{\prime \prime} v^{\prime \prime}}}{u_\tau^2} \frac{\bar{\rho}}{\bar{\rho}_w}.
\end{equation}
The profiles are shown as a function of the wall-normal distance in semilocal scaling $y^*$ \citep{huang1995compressible}, considering its ability to collapse compressible profiles of different Mach numbers and wall temperature conditions, in particular with respect to the peak positions \citep{Zhang2018,wan2022wall}. Right panels of figures \ref{fig:velfluc_t} and \ref{fig:velfluc_m} show the corresponding turbulent kinetic energy budget terms,
(being $k=\widetilde{u_i^{\prime \prime} u_i^{''}}/2$ the turbulent kinetic energy, TKE)
 according to the derivation of \citet{Zhang2018}, where $P$ is the production term, $TT$ represents the turbulent transport, $\Pi$ includes the pressure diffusion and dilatation, $-\phi$ is the viscous dissipation and $D$ is the viscous diffusion.
 For these results, semilocal scaling is also employed in the normalisation of budget terms (refer to \citet{Zhang2018}) and for the wall-normal distance $y^*$, enabling a good collapse between different profiles \citep{Zhang2018,cogo2022direct}. 
The effect of wall-cooling on velocity fluctuations, shown in the left panels of figure \ref{fig:velfluc_t}, is apparent as an increase in the peak of the streamwise component located at $y^*\approx 15$ that is more prominent at high Mach numbers. In contrast, the spanwise component of highly cooled cases shows the opposite behaviour, being reduced in intensity compared to the adiabatic reference. 
This implies an increase in the anisotropy of normal components of Reynolds stresses in the near-wall region, which is discussed in more detail at the end of this section.
The semilocal scaling provides an excellent collapse of the peak positions for all cases, preventing the outward shift that is present for cold cases when plotted in wall units (not shown). 
This is also true for the position of the turbulent production peak (right panels of figure \ref{fig:velfluc_t}), which would move farther from the wall if displayed in wall units. 
In general, the effect of wall-cooling on the turbulent kinetic energy budget is marked in the very near-wall region, especially at high Mach numbers, while all profiles progressively collapse in the outer layer.

%
The effect of the Mach number on velocity fluctuations is reported in the left panels of figure \ref{fig:velfluc_m}, where an increase of the streamwise component peak with the Mach number is apparent, while the other normal components intensities are observed to weakly decrease until $y^*<40$.
Unlike the wall-cooling effect, all normal components increase in the log layer as $M_{\infty}$ increases.
We note that this effect could be reduced at the BL edge by matching the semilocal friction Reynolds number $Re_{\tau}^*$ in place of the conventional definition (see table \ref{Tab:table1}), which would allow all profiles to collapse when $y^* \approx Re_{\tau}^*$.
In fact, $Re_{\tau}^*$ has been shown by several authors to better incorporate compressibility and wall-cooling effects on the separation of scales in highly-compressible flows (e.g. \citet{griffin2021velocity,hirai2021effects}).
This suggests that compressibility acts in the direction of increasing the scale separation in the outer layer, while wall-cooling has the opposite effect \citep{fan2022energy}.
For all values of $\mathit{\Theta}$, the turbulent kinetic energy budget (right panels of figure \ref{fig:velfluc_m}), shows an increase of the production term $P$ in the buffer and log layers as the Mach number increases and a corresponding decrease of diffusion $D$ and turbulent transport $TT$ in the same regions, consistently with \citet{cogo2022direct}, who noted the presence of this effect also in the outer region at higher $Re_{\tau}$.

While the effect of wall-cooling on the TKE budget seems confined in the near-wall region, the influence of Mach number is more prominent after the peak of production and throughout the log layer.

Further insights on the mechanism of redistribution of turbulent kinetic energy in the near-wall region can be gained by looking at the ratio between the streamwise component of the pressure-strain term and the streamwise component of turbulent production \citep{Duan2010}:
\begin{equation}
\mathcal{R} = \left(\overline{p^{'} \frac{\partial u^{''}}{\partial x}}\right) /\left(\overline{\rho u^{''} v^{''}}\frac{\partial \tilde{u} }{\partial y}\right)
\end{equation}
which is a measure of the energy transfer from the streamwise velocity fluctuations to the others.
\begin{figure}  
	\centering
	\subfigure[PARAMETRI-1][$\mathit{\Theta}=1.0$ \label{fig:R_a}]{\includegraphics[width=0.48\textwidth]{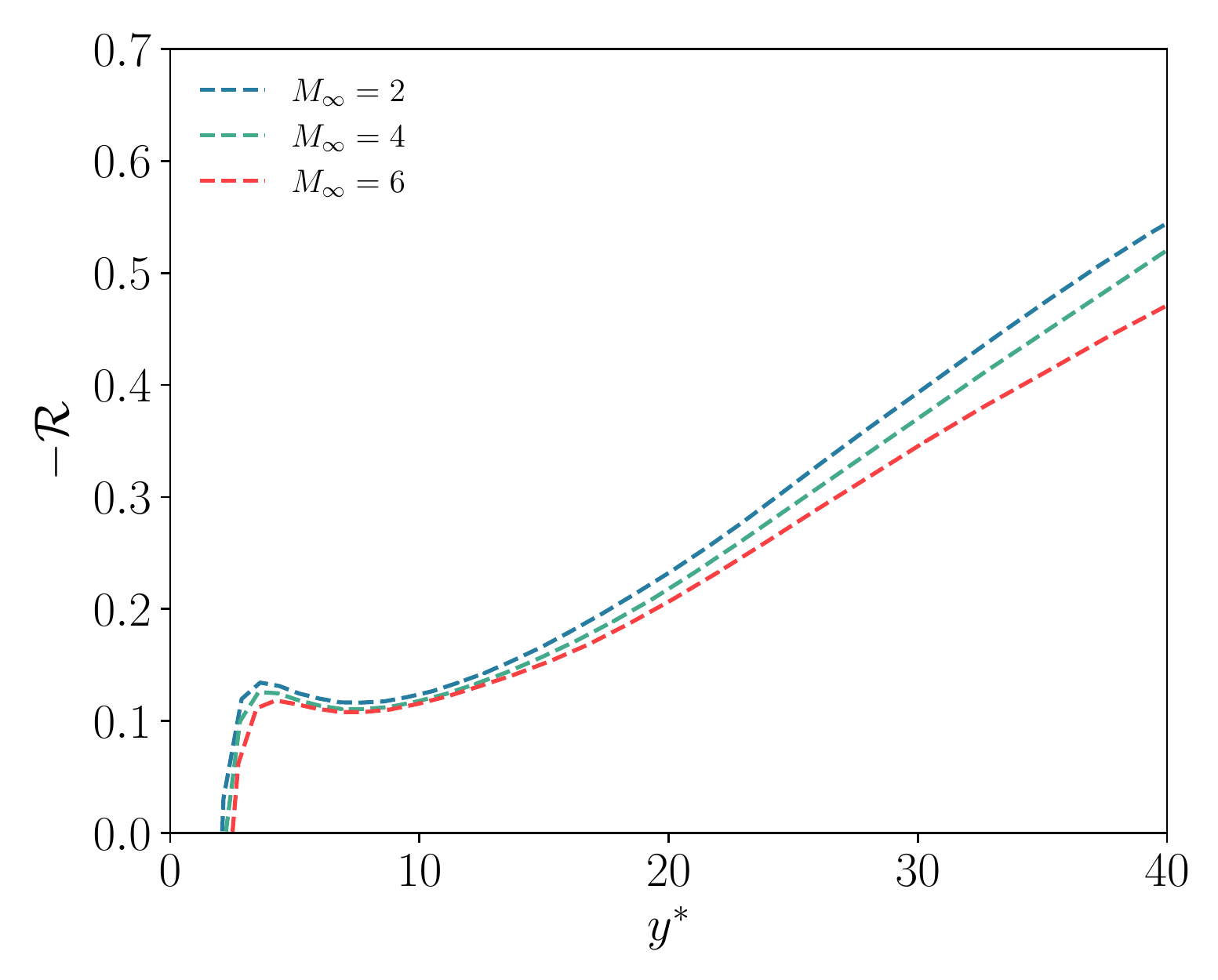}}
	\subfigure[PARAMETRI-1][$M_{\infty}=6$ \label{fig:R_b}]{\includegraphics[width=0.48\textwidth]{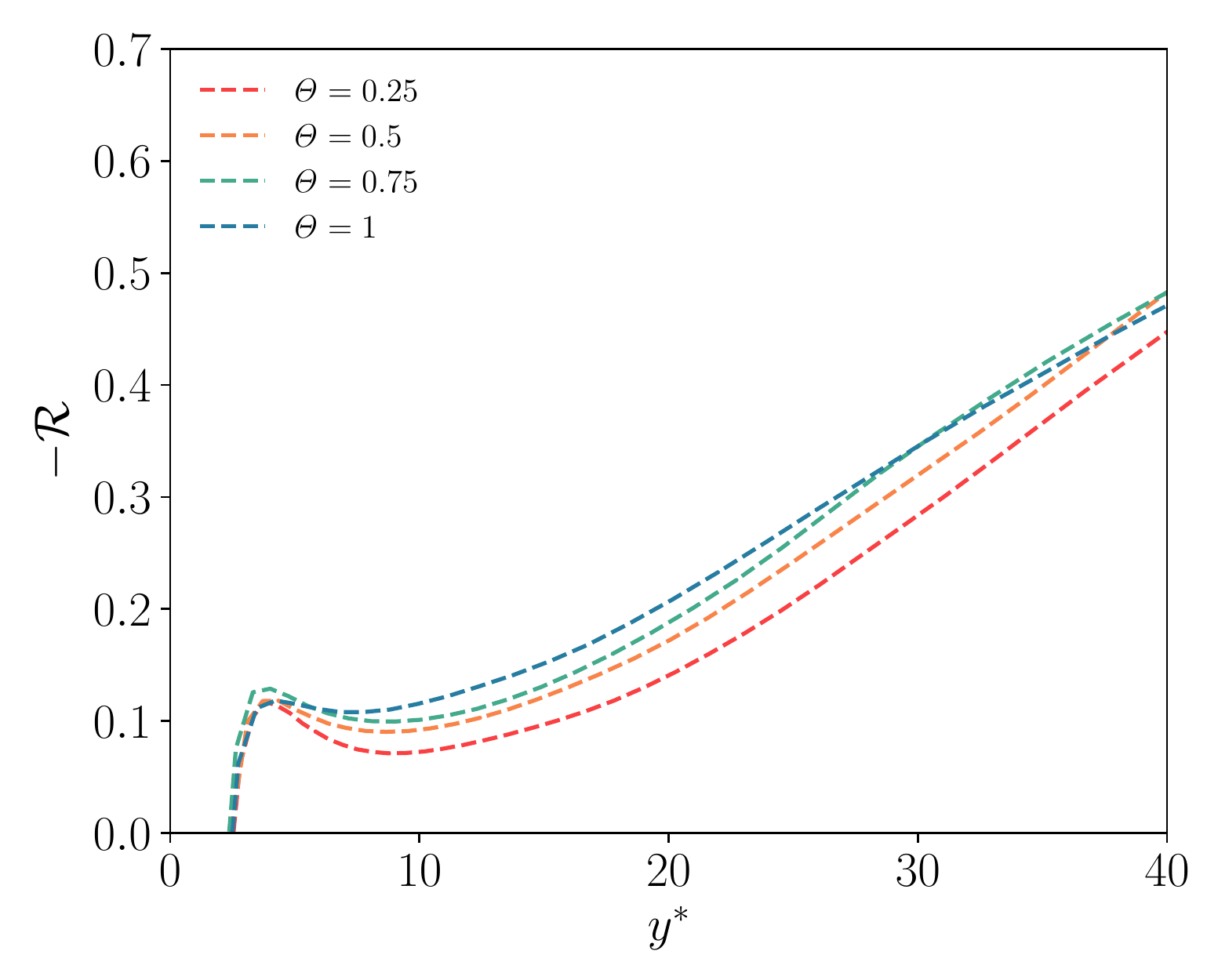}}
	\vspace{-0.2cm}
  \caption{Ratio of streamwise components of pressure-strain and turbulent production terms for cases at (a) $\mathit{\Theta}=1.0$ and (b) $M_{\infty}=6$, as function of the wall-normal distance in semilocal units. \label{fig:Ratio}}
\end{figure}
The role of the pressure-strain term in increasing turbulence anisotropy was also noted for other flows (e.g. \citet{foysi2004compressibility}).
To gauge the respective effects of the Mach number and the wall temperature condition, figure \ref{fig:Ratio} compares $-\mathcal{R}$ for cases at $\mathit{\Theta}=1.0$ (figure \ref{fig:R_a}) and at $M_{\infty}=6$ (figure \ref{fig:R_b}). 
In panel \ref{fig:R_a}, profiles of $-\mathcal{R}$ are reduced in magnitude as compressibility increases, with greater intensity farther from the wall.

This is consistent with the less efficient redistribution of turbulent kinetic energy discussed before, and is attributed to the absence of a solenoidal condition for the velocity field for highly-compressible cases preventing an efficient energy transfer between velocity components.
Looking at panel \ref{fig:R_b}), we observe that wall-cooling acts similarly to an increase of compressibility, strongly decreasing the profiles of $-\mathcal{R}$, but its effect is localised in the near-wall region and strongly reduced after the buffer layer.

We attribute this effect to a localised stratification of flow properties in the near-wall region.
As wall-cooling is increased and the location of the mean temperature peak approaches the buffer layer, the flow above and below the peak location is relatively colder and denser. This is true for all Mach numbers in our database (although with different intensities), since the temperature peak location remains unaffected (see figure \ref{fig:t_peaks}).
This localised stratification forces turbulent fluctuations to be active almost only in the streamwise direction, while the other components tend to be suppressed.
\begin{figure}  
	\centering
	\subfigure[PARAMETRI-1][$\mathit{\Theta}=1.0$ \label{fig:tri_a}]{\includegraphics[width=0.48\textwidth]{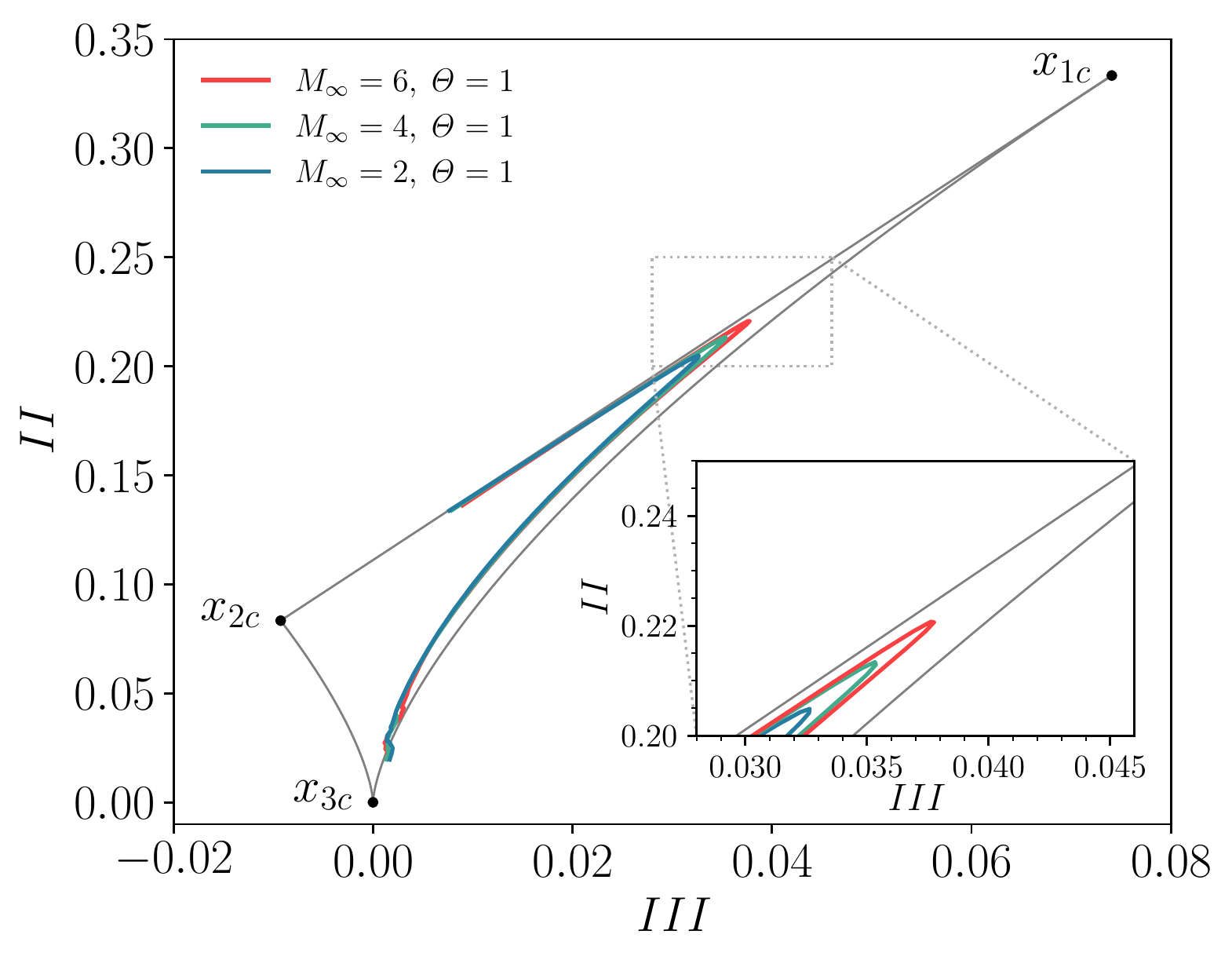}}
	\subfigure[PARAMETRI-1][$M_{\infty}=6$ \label{fig:tri_b}]{\includegraphics[width=0.48\textwidth]{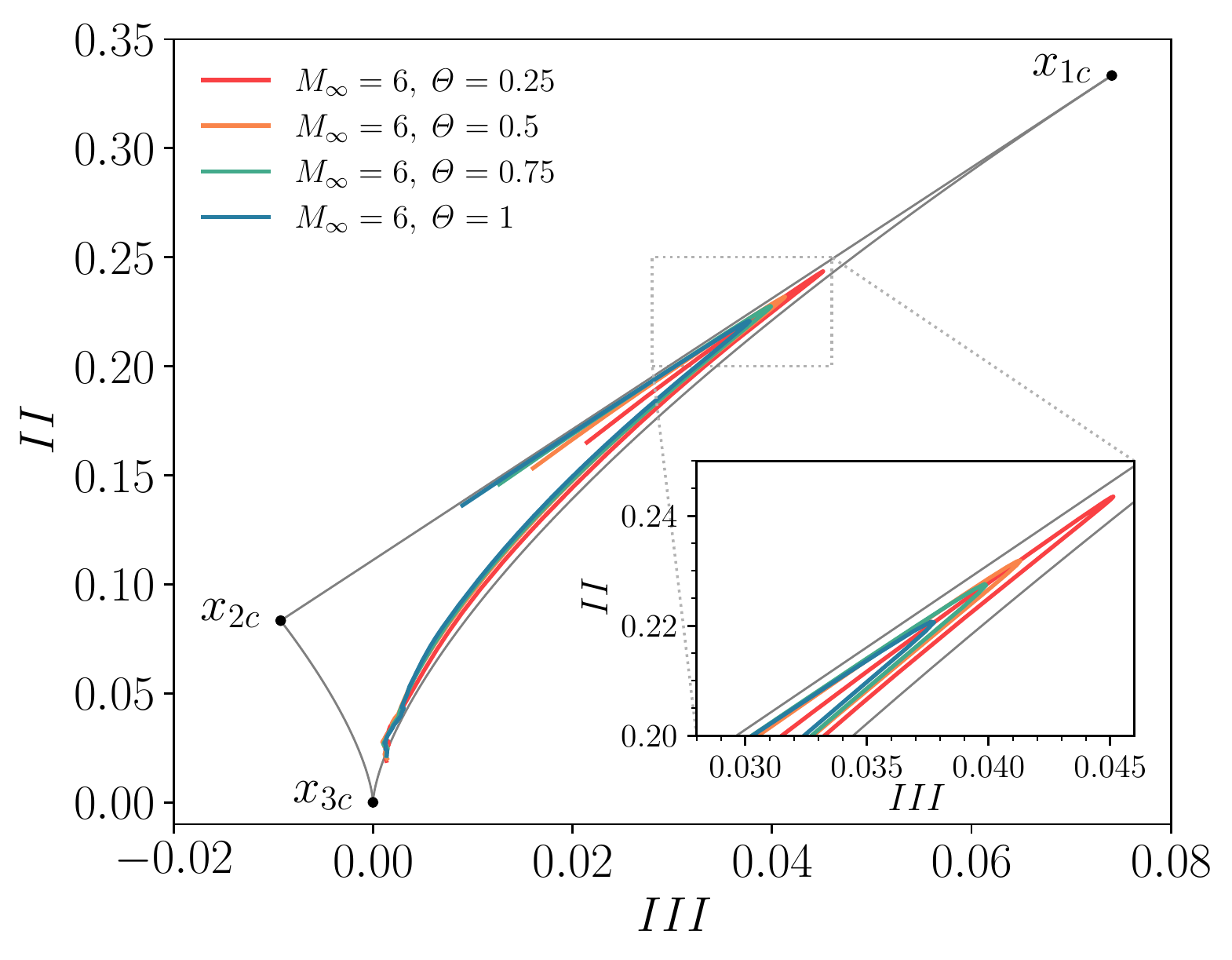}}
	\vspace{-0.2cm}
  \caption{Diagram of the anisotropy invariant map of \citet{lumley1977return} for cases at (a) $\mathit{\Theta}=1.0$ and (b) $M_{\infty}=6$. \label{fig:ltri}}
\end{figure}
This effect is quantified in figure \ref{fig:ltri} by showing the anisotropic invariant map of \citet{lumley1977return}, also known as the Lumley triangle, which uses the second ($II$) and third ($III$) principal components of turbulence anisotropy.
The invariant map is composed of three limiting states: one-component ($x_{1c}$), axisymmetric two-component ($x_{2c}$) and isotropic ($x_{3c}$); which are representative of the relative strengths of the fluctuating velocity components.
Looking at figure \ref{fig:tri_a}, we note that the cusp point, which coincides approximately with the peak of velocity fluctuations in the buffer layer, shifts towards a one-component behaviour ($x_{1c}$) as $M_{\infty}$ increases.
This effect is strongly enhanced by wall cooling, figure \ref{fig:tri_b}, which further promotes the one-dimensional state of the flow.

We note that although this effect resembles a promotion of compressibility, as noted by several authors \citep{duan2011direct,chu2013effect}, the underlying mechanism is strongly different and relevant only when $M_{\infty}$ is high.
In fact, different wall-cooling and compressibility signatures are clearly noted for other effects, such as their effect on scale separation and their region of influence through the BL.

\begin{figure}  
	\centering
	\subfigure[PARAMETRI-1][$\mathit{\Theta}=1.0$ \label{fig:l_a}]{\includegraphics[width=0.48\textwidth]{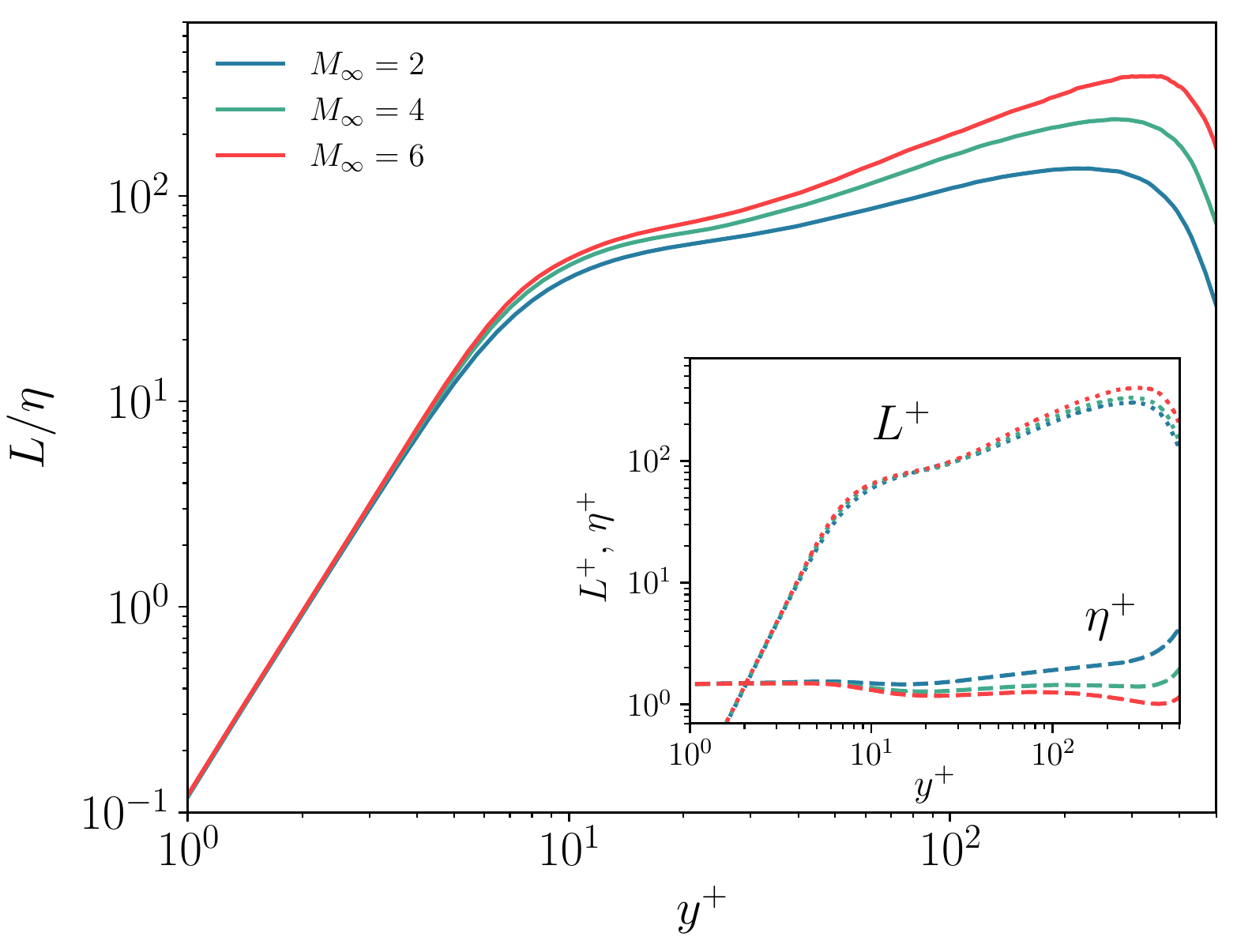}}
	\subfigure[PARAMETRI-1][$M_{\infty}=6$ \label{fig:l_b}]{\includegraphics[width=0.48\textwidth]{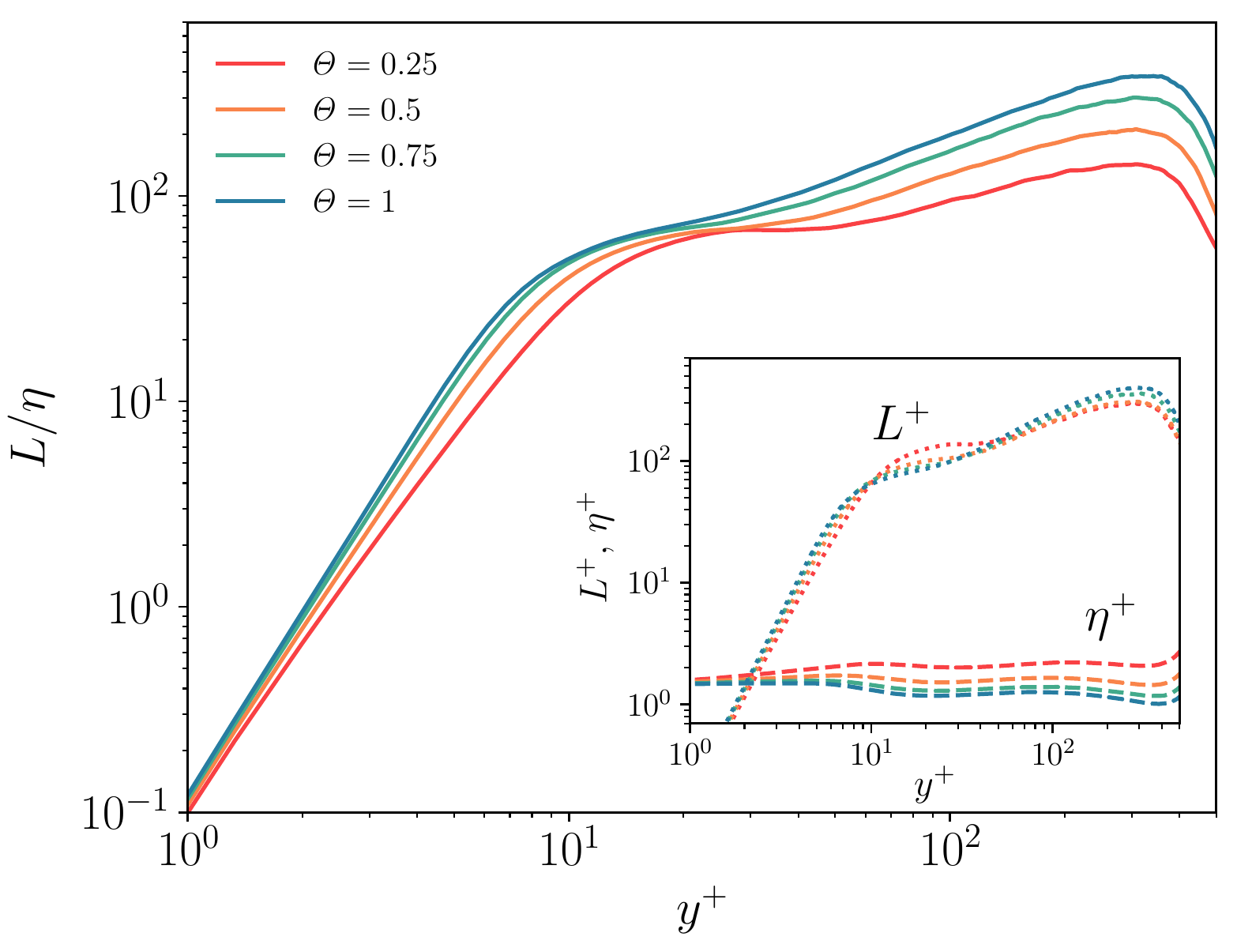}}
	\vspace{-0.2cm}
  \caption{Ratio of integral length scale $L$ and Kolmorov scale $\eta$ for cases at (a) $\mathit{\Theta}=1.0$ and (b) $M_{\infty}=6$, as function of the wall-normal distance $y^+$. The inset shows separately $L^+$ and $\eta^+$, normalized with the viscous length $\delta_{\nu}$. \label{fig:lscales}}
\end{figure}

To provide further insights on these differences, we analyse the characteristic turbulent lengths.
We consider the length scale characterising large eddies as $L=\bar{\rho} k^{3/2}/ \phi$ \citep{pope2000turbulent}, and the Kolmogorov length scale $\eta=\left[(\bar{\mu} / \bar{\rho})^3 /(\phi / \bar{\rho})\right]^{0.25}$ for the smallest ones, with $\phi$ being the local dissipation rate of TKE.
The ratio of these two scales, reported in figure \ref{fig:lscales}, measures the separation between large and small scales, which in our discussion can be ascribed to the effect of $M_{\infty}$ and $\mathit{\Theta}$ numbers (since $Re_{\tau}$ is fixed).
In agreement with previous observations, figure \ref{fig:l_a} shows that the separation of scales in the outer layer increases with the Mach number, while the opposite behaviour is found reducing the diabatic parameter $\mathit{\Theta}$, see figure \ref{fig:l_b}.
The insets in figures \ref{fig:l_a} and \ref{fig:l_b} show the individual change of $L^+=L/ \delta_{\nu}$ and $\eta^+=\eta/ \delta_{\nu}$, revealing that $M_{\infty}$ and $\mathit{\Theta}$ strongly affect the Kolmogorov length $\eta^+$, with a minor impact on large scales $L^+$, influencing the separation of large to small scales $L/ \eta$ in the outer layer.

On this aspect, we remark that while an increase in compressibility, i.e. $M_{\infty}$, reduces the Kolmogorov length, the opposite holds decreasing the wall temperature, i.e. $\mathit{\Theta}$.

The variation of $L/\eta$ in the outer layer is effectively captured by the change of $Re_{\tau}^* = \mu_w / \mu_{\infty}  \sqrt{\rho_{\infty} / \rho_w} Re_{\tau}$ (see table \ref{tab:stat}), which better account for density and viscosity variations in the outer layer.
It should be noted, however, that the definition of a single similarity parameter among different flow cases concerning the scale separation is prevented by the strong change of flow properties across the BL.
In particular, while $Re_{\tau}$ essentially regulates the outer-inner scale separation, i.e. $L^+$, $Re_{\tau}^*$ controls the large-small scale separation in the outer layer, i.e. $L/ \eta$. These two variables are strongly related in incompressible flows and both growing functions of $y^+$ in the log-layer \citep{pope2000turbulent}, while they appear to be decoupled for highly compressible flows due to the influence of $M_{\infty}$ and $\mathit{\Theta}$ numbers.
For this reason, specific flow features associated with the outer-inner scales separation, such as the enhancement of outer layer motions at high $Re_{\tau}$ \citep{cogo2022direct}, are not visible here, even though $L/ \eta$ actually increases in the outer layer.

\subsection{Thermodynamic quantities}\label{sec:thermo_fluc}
\begin{figure} 
	\centering
	\subfigure[PARAMETRI-1][$M_{\infty}=2$]{\includegraphics[width=0.48\textwidth]{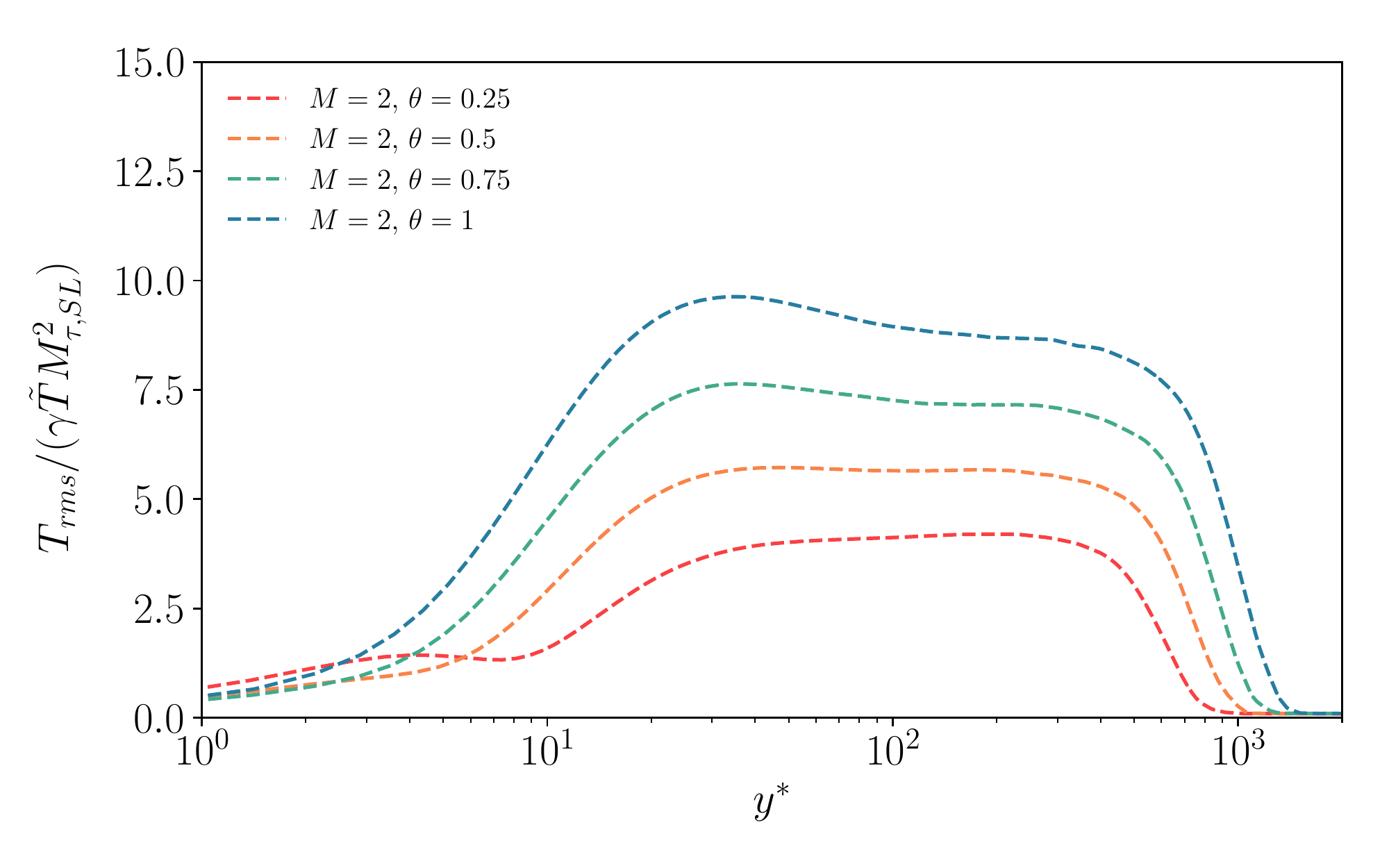}}
	\subfigure[PARAMETRI-1][$M_{\infty}=2$]{\includegraphics[width=0.48\textwidth]{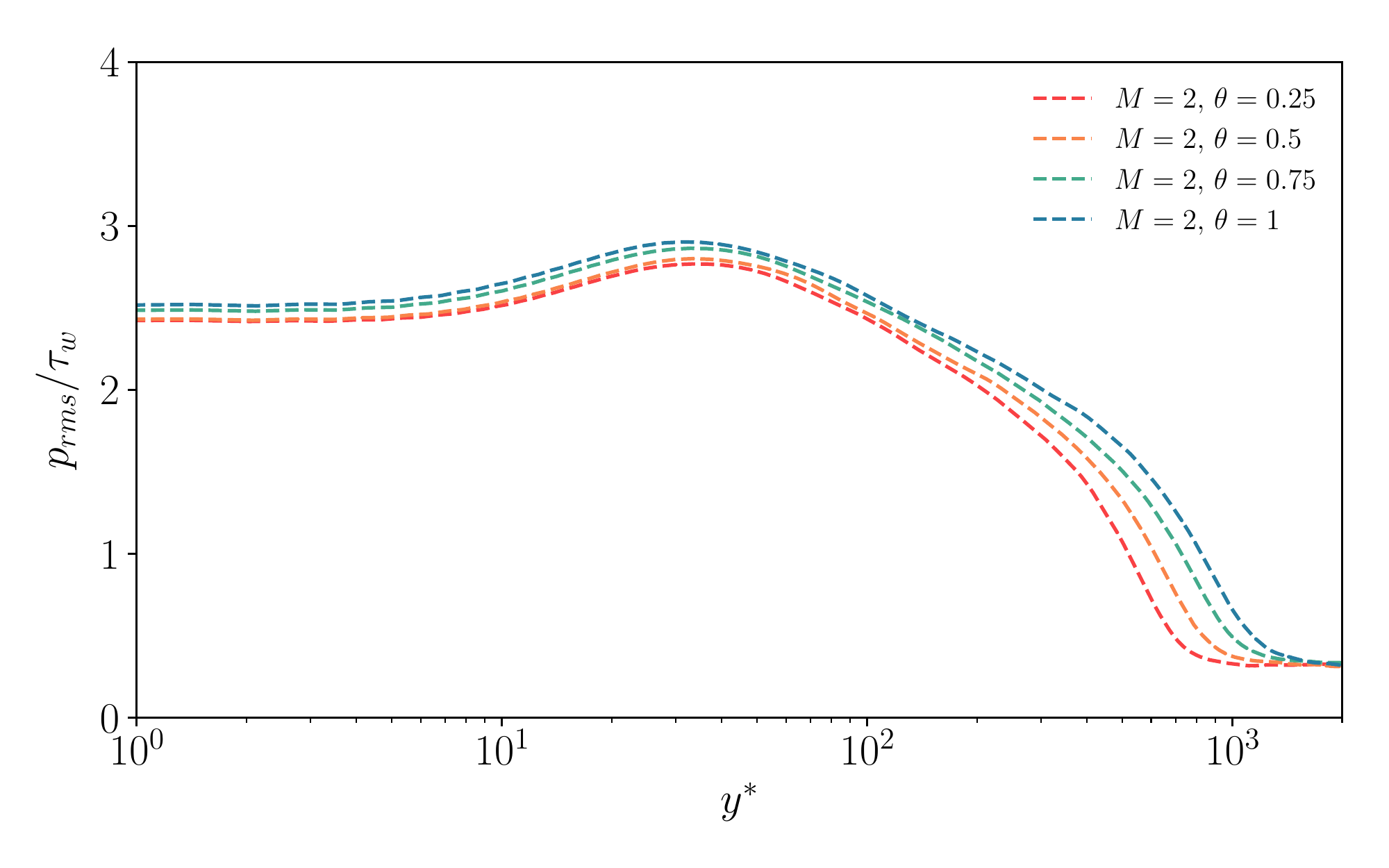}}\\
	\subfigure[PARAMETRI-2][$M_{\infty}=4$]{\includegraphics[width=0.48\textwidth]{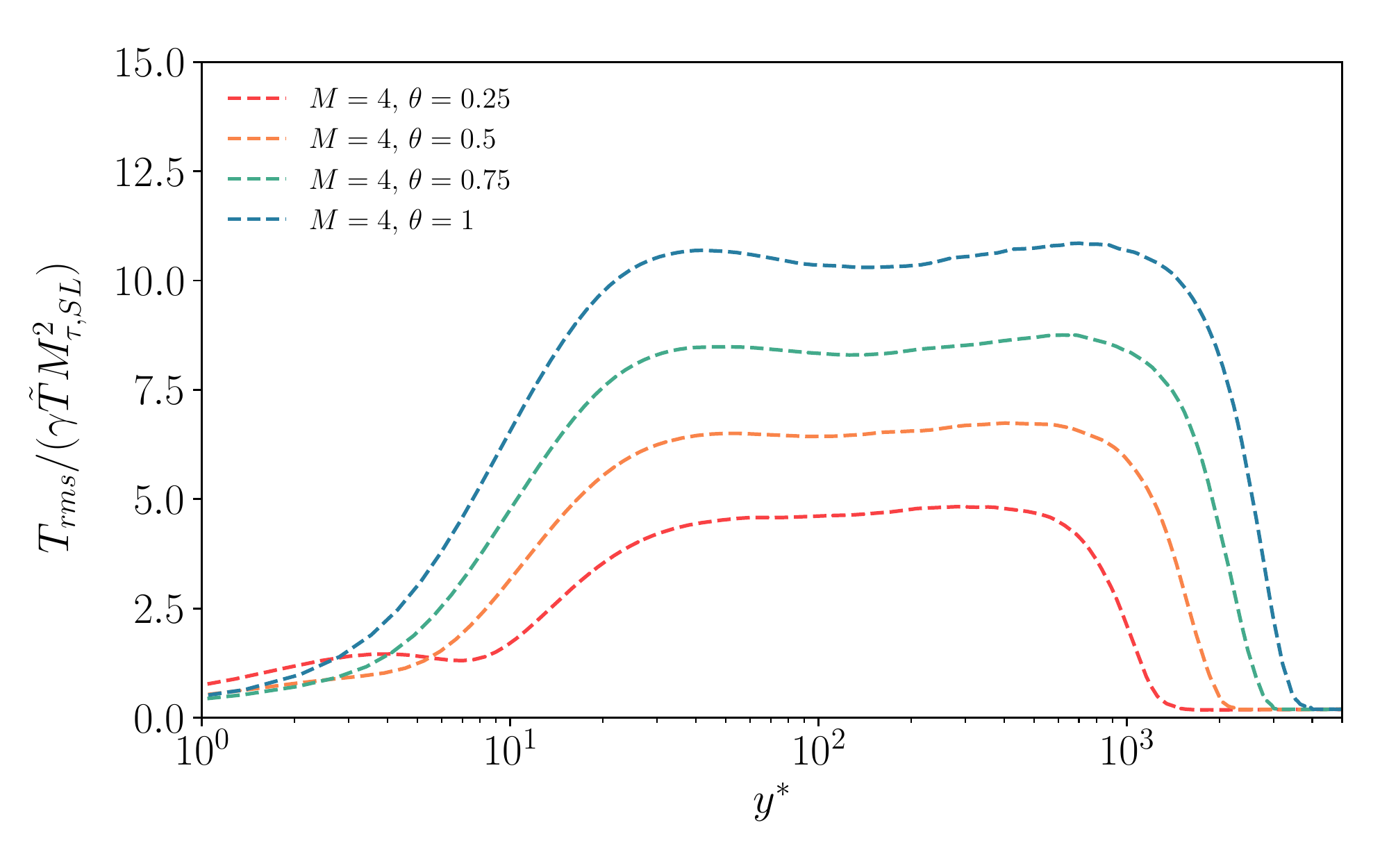}}
	\subfigure[PARAMETRI-2][$M_{\infty}=4$]{\includegraphics[width=0.48\textwidth]{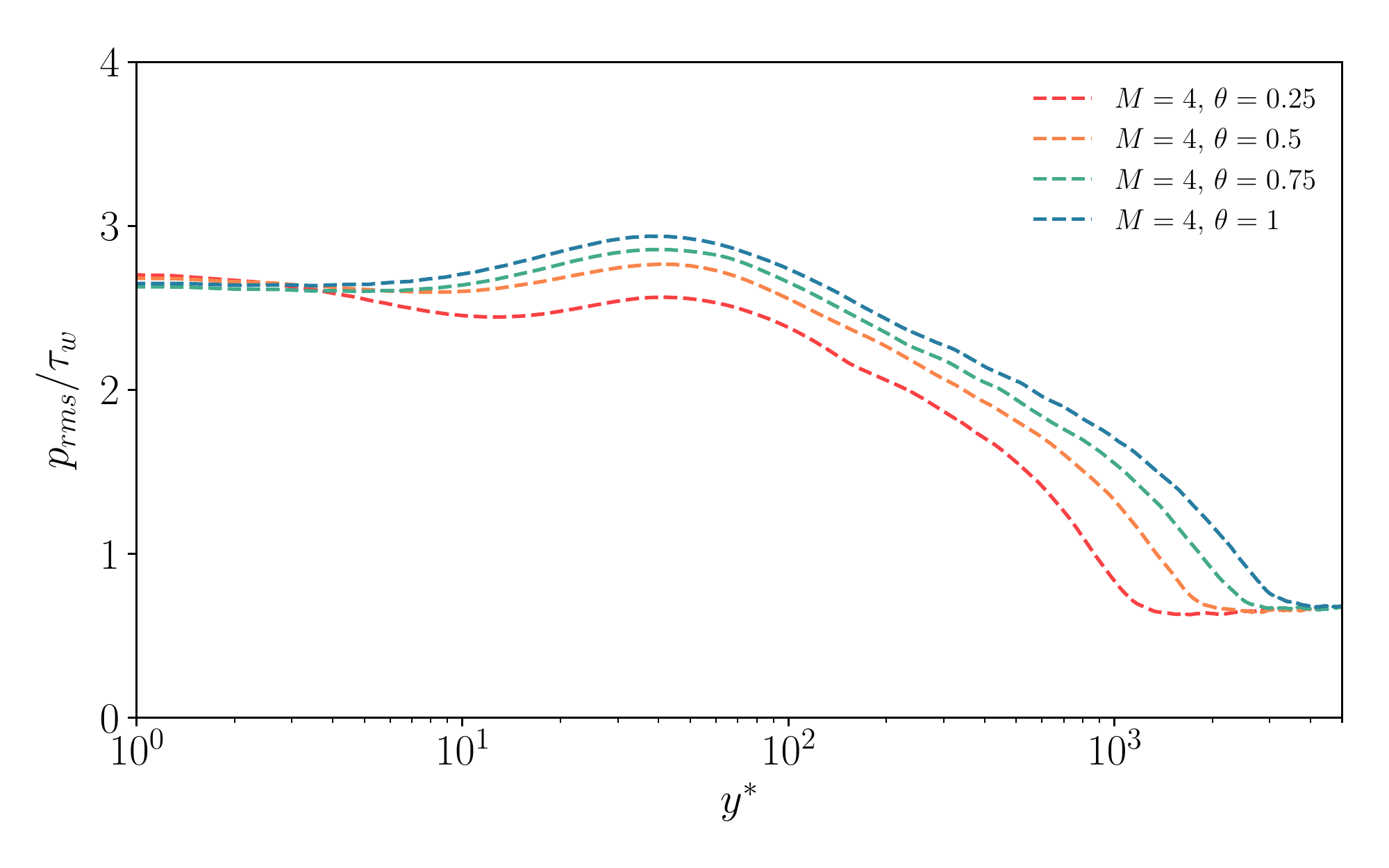}}\\
	\subfigure[PARAMETRI-2][$M_{\infty}=6$]{\includegraphics[width=0.48\textwidth]{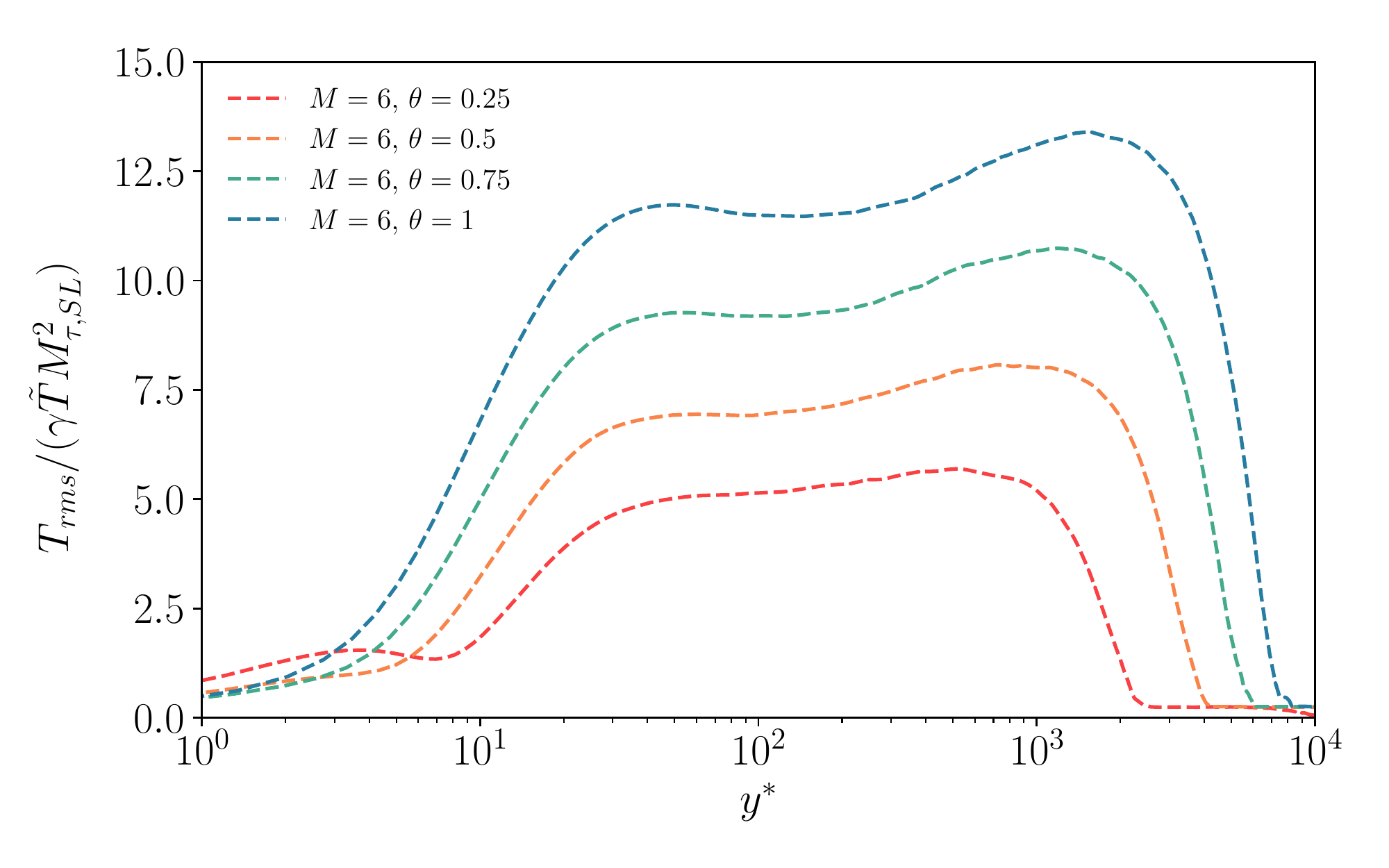}}
	\subfigure[PARAMETRI-2][$M_{\infty}=6$]{\includegraphics[width=0.48\textwidth]{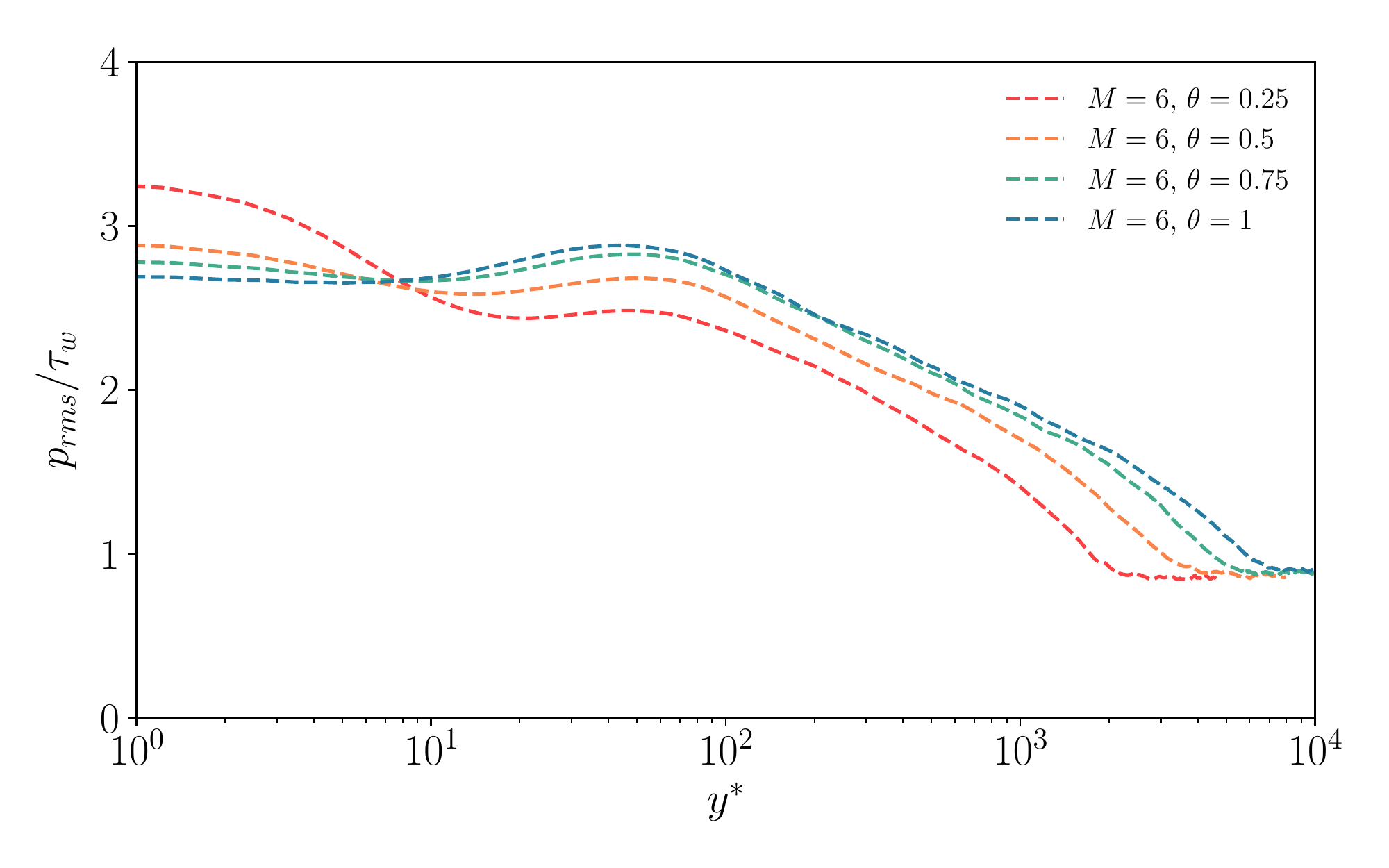}}
	\vspace{-0.2cm}
	\caption{Profiles of RMS temperature (left) and pressure (right) in semilocal scaling. Here, different diabatic parameters $\mathit{\Theta}$ are compared at a given Mach number $M_{\infty}$. \label{fig:thermo_m}}
\end{figure}

\begin{figure} 
	\centering
	\subfigure[PARAMETRI-1][$\mathit{\Theta}=0.25$]{\includegraphics[width=0.48\textwidth]{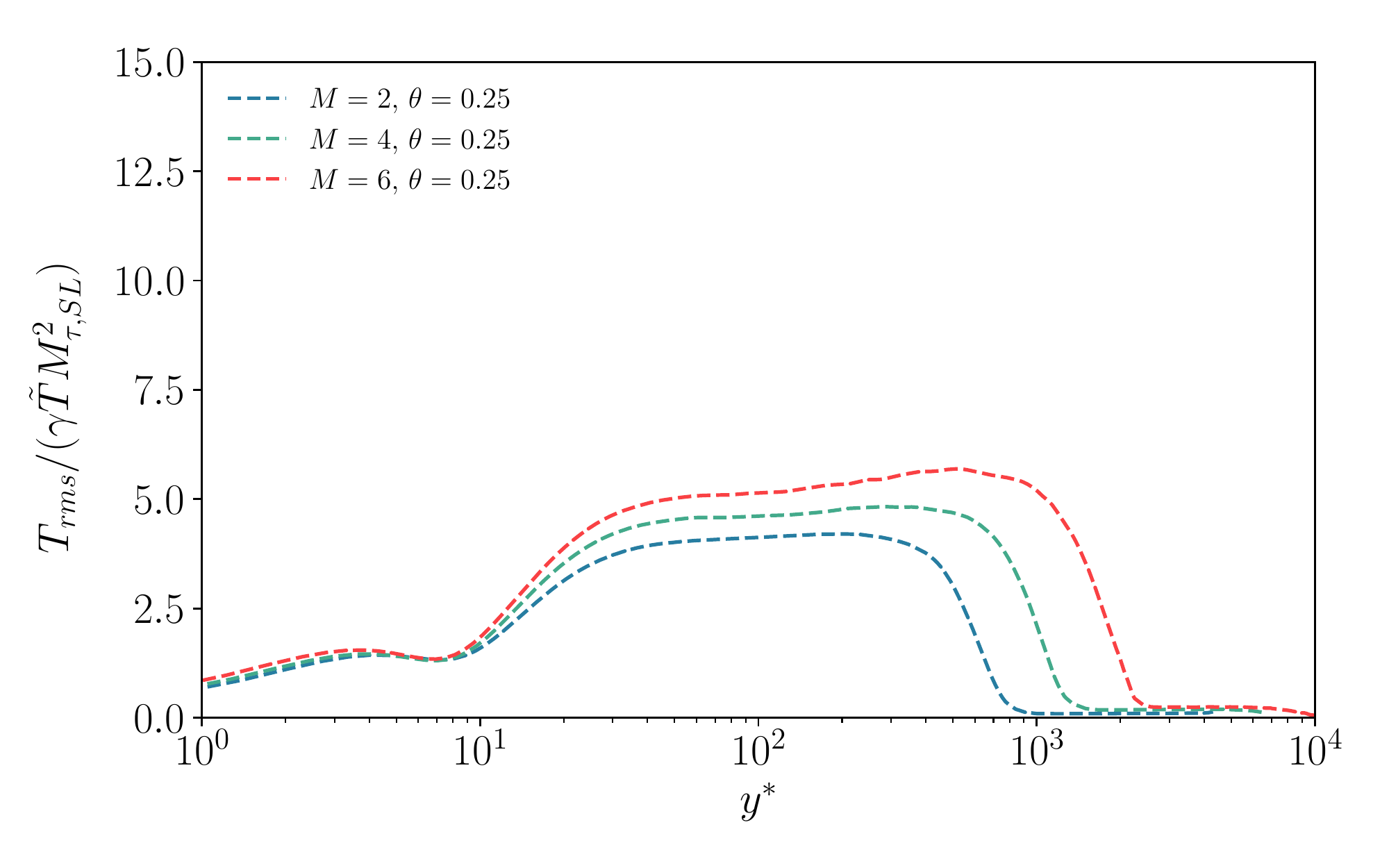}}
	\subfigure[PARAMETRI-1][$\mathit{\Theta}=0.25$]{\includegraphics[width=0.48\textwidth]{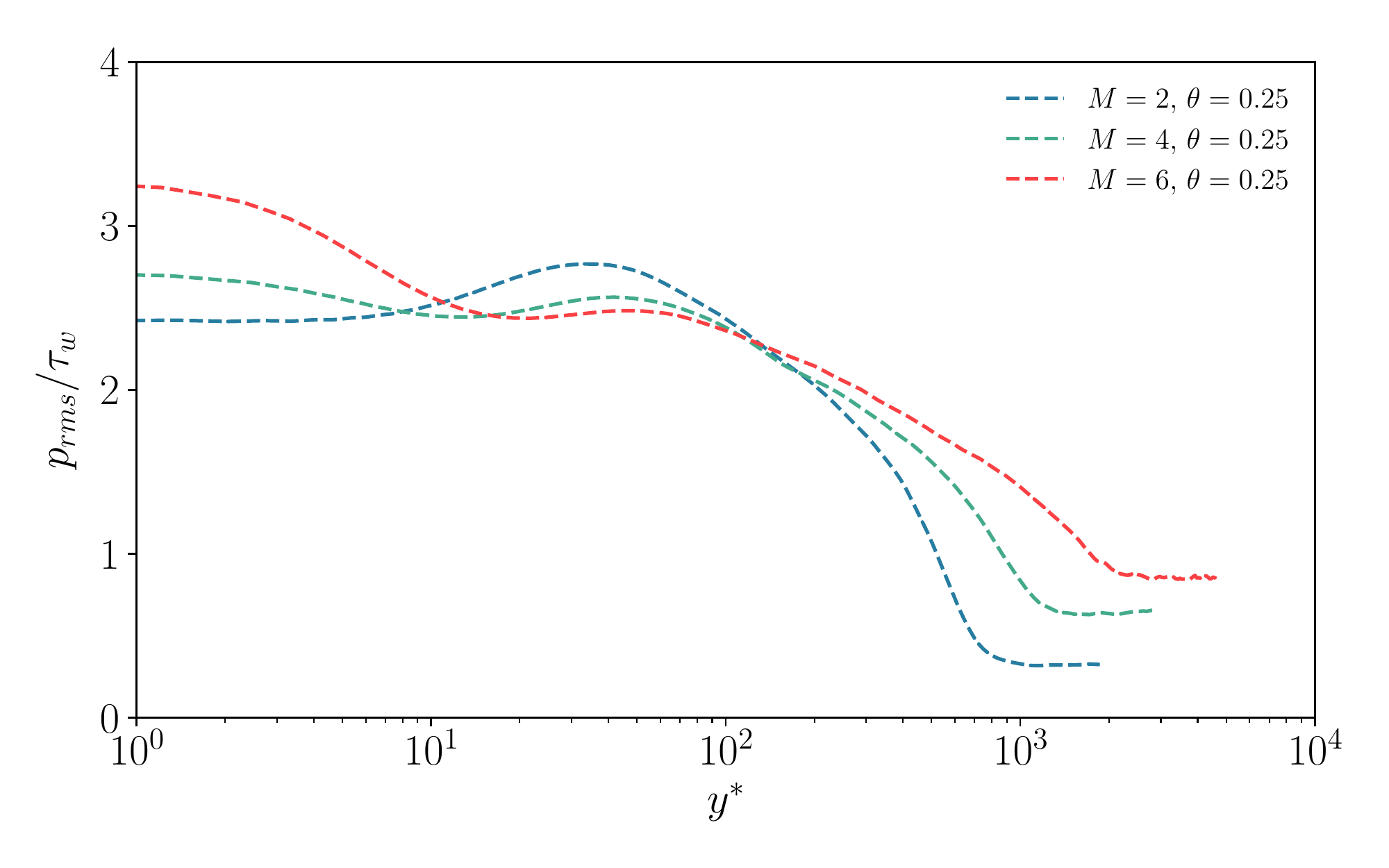}}\\
	\subfigure[PARAMETRI-2][$\mathit{\Theta}=0.5 $]{\includegraphics[width=0.48\textwidth]{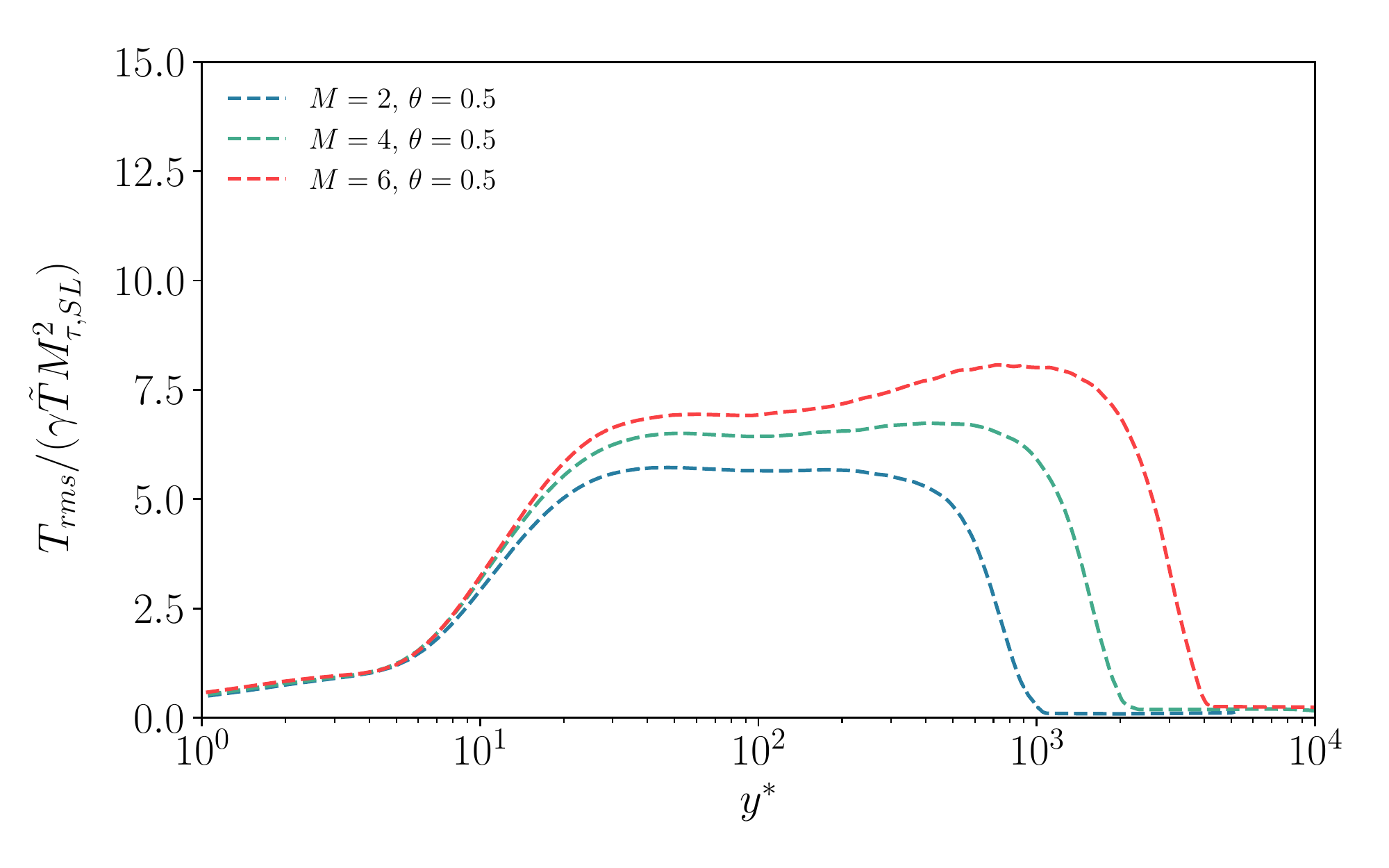}}
	\subfigure[PARAMETRI-2][$\mathit{\Theta}=0.5 $]{\includegraphics[width=0.48\textwidth]{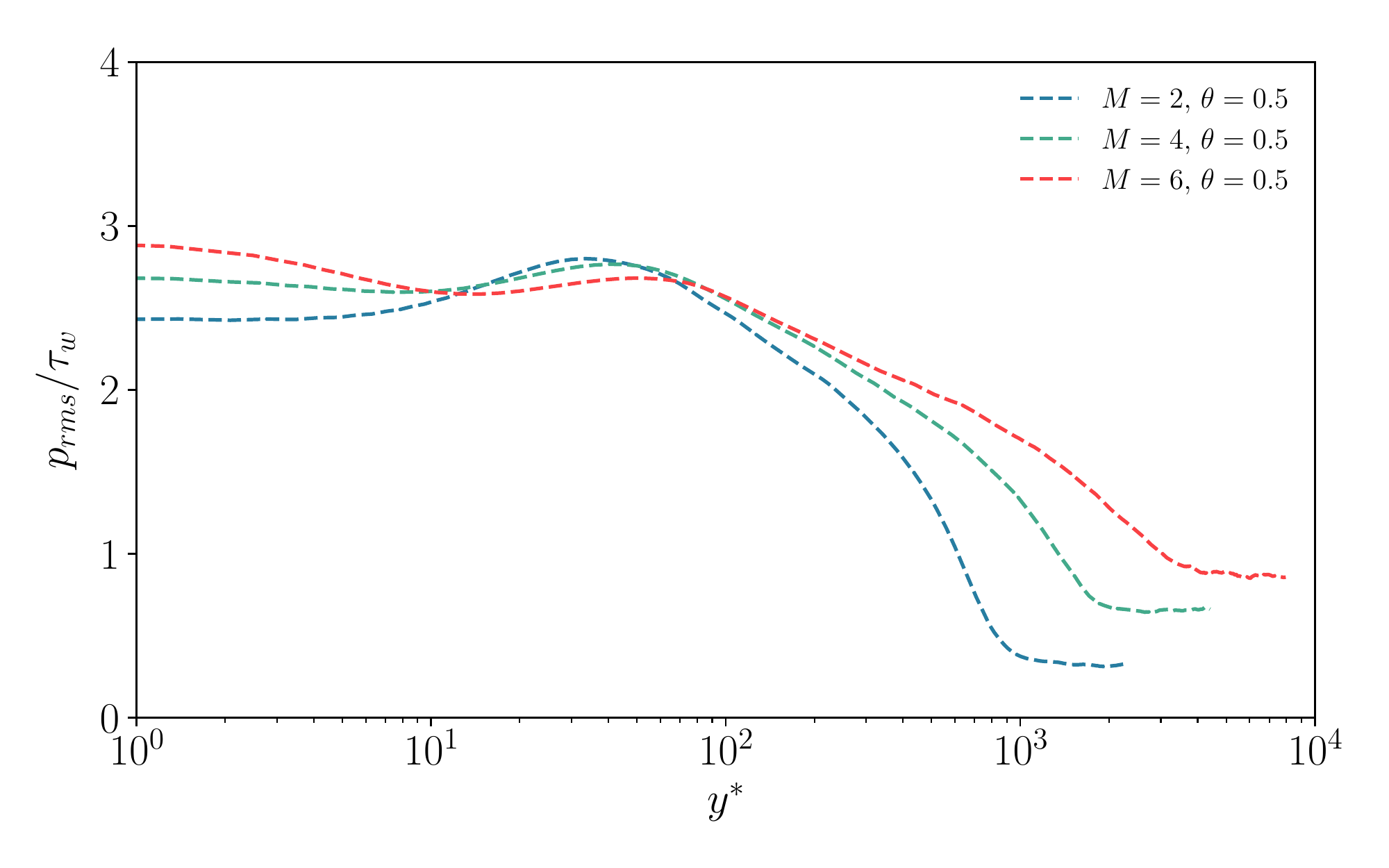}}\\
	\subfigure[PARAMETRI-2][$\mathit{\Theta}=0.75$]{\includegraphics[width=0.48\textwidth]{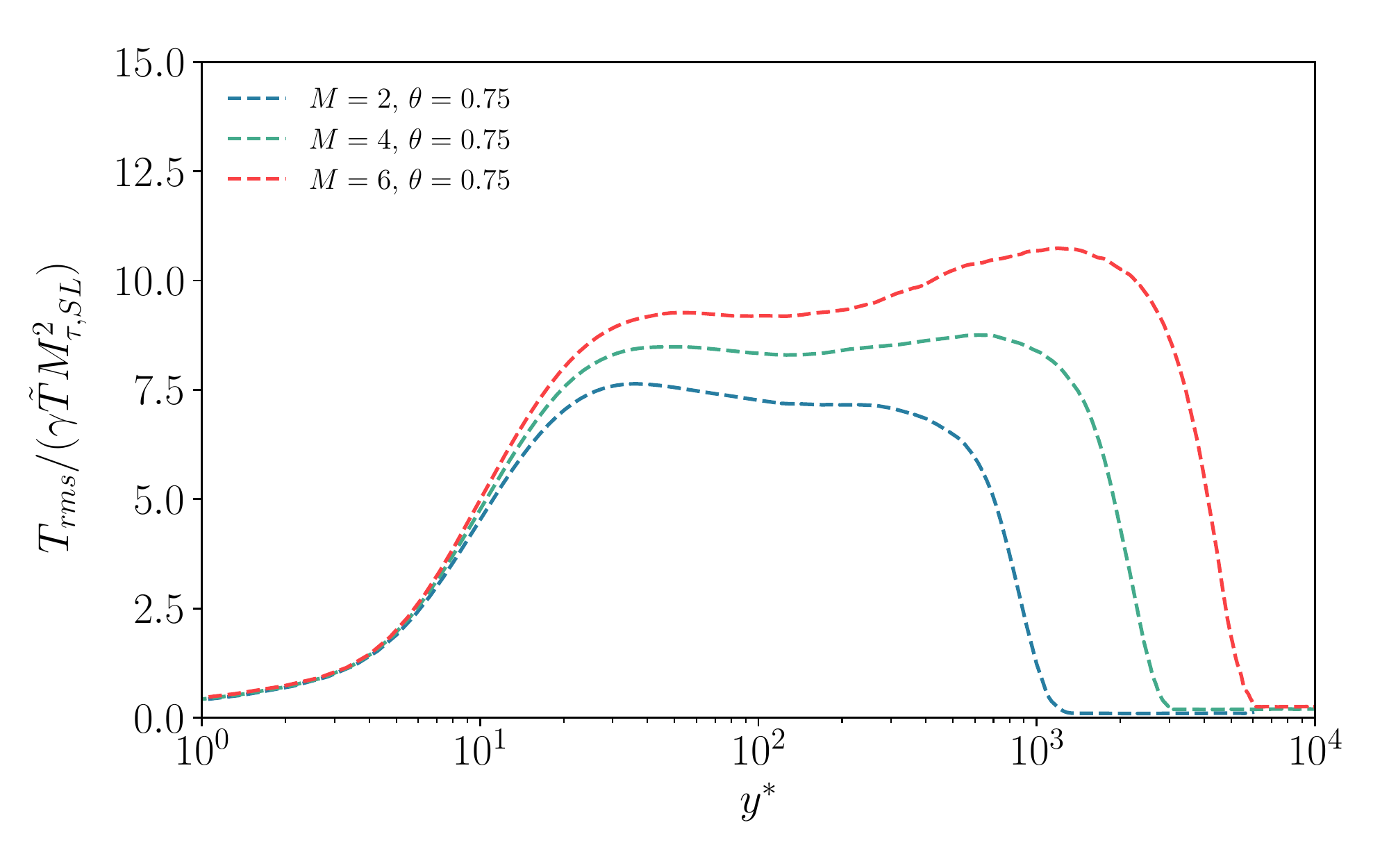}}
	\subfigure[PARAMETRI-2][$\mathit{\Theta}=0.75$]{\includegraphics[width=0.48\textwidth]{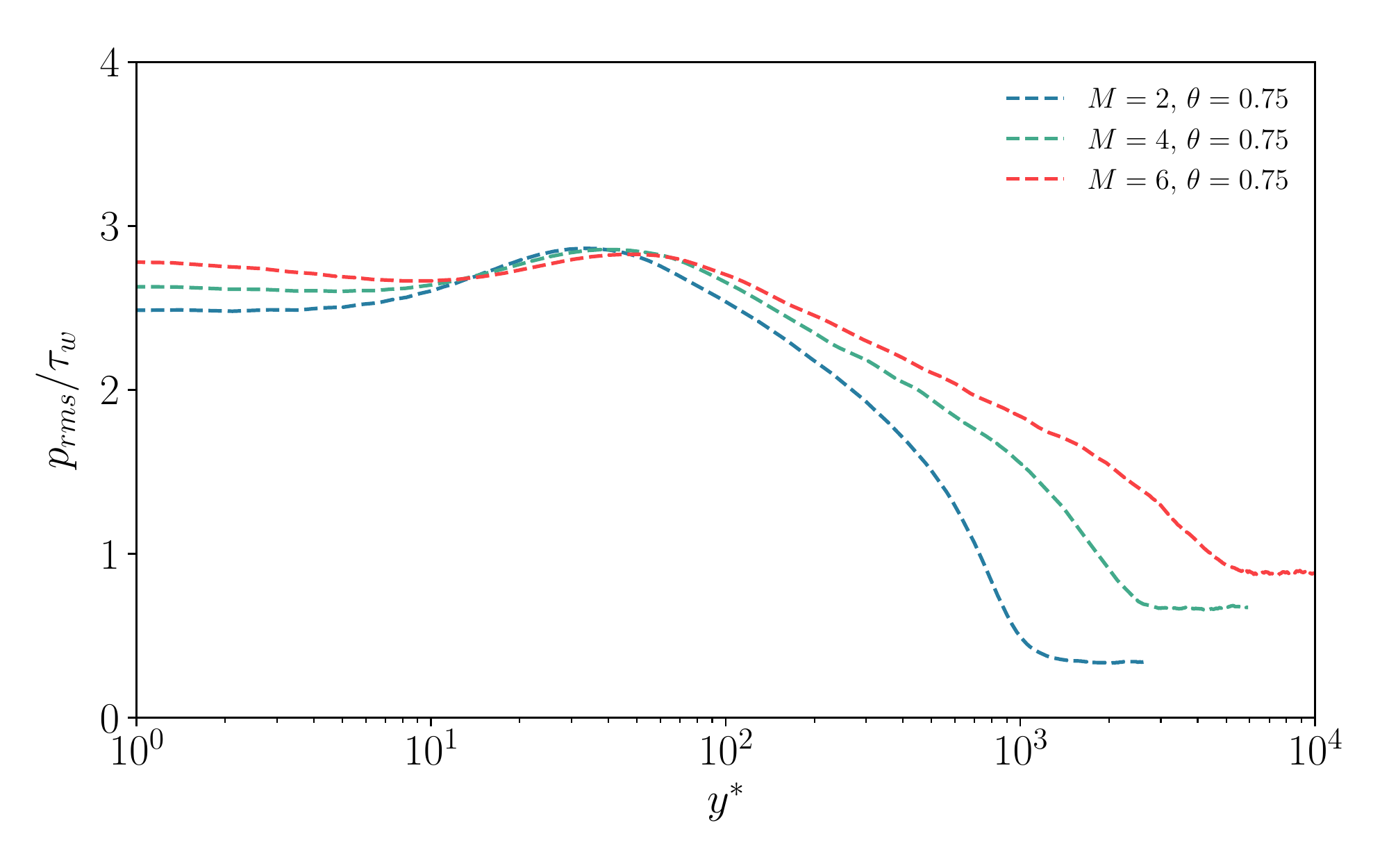}}\\
	\subfigure[PARAMETRI-2][$\mathit{\Theta}=1.0 $]{\includegraphics[width=0.48\textwidth]{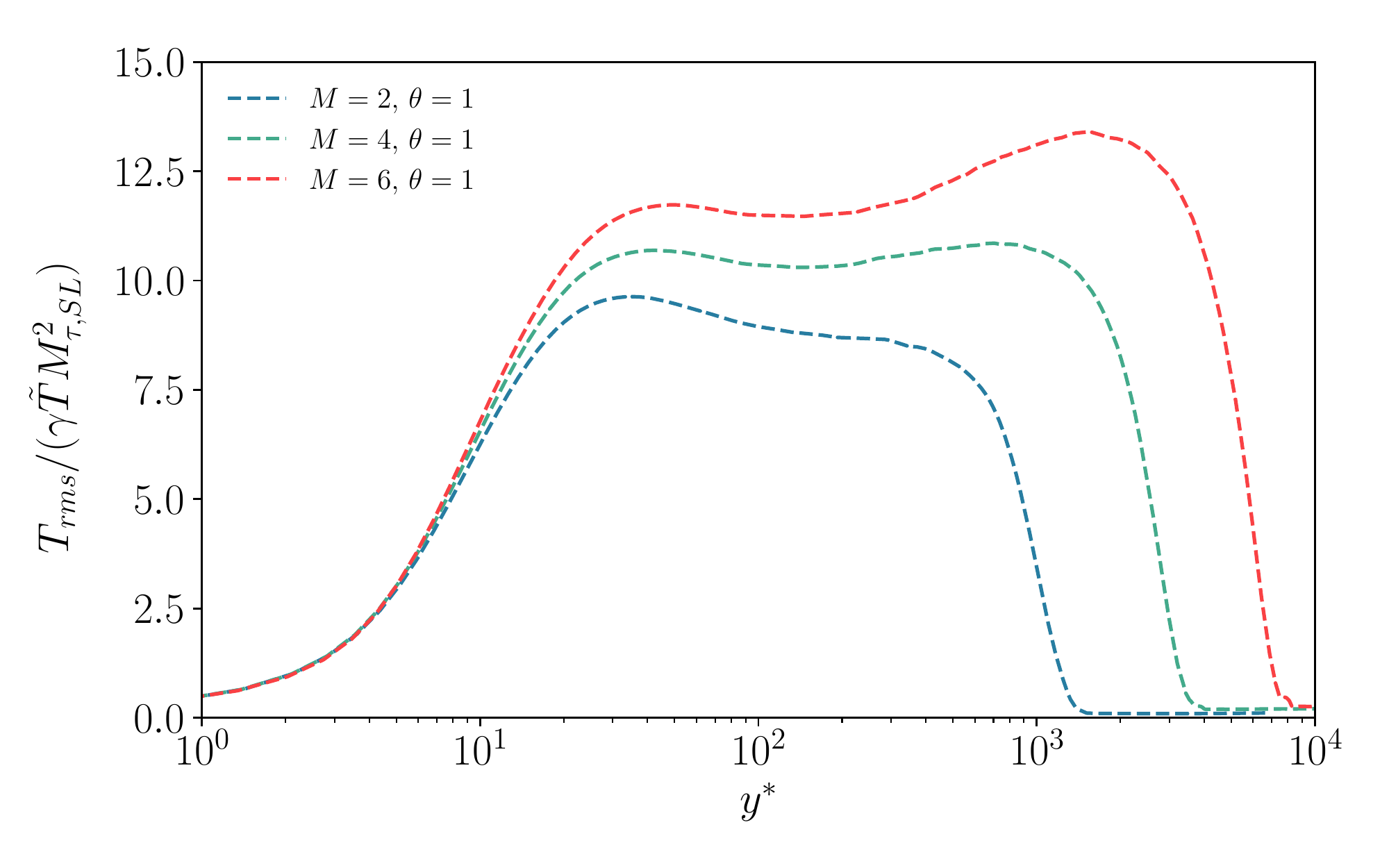}}
	\subfigure[PARAMETRI-2][$\mathit{\Theta}=1.0 $]{\includegraphics[width=0.48\textwidth]{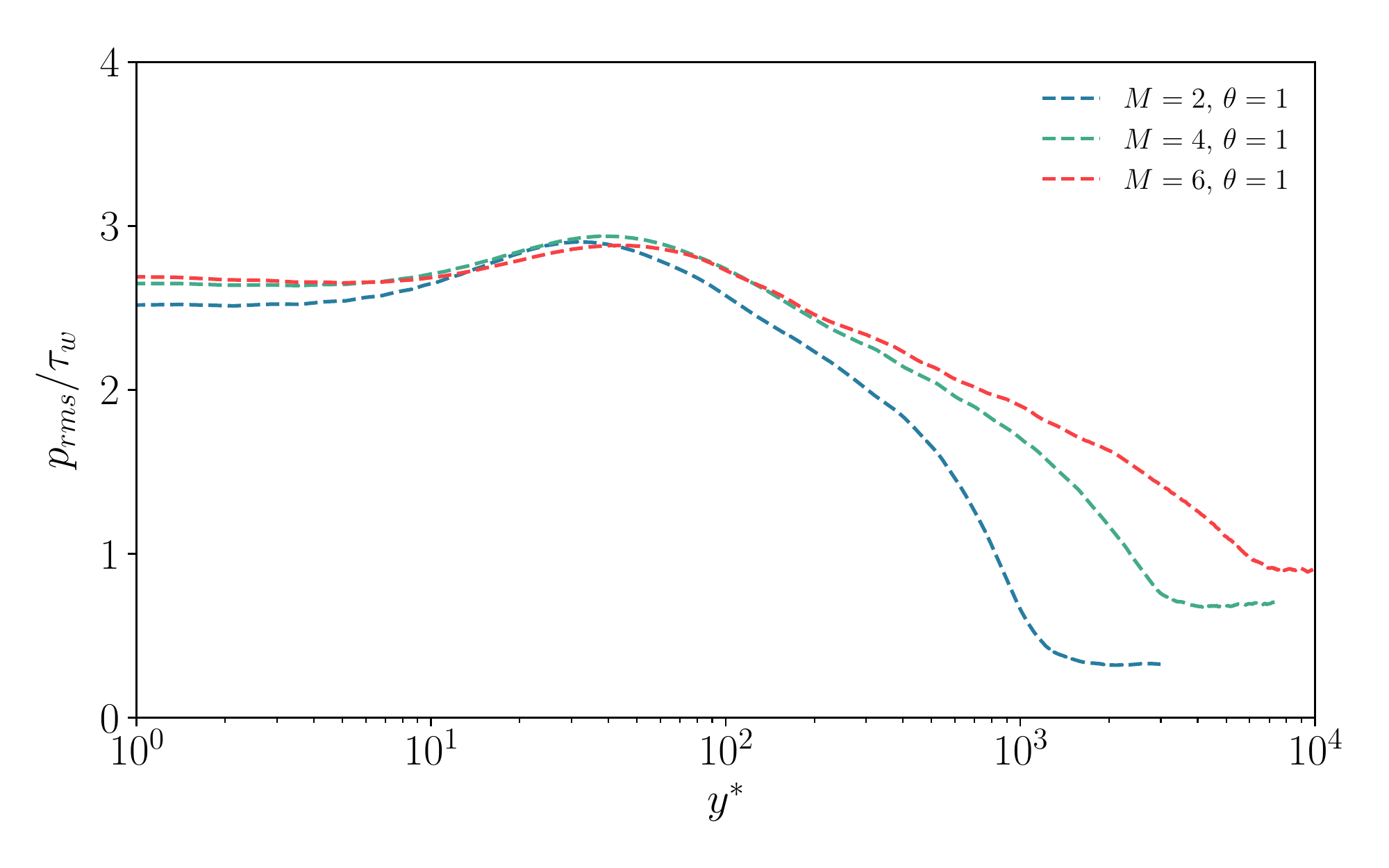}}
	\vspace{-0.2cm}
	\caption{Profiles of RMS temperature (left) and pressure (right) in semilocal scaling. Here, different Mach number $M_{\infty}$ are compared at a given diabatic parameter $\mathit{\Theta}$. \label{fig:thermo_t}}
\end{figure}

\begin{figure}  
	\centering
	\subfigure[PARAMETRI-1][$M_{\infty}=2$]{\includegraphics[width=0.45\textwidth]{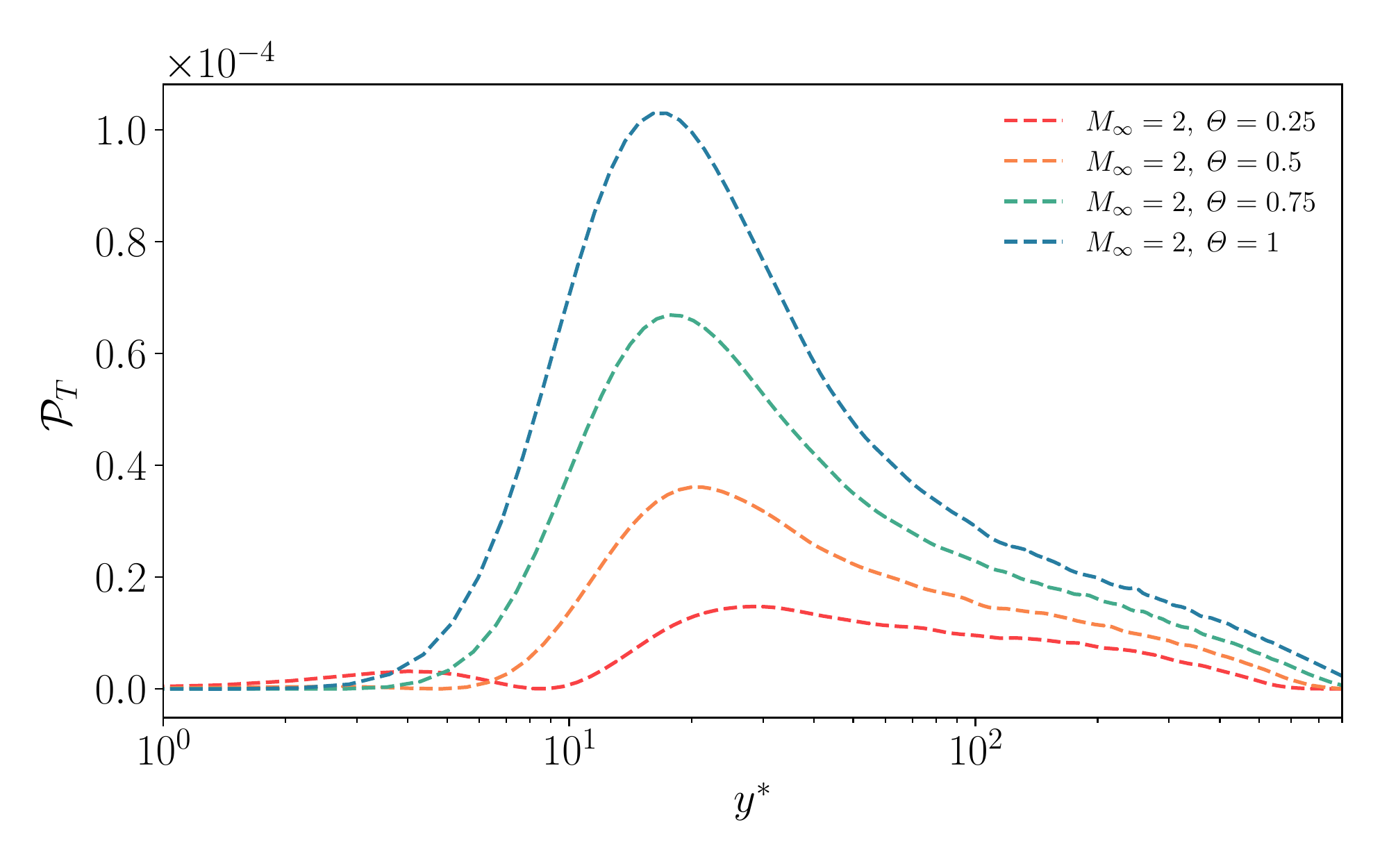}}
	\subfigure[PARAMETRI-2][$M_{\infty}=4$]{\includegraphics[width=0.45\textwidth]{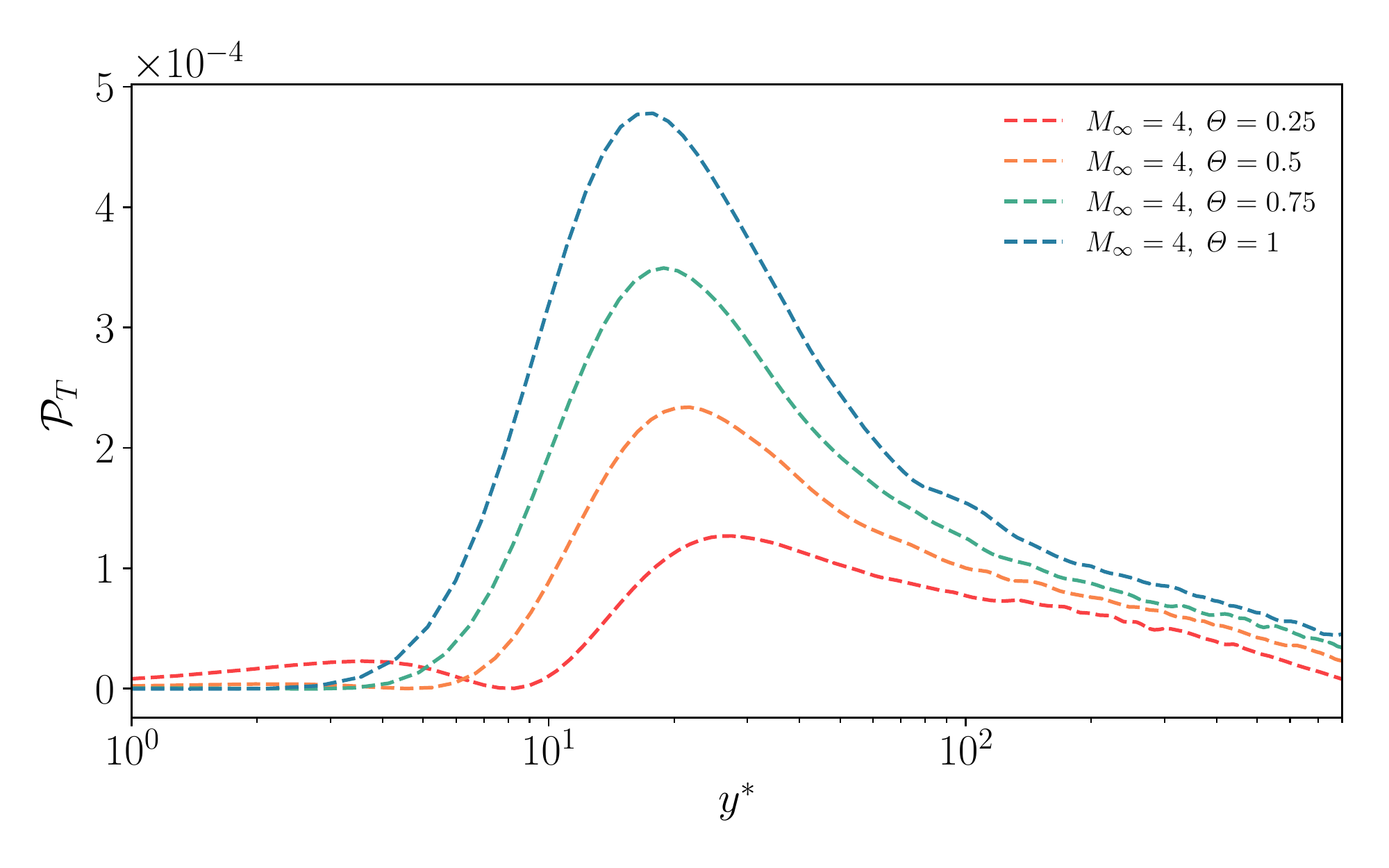}}\\
	\subfigure[PARAMETRI-2][$M_{\infty}=6$]{\includegraphics[width=0.45\textwidth]{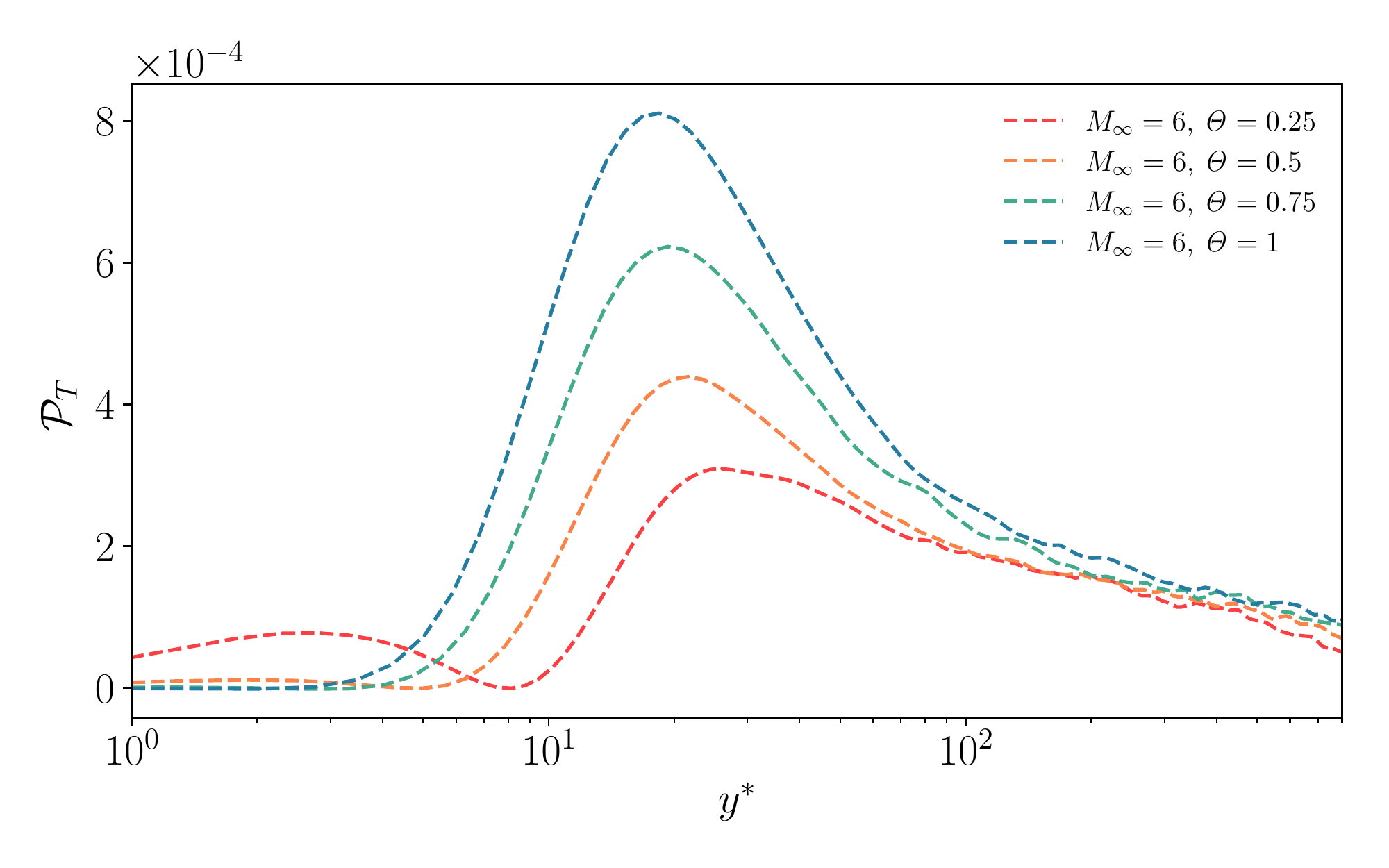}}
	\subfigure[PARAMETRI-2][$M_{\infty}=6$ \label{fig:comp}]{\includegraphics[width=0.45\textwidth]{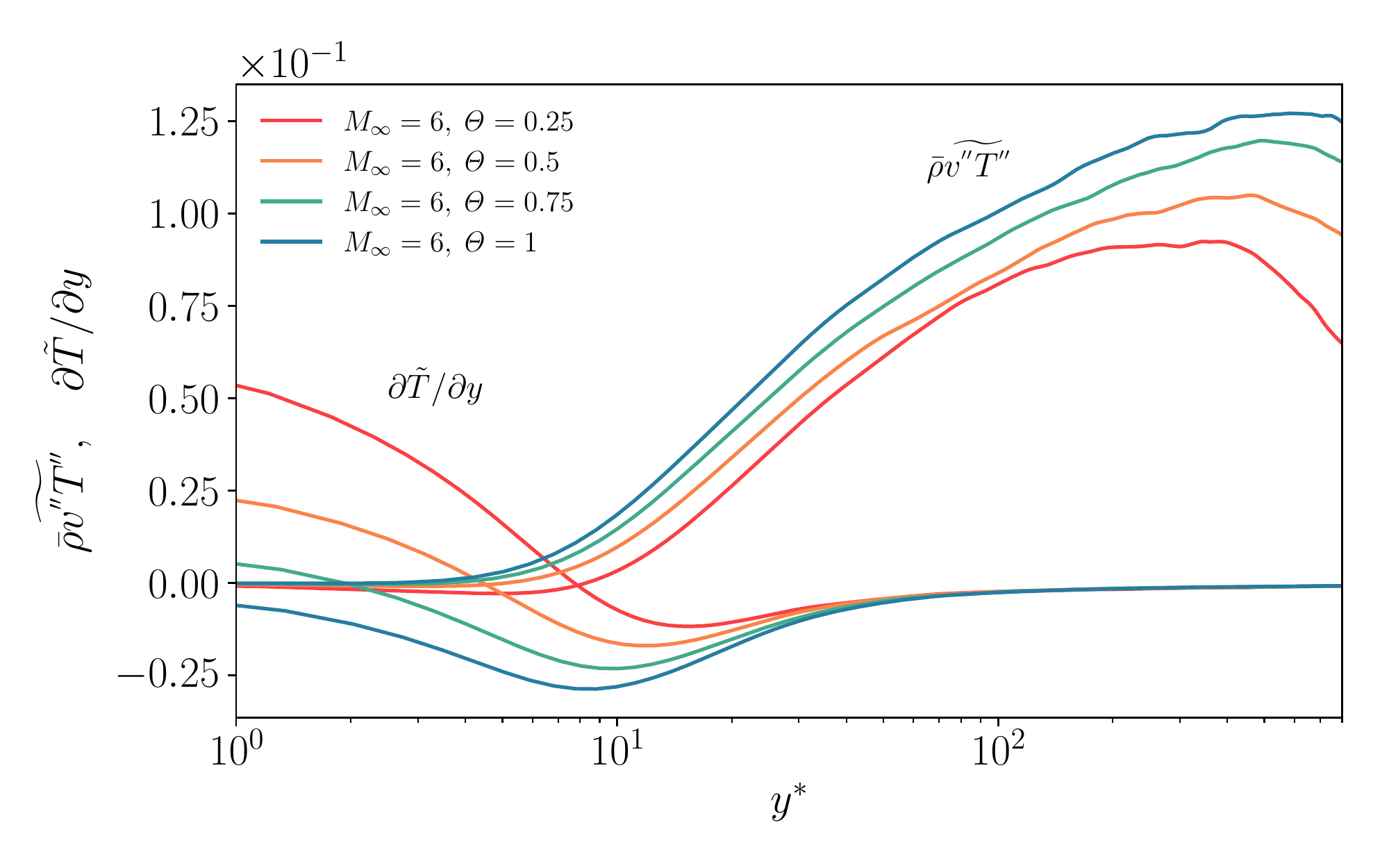}}
	\vspace{-0.2cm}
  \caption{Panels (a$\,\to\,$c): Production of temperature variance $\mathcal{P}_T$ as function of $y^*$ and scaled by $ \bar{\rho} u_{\tau, SL} \widetilde{T}^2 / \delta_{\nu, SL}$. Here, different wall-cooling conditions are compared for each Mach number. Panel (d): Turbulent $\bar{\rho} \widetilde{v^{''} T^{''}} $ and conductive $\partial \widetilde{T} / \partial y$ heat transfer terms in the thermal production. Here, different wall-cooling conditions are compared for case $M_{\infty}=6$.  \label{fig:prod1}}
\end{figure}

Important insights into the respective roles of Mach number and wall-cooling can also be attained by looking at root-mean-square profiles of temperature and pressure fluctuations shown in figures \ref{fig:thermo_m} and \ref{fig:thermo_t}. The semilocal scaling is used to better account for fluid property variations across the boundary layer and rms quantities are scaled accordingly. In particular, rms profiles of pressure are scaled with the wall shear stress $\tau_w$, while the resulting scaling for temperature is obtained using the ideal gas law $P=R \rho T$:
\begin{equation}
\frac{\tau_w}{R \bar{\rho}}=\frac{\bar{\rho} u_{\tau, S L}^2}{R \bar{\rho}}=\frac{u_{\tau, SL}^2}{R}=\gamma \bar{T} \frac{u_{\tau, S L}^2}{R \gamma \bar{T}}=\gamma \bar{T} M_{\tau, S L}^2
\end{equation}
being $u_{\tau, S L}= \sqrt{\tau_w/\bar{\rho}}$ the semilocal friction velocity and $M_{\tau, SL}=u_{\tau, SL} / \sqrt{\gamma R \bar{T}}$ the semilocal friction Mach number.
First, the effect of $\mathit{\Theta}$ at a given Mach number is presented in figure \ref{fig:thermo_m}. Considering the region starting from $y^*>10$, both temperature and pressure fluctuations show a reduction in intensity as $\mathit{\Theta}$ decreases, although more intense for the temperature. In particular, the suppression of temperature fluctuations by wall-cooling forms a plateau for the coldest case that is due to the great attenuation of the turbulent heat flux in the log-layer, consistently with \citet{fan2022energy}.
Around $y^* \approx 10$, the aforementioned attenuation of temperature fluctuations reaches its maximum for highly cooled cases ($\mathit{\Theta}=0.25$), which is the point where mean temperature gradients are close to zero.
In the near-wall region ($y^*<10$), strongly cooled cases exhibit a peculiar behaviour, which goes in direct contrast to the monotonic attenuation of adiabatic profiles.
In fact, in this region, there is an increase in the intensity of the temperature fluctuations that forms a local peak.
We attribute this phenomenon to the large increase of conductive heat flux close to the wall, which is able to overcome the expected attenuation of turbulent heat flux that concurs with the generation of thermal production (see figure \ref{fig:comp}). This is due to the large increase of near-wall temperature gradients that generate steeper mean profiles and for a wider region of $y^*$ values (before reaching the temperature peak), as visible in figure \ref{fig:tmean_yd} of section \S \ref{sec:mean}. The increase in pressure fluctuations in this region is shared only by the high-Mach number case, showing that additional physical interpretations are needed on the distinct role of vorticity and acoustic modes, for which we remind to the recent study of \citet{wan2022wall}.
Figure \ref{fig:thermo_t} shows the effect of Mach number at a given $\mathit{\Theta}$. Here, temperature fluctuation profiles are very similar up to roughly $y^*<15$, while the main differences are present in the outer layer, where at higher Mach numbers a peak is formed. This result indicates that $\mathit{\Theta}$ is an adequate parameter to recover the same general behaviour with respect to wall-cooling at different Mach numbers, as noted by \citet{wenzel2022influences} (we remark the similarity between $\mathit{\Theta}$ and Eckert number). 
As discussed for velocity fluctuations in section \S \ref{sec:velfluc}, we note the tendency of compressibility to increase the separation of scales (figure \ref{fig:thermo_m}), while the opposite is true for enhanced wall-cooling (figure \ref{fig:thermo_t}). This effect would be greatly reduced if profiles were compared at the same $Re_{\tau}^*$, which incorporates these effects (not shown).
Pressure fluctuations exhibit a good collapse at the peak location around $y^*\approx 30$, in accordance with \citet{wan2022wall}, but do not share the collapse between profiles in the near-wall region noted for temperatures. 

Further insights on the sources of production of temperature fluctuations, which are highly influenced by wall-cooling, can be gained by considering the temperature variance budget $K_T=\widetilde{T^{''2}}$, which can be written as (ref. \citet{gatski2013compressibility}):
\begin{equation}
\begin{aligned}
\bar{\rho} \frac{D K_T}{D t}=&-\bar{\rho} \widetilde{u_k^{\prime \prime} T^{\prime \prime}} \frac{\partial \widetilde{T}}{\partial x_k}-\frac{\partial}{\partial x_k}\left(\frac{\bar{\rho} \widetilde{u_k^{\prime \prime} T^{\prime \prime2}}}{2}\right)+\gamma \overline{T^{\prime \prime}} \frac{\partial}{\partial x_k}\left(\frac{\bar{k}_T}{c_p} \frac{\partial \bar{T}}{\partial x_k}\right) \\
&+\bar{\rho} D_T-\bar{\rho} \varepsilon_T+\bar{\rho} C_T,
\end{aligned}
\end{equation}
where the terms on the right-hand side are in order of appearance: thermal production, turbulent velocity transport, mean thermal conduction, thermal diffusion, thermal dissipation rate, and contributions due to pressure–dilatation and viscous dissipation, respectively. Details on the composition of each term can be found on \citet{gatski2013compressibility}.
Here, we analyse the thermal production term, which acts in a similar way to turbulent production, transferring internal energy from the mean field to the fluctuating one \citet{fan2022energy}.
For turbulent boundary layers, its wall-normal component is the main contributor, especially as we approach the wall, which we refer to as $ \mathcal{P}_T $:
\begin{equation}\label{eq:thermprod}
  \mathcal{P}_T = -\bar{\rho} \widetilde{v^{''} T^{''}} \frac{\partial \widetilde{T}}{\partial y}
\end{equation}
 Here, two terms concur to the heat exchange between different flow regions by two distinct processes: the first part $\bar{\rho} \widetilde{v^{''} T^{''}} $ is dominated by turbulence with the velocity-temperature fluctuations correlation, while $\partial \widetilde{T} / \partial y$ represent the conductive part and is related to the mean temperature gradient. 
Profiles of $\mathcal{P}_T$ are reported in figures \ref{fig:prod1}, showing the effect of wall-cooling at different Mach numbers. Similarly to temperature fluctuations, cold profiles behave differently before and after $y^*\approx 10$, where mean temperature gradients change after the peaks. While adiabatic profiles monotonically rise from zero to a clear peak at around $y^*\approx 15$, proving their coupling with velocity fluctuations, cold cases progressively exhibit a reduction and outward shift of the main peak, with the creation of another peak in the viscous sub-layer. 
An insight to understand this process, which is more apparent at $M_{\infty}=6$, can be gained by analysing the individual behaviour of turbulent and convective heat exchange terms in thermal production \citep{fan2022energy}, which are shown in figure \ref{fig:comp}.
Here, it can be seen that while the turbulent term is significantly far from the wall with reduced intensity for cold cases, convective heat exchange dominates the near-wall region as wall-cooling increases. In this region, even though $ \bar{\rho}\widetilde{v^{''} T^{''}} $ is close to zero for all cases, the temperature gradient raises considerably for $\mathit{\Theta}=0.25$ which result in a non-zero product that is visible in plots of thermal production. 
Thus, the formation of a peak of thermal fluctuation production in the viscous sublayer is promoted. 
The vanishing mean temperature gradient in the buffer layer reduces the production of temperature fluctuations and promotes a decorrelation with velocity fluctuations, as discussed in the previous sections.

\begin{figure}  
	\centering
	\subfigure[PARAMETRI-1][$M_{\infty}=6, \mathit{\Theta}=0.25$\label{fig:scatter025}]{\includegraphics[width=0.48\textwidth]{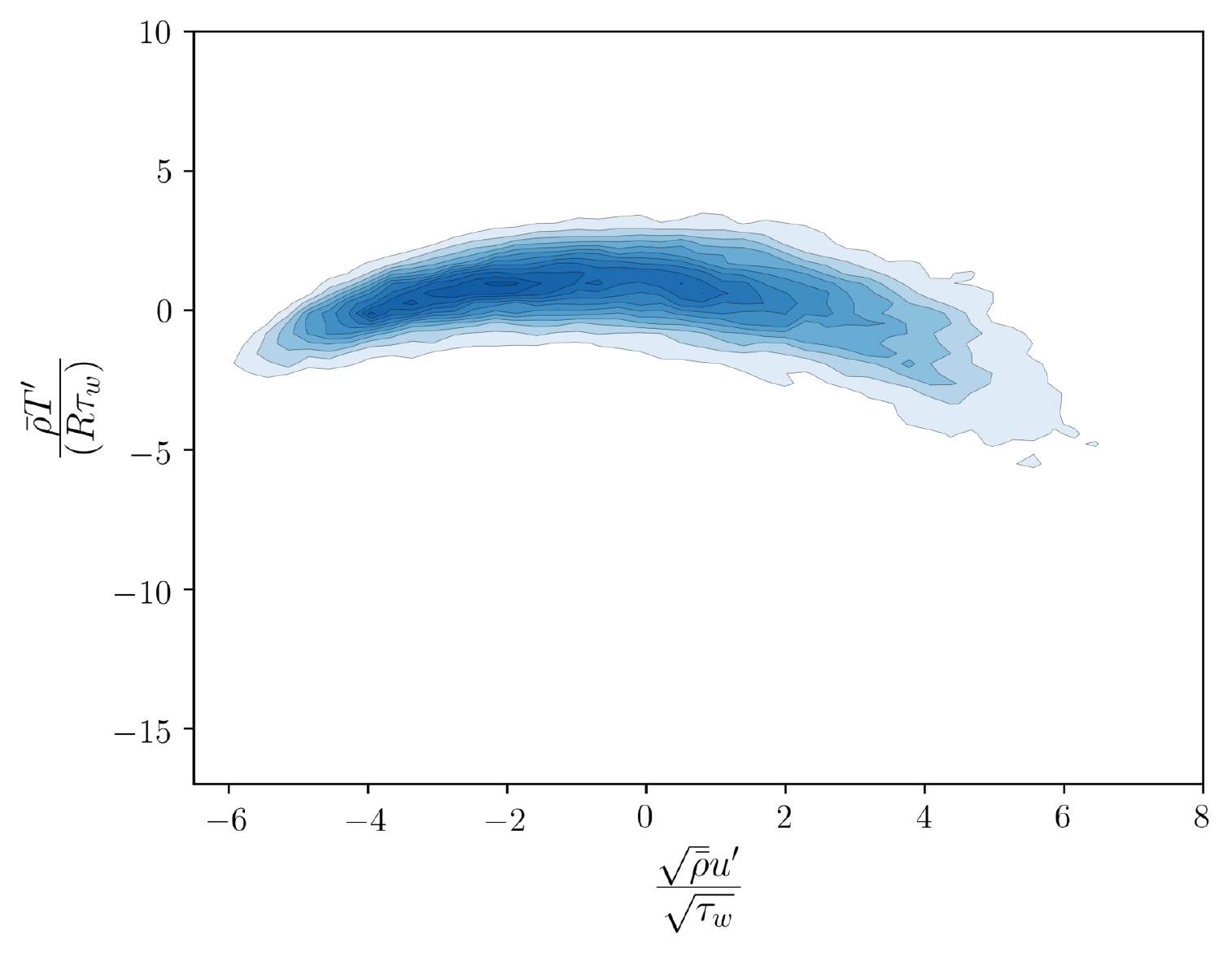}}
	\subfigure[PARAMETRI-1][$M_{\infty}=6, \mathit{\Theta}=1.0 $\label{fig:scatter100}]{\includegraphics[width=0.48\textwidth]{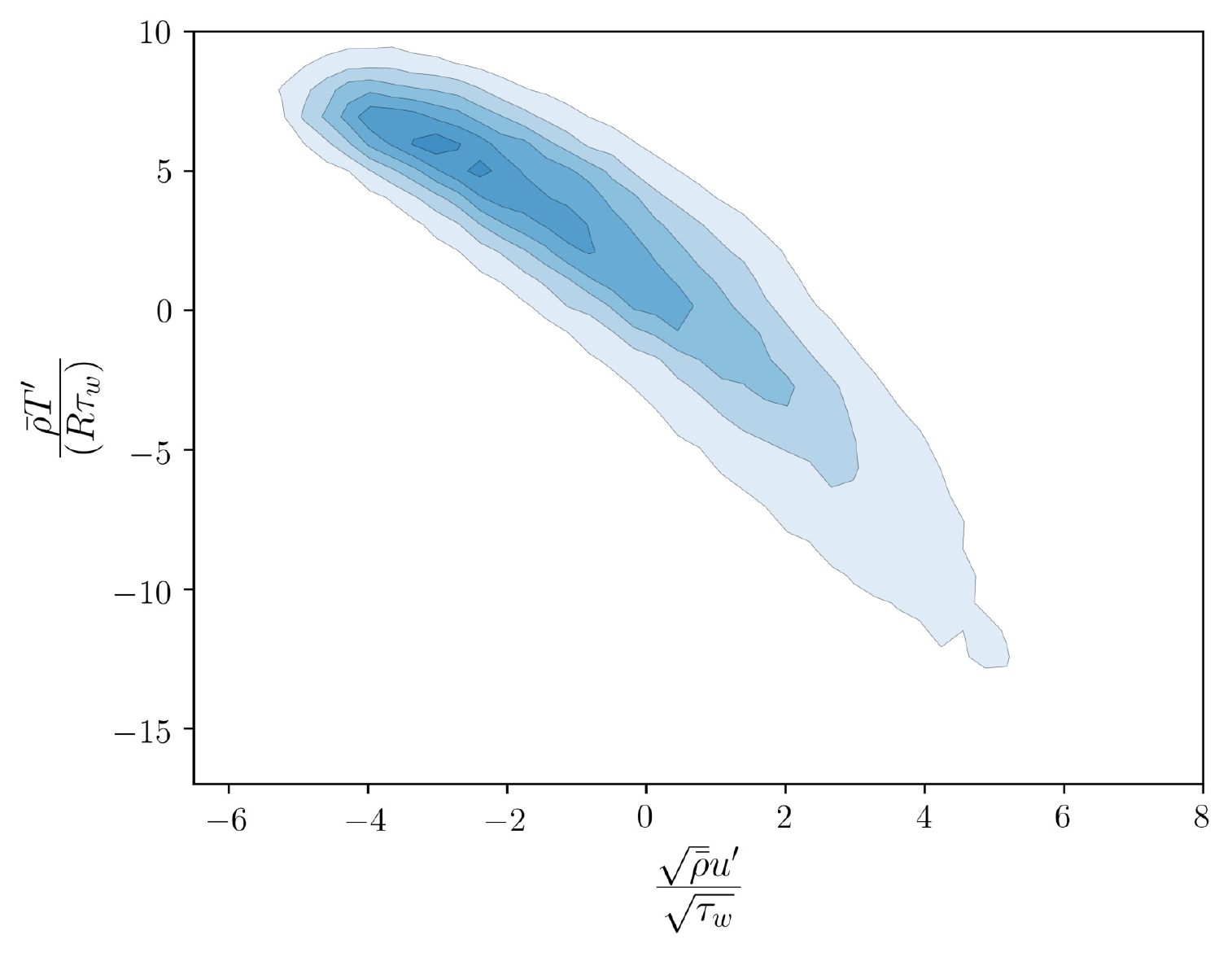}}
	\vspace{-0.2cm}
  \caption{Scatter plot of $\bar{\rho} T^{''}/ \tau_w$ vs $\sqrt{\bar{\rho}} u^{''}/\sqrt{\tau_w}$ . Here, only the cases M6T025 and M6T100 are shown. Data were collected in the same plane shown in figure \ref{fig:slice_xz} ($y^*\approx10$).\label{fig:scatter}}
\end{figure}

At this point, it is possible to reconsider the qualitative results presented in figure \ref{fig:slice_xz} in a more quantitative way. Wall parallel slices of velocity and temperature fluctuations were taken approximately at $y^*=10$, where $\partial \tilde{T}/\partial y \approx 0$ for extremely cold cases.
It is now apparent that the decorrelation between $\bar{\rho} T^{''}/ \tau_w$ and $\sqrt{\bar{\rho}} u^{''}/\sqrt{\tau_w}$ can be explained with the interplay of the mean temperature gradient and $ \bar{\rho} \widetilde{v^{''} T^{''}} $, which entirely damps the production of temperature fluctuations. 
This is also visible in figure \ref{fig:scatter}, which shows the joint probability density function between velocity and temperature fluctuations only for extreme cases at $M_{\infty}=6$ (other cases are similar).
Here, a direct contrast is present between figures \ref{fig:scatter025} and \ref{fig:scatter100}).
While the latter (M6T100) shows a good correlation between the two fields, supporting their similarity, the former (M6T025) shows a strong decorrelation, especially when velocity fluctuations are negative, which is due to the influence of wall-cooling.

\section{Conclusions}
In this paper, we have presented a systematic study on the effect of the Mach number and wall-cooling on zero-pressure-gradient TBLs using direct numerical simulations.
A total of 12 computations have been carried out spanning three Mach numbers and four values of the diabatic parameter $\mathit{\Theta}$, while the friction Reynolds number has been kept constant.
In this parameter space, we put emphasis on the choice of the wall-cooling parameter $\mathit{\Theta}$, first proposed by \citet{zhang2014generalized}, which can better incorporate the indirect effects of the Mach number on wall-cooling, yielding the same integral behaviour between different cases.
It is worth noting that $\mathit{\Theta}$ can be directly related to the Eckert number $Ec$, whose relevance has recently been discussed by \citet{wenzel2022influences}.
These parameters show an improved ability to account for the wall-cooling effect at different Mach numbers with respect to the
more classically used wall-to-recovery temperature ratio $T_w/T_r$, which has been shown to produce vastly different effects of wall temperature on the flow dynamics in the near-wall region across different Mach numbers (e.g. \citet{cogo2022direct,wan2022wall}).

A summary of the most important remarks is presented below.

\begin{enumerate}
    \item \ The instantaneous flow organisation of temperature fluctuations near the wall, which for adiabatic cases is clearly discernible with the presence of near-wall elongated streaks highly correlated to streamwise velocity, breaks down as the wall temperature is progressively lowered, showing an isotropic behaviour for extremely cold cases without organised patterns. Nevertheless, a similar flow organisation is attained at different Mach numbers when $\mathit{\Theta}$ is fixed, a first sign of the aptness of this parameter to yield the same wall-cooling effects across different $M_{\infty}$. \label{item1} \\
    \item \ The recent compressibility transformations of \citet{volpiani2020data} and \citet{griffin2021velocity} correctly collapse all mean velocity profiles of our database to the incompressible laws of the wall. 
    Similarly, \citet{zhang2014generalized} mean velocity-temperature relations are able to capture non-adiabatic and compressibility effects in an excellent manner. 
    When this relation is approximated with the computed mean value of the Reynolds analogy factor $s=0.78\pm0.03$ (which is close to the fit of \citet{zhang2014generalized}), an excellent estimate  is recovered, with maximum errors of $5\%$ from the DNS data. \label{item2} \\
    
    \item \ As the Mach number increases, we observe an increased separation between large and small scales in the outer layer measured by the ratio $L/ \eta$, which is mainly regulated by the strong reduction of the Kolmogorov length $\eta$, and only weakly by a growth of the largest scale $L$. This effect can be effectively described by the growth of the semilocal friction Reynolds number $Re_{\tau}^*$, even though the resulting flow dynamics is different from a pure increase of the friction Reynolds number $Re_{\tau}$, the latter also leading to an increase of the inner-outer scale separation $L^+$, feeding outer layer motions.
    In the near-wall region, compressibility enables a less efficient redistribution of turbulent kinetic energy, which results in a promotion of the peak of the streamwise velocity component and a decrease of the others. \label{item3}\\

    \item \ The enhancement of wall-cooling appears as a reduction of the large-small scale separation in the outer layer (as opposed to the effect of Mach number), which is mainly due to an increase of the Kolmogorov length scale $\eta$ that occurs throughout the whole BL thickness.
    Lower wall temperatures force the rise of the mean temperature peak, inducing a stratification of flow properties localised around the buffer layer.
    This effect is visible as an apparent promotion of compressibility, since velocity fluctuations are enhanced in the streamwise direction, while the other components are  damped. \label{item4}\\
    
    \item \ In the near-wall region, a dominant effect of wall-cooling is present in the RMS temperature profiles and TKE budget, while the Mach number exerts its influence mainly through the buffer and log layers. 
    When the diabatic parameter $\mathit{\Theta}$ is kept constant, the RMS temperature profiles at different $M_{\infty}$ collapse into each other near the wall, displaying a similar wall-cooling effect. \label{item5} \\
    
    \item \ For extremely cold cases (in our database $\mathit{\Theta}=0.25$), the effect of wall-cooling is so marked that temperature fluctuations are massively damped at the point where the mean temperature gradient is zero (thus thermal production is also zero), and a second (minor) peak arises in the viscous sublayer.
    This phenomenon completely decorrelates velocity and temperature fields in the near-wall region, and is more pronounced at high Mach numbers.
    The different behaviour of thermal production for cold cases can be explained by looking at the mean temperature gradient, which persists with a positive value for a wider wall-normal region (before reaching the mean temperature peak at $y^*_{peak}$), and with a much stronger intensity than for adiabatic cases. \label{item6} \\

\end{enumerate}

It should be noted that points (\ref{item1}), (\ref{item5}), and (\ref{item6}) are closely connected since they are representative of the same physical phenomenon (aerodynamic heating), which is a key point in this study.
Moreover, points (\ref{item3}) and (\ref{item4}) consist in novel observations on the modulation of scales separation in compressible flows that require future studies.
In the authors' opinion, the observed decorrelation between velocity and temperature fields due to wall-cooling poses a major challenge for the development of simplified wall models, since any similarity between the two fields is lost.
In this context, we believe that the present database and the relative discussion can help to gain physical insights on the important mechanisms that can be captured by a physics-informed model for Reynolds-averaged Navier-Stokes (RANS) or large-eddy simulation (LES), which is certainly of great interest to the scientific community given the recent advancements on the topic (e.g. \citet{kawai2010dynamic,Yang2018}).

\clearpage


\bibliographystyle{unsrtnat}
\bibliography{hyp_bib}


\end{document}